\documentclass[aps, prd, superscriptaddress, nofootinbib, preprintnumbers]{revtex4}
\usepackage{graphicx}
\usepackage{color}
\usepackage{amsmath,amssymb}
\usepackage{array,multirow}

\newcommand{\TeV}{\text{TeV}}
\newcommand{\GeV}{\text{GeV}}

\newcommand{\gtwo}{I\kern-.1em I\,}
\newcommand{\be}{\begin{equation}}
\newcommand{\ee}{\end{equation}}
\newcommand{\beq}{\begin{eqnarray}}
\newcommand{\eeq}{\end{eqnarray}}
\newcommand{\bpm}{\begin{pmatrix}}
\newcommand{\epm}{\end{pmatrix}}

\begin{document}

\title{Conformal Barrier and Hidden Local Symmetry Constraints: \\
Walking Technirhos in LHC Diboson Channels
}

\author{Hidenori S. Fukano}
\thanks{\tt fukano@kmi.nagoya-u.ac.jp}
      \affiliation{ Kobayashi-Maskawa Institute for the Origin of Particles and 
the Universe (KMI) \\ 
 Nagoya University, Nagoya 464-8602, Japan.}
\author{Shinya Matsuzaki}\thanks{\tt synya@hken.phys.nagoya-u.ac.jp}
      \affiliation{ Institute for Advanced Research, Nagoya University, Nagoya 464-8602, Japan.}
      \affiliation{ Department of Physics, Nagoya University, Nagoya 464-8602, Japan.} 
\author{Koji Terashi}
\thanks{\tt Koji.Terashi@cern.ch}
\affiliation{The University of Tokyo, International Center for Elementary Particle Physics and Department of Physics, 
7-3-1 Hongo, Bunkyo-ku, JP - Tokyo 113-0033, Japan.}     
\author{{Koichi Yamawaki}} \thanks{
      {\tt yamawaki@kmi.nagoya-u.ac.jp}}
      \affiliation{ Kobayashi-Maskawa Institute for the Origin of Particles and the Universe (KMI) \\ 
 Nagoya University, Nagoya 464-8602, Japan.}

\date{\today}

\begin{abstract} 
We expand the previous analyses of the conformal barrier on the walking technirho for the 2 TeV diboson excesses 
reported by the ATLAS collaboration, 
with a special emphasis on the hidden local symmetry (HLS) constraints.  
We first show that the Standard Model (SM) Higgs Lagrangian is equivalent to the scale-invariant nonlinear chiral Lagrangian, 
which is further gauge equivalent to the scale-invariant HLS model, with
the scale symmetry realized nonlinearly via SM Higgs as a (pseudo-) dilaton. 
The scale symmetry forbids the new vector boson    
decay to the 125 GeV Higgs plus $W/Z$ boson,  
in sharp contrast to the conventional ``equivalence theorem'' which is invalidated by the conformality.   
The HLS forbids mixing between the iso-triplet technirho's, $\rho_{\Pi}$ and $\rho_{P}$, 
of the one-family walking technicolor (with four doublets $N_D=N_F/2=4$), 
which, without the HLS, would be generated when switching on the standard model gauging. 
We also present updated analyses of the walking technrho's for the diboson excesses  
by fully incorporating the constraints from the conformal barrier and the HLS 
as well as possible higher order effects: still characteristic of the one-family walking technirho is 
its smallness of the decay width,  
roughly of order $\Gamma/M_\rho \sim [3/N_C\times 1/N_D] \times [\Gamma/M_\rho]_{\rm QCD} \, \simeq 70\, {\rm GeV}/2\,{\rm TeV}$ 
($N_D= N_C=4$), in perfect agreement with the expected diboson resonance with $\Gamma<100\, {\rm GeV}$.   
The model is so sharply distinguishable from other massive spin 1 models 
without the conformality and HLS that it is clearly testable at the LHC Run II. 
If the 2 TeV boson decay to $WH/ZH$ is not observed in 
the ongoing Run II, then the conformality is operative on the 125 GeV Higgs, strongly suggesting that  
the 2 TeV excess events are responsible for the walking technirhos and the 125 GeV Higgs is the technidilaton. 
\end{abstract}

\maketitle

\section{Introduction}

The Higgs boson was discovered at LHC, which is so far 
consistent with the standard model (SM)~\cite{Aad:2012tfa}, having no obvious  hints for the new physics beyond the SM, as far as the Higgs decays are concerned.  However, the origin of the mass, particularly of the Higgs itself, is a mere input parameter of  the SM to be revealed 
by new physics at deeper level beyond SM.

One of the candidates for such new physics towards the origin of mass is the walking technicolor, which has an approximate scale symmetry and as such 
produces
a large anomalous dimension $\gamma_m\simeq 1$ and a {\it light composite Higgs as the technidilaton}, a pseudo Nambu-Goldstone (NG) boson of the
approximate scale symmetry~\cite{Yamawaki:1985zg}, in sharp contrast to the original technicolor as  a QCD scale up \cite{Weinberg:1975gm} which was already excluded long ago by the large flavor-changing neutral currents and large $S$ parameter, as well as most recently and dramatically  by the absence of the light 125 GeV Higgs.

It was in fact shown~\cite{Matsuzaki:2012gd,Matsuzaki:2012xx,Matsuzaki:2015sya} that the technidilaton in the walking technicolor of the one-family model \cite{Farhi:1979zx}, with four weak-doublets $N_D=N_F/2=4$ in the 
$SU(N_C)$ gauge
group, can be nicely fit to the current 125 GeV Higgs data for $N_C=4$. It was further shown \cite{Matsuzaki:2015sya} that the anti-Veneziano limit, $N_C \rightarrow \infty$ with $N_C\alpha$=const. 
and $N_F/N_C$=fixed $\gg 1$, yields the theory becoming walking with infrared conformality, in such a way that the technidilaton mass and couplings vanish in the limit. Numerically, the one-family model  
with $N_F=8$ and $N_C=4$ is already close to the anti-Veneziano limit 
picture  so to have a good approximate scale symmetry for the technidilaton becoming naturally light, as light as 125 GeV, and moreover its coupling  {\it even  weaker
than the SM Higgs} \cite{Matsuzaki:2015sya}, thus justifying the numerical agreement with the current LHC Higgs data. Recent lattice results in fact suggest that
the theory with $N_F=8$ and $N_C=3$ has walking signals with anomalous dimension $\gamma_m\simeq 1$~\cite{Aoki:2013xza}  and moreover has 
 a light flavor-singlet scalar bound state as a candidate for the technidilaton \cite{Aoki:2014oha} (There exists a light flavor-singlet scalar also in 
the case of $N_F=12$~\cite{Aoki:2013zsa}).     
 This is in sharp contrast to a folklore that the strongly coupled theory would not produce light weakly coupled composites, which is merely
a prejudice  based on the analogy with the QCD having no scale symmetry.

Crucial issue is that the walking technicolor will give us not just the Higgs but
a plenty of other bound states as 
new phenomena beyond the SM. Typical of such is the walking technirho, which is described by the effective theory based on the hidden local symmetry (HLS) model 
successful for the QCD rho meson \cite{Bando:1984ej,Bando:1987ym,Harada:2003jx}  so as to be made scale-invariant via nonlinear realization (``s-HLS'' model) \cite{Kurachi:2014qma}  in accord with the (spontaneously broken) scale symmetry of 
the underlying walking technicolor. Also used was a straightforward application of the loop expansion, the HLS chiral perturbation theory \cite{Tanabashi:1993np,Harada:2003jx} (usual chiral perturbation theory extended to incorporating the HLS gauge bosons),  to the present scale-invariant version, s-HLS model.
We thus may expect that the breakthrough may take place more drastically in
somewhat different channels than the Higgs decay processes.

In fact, the ATLAS collaboration~\cite{Aad:2015owa} has recently reported interesting
excesses about 2.5 sigma (at global significance)
at mass around 2 TeV in the diboson channels~\footnote{
Small excesses about $\sim 2$ sigma in the same mass region have been seen 
also in the CMS diboson analysis~\cite{Khachatryan:2014hpa}. }. 
No doubt this could be an outstanding signature of new physics, more dramatic show-up than the possible deviations of the Higgs modes from the SM.
Since it would definitely be the phenomenon in the TeV region, it should be deeply connected with the
long-standing mystery of mass, such as the naturalness and  the dynamical
origin of the  Higgs itself.

In the previous paper \cite{Fukano:2015hga} we showed that these excess events can 
easily be identified with the Drell-Yan produced walking technirho decaying into the diboson channel $WW,WZ$
based on a benchmark model of  the one-family walking technicolor model \cite{Kurachi:2014qma}. 
As far as the longitudinal $W/Z$ are concerned, there are only two parameters, $F_\rho$ and $g_{\rho\pi\pi}$, 
relevant to the processes, which successfully reproduced the ATLAS excess events 
in a way to satisfy all other current LHC experiments.  
In particular, because of the scale-invariant form of the couplings, 
the model has no technirho couplings to 125 GeV Higgs ($H$) (technidilaton)  
and hence no decay to $WH, ZH$, thus is 
free from the LHC constraints on these processes~\cite{Aad:2015yza,Khachatryan:2015bma}. 
Another characteristic feature  comes from  the very nature of the one-family walking technicolor with $N_D=N_F/2=N_C=4$, 
which naturally accounts for the  narrowness of the  reported width $\Gamma <100\, {\rm GeV}$, 
since the decay width is scaled as $\Gamma/M_\rho \simeq \frac{3}{N_C}\frac{1}{N_D} \times (\Gamma/M_\rho)|_{\rm QCD}\simeq (70 \, {\rm GeV})/(2\, {\rm TeV})$.

We further showed \cite{Fukano:2015uga} that  the absence of the walking technirho decays to $WH$ and $ZH$ 
is a generic feature of the (nonlinearly realized) scale symmetry,
what we called ``conformal barrier'', which is not just our model but the universal feature of the scale-symmetric massive-vector model.
This is in sharp contrast to other vector boson models on 
the market~\cite{Hisano:2015gna,
Franzosi:2015zra,
Cheung:2015nha,
Dobrescu:2015qna,Alves:2015mua,Gao:2015irw,Thamm:2015csa,Brehmer:2015cia,Cao:2015lia,Cacciapaglia:2015eea,Abe:2015jra,Heeck:2015qra,Abe:2015uaa,Carmona:2015xaa,Dobrescu:2015yba,Anchordoqui:2015uea,Bian:2015ota,
Low:2015uha,Terazawa:2015bsa,Arnan:2015csa,Dev:2015pga,Dobado:2015hha,Deppisch:2015cua,Aydemir:2015nfa,Dobado:2015nqa,Li:2015yya}
which, having no scale symmetry,  
yield the ratio of the vector boson ($V$) decay rates 
of almost one, $\Gamma(V \to WW/WZ)/\Gamma(V \to WH/ZH) \simeq 1$, 
according to the 
 ``equivalence theorem''~\footnote{
The ``equivalence theorem'' of this kind customarily implies a relation obtained only when
the usual Goldstone equivalence theorem 
for 
the processes of the new vector boson $V$ decay to $W_L$ and $Z_L$ is combined with 
an additional specific assumption 
that   
the Higgs boson $H$ is embedded  in 
the electroweak doublet $h$  in Eq. (\ref{hrepr}),  as $\hat{\sigma} =v + H$ 
together with the would-be Nambu-Goldstone bosons $\hat{\pi}_i$ (eaten by $W_L$  $Z_L$) as 
the chiral partner. This additional assumption is specific to the coordinates of the field components $(\hat{\sigma},\hat{\pi})$ and not
general,
in sharp contrast to the polar decomposition Eq.(\ref{Polar}).   
}, see e.g., \cite{Hisano:2015gna}. 
We in fact demonstrated that the ``equivalence theorem'' 
is realized in way incompatible with the scale symmetric limit.
We then proposed a novel way to identify the dynamical 
origin of the 125 GeV Higgs through checking the possible
decays of the 2 TeV new bosons:   
If the 2 TeV new bosons have no decays to the SM gauge bosons plus the 125
GeV Higgs,  then it is suggested that   the 125 GeV Higgs is a dilaton, 
pseudo NG boson of the spontaneously broken conformality/scale symmetry of some
underlying new physics. 
One such an explicit example of the underlying theory
is the walking technicolor~\cite{Yamawaki:1985zg}
where the 125 GeV Higgs 
and the new bosons 
have been successfully identified with the technidilaton~\cite{Matsuzaki:2012gd,Matsuzaki:2012xx}  
and the walking technirho~\cite{Fukano:2015hga}, respectively.

In this paper, 
we first show that the SM Higgs Lagrangian is nothing but a scale-invariant nonlinear chiral Lagrangian, 
which is further gauge equivalent to the scale-invariant HLS model, with
the scale symmetry realized nonlinearly via SM Higgs as a (pseudo-)dilaton. Thus the SM Higgs Lagrangian itself, when incorporating the vector mesons via HLS, 
forbids the HLS vector bosons to decay into the $W/Z$ plus the SM Higgs as a (pseudo-)dilaton. Conformal barrier is already operative for the SM Higgs, and so is the
walking technirho embedded into a chiral group $SU(N_F)_L \times SU(N_F)_, (N_F>2)$ larger than that of the SM Higgs. 
We further discuss in details the consequences of the conformal barrier on the LHC 2 TeV diboson excesses, 
including the higher order effects through mixing and transverse $W, Z$ effects,  
which were not considered before~\footnote{
In the one-family walking technicolor Ref.~\cite{Kurachi:2014qma} discussed the coupling and decay of the color-octet iso-singlet technirho 
($\rho^0_{\theta_a}$, ``coloron'') to the gluon plus Higgs, 
which actually is forbidden by the conformal barrier, 
but more fundamentally, even without the scale invariance, 
by the $SU(3)_c$ color gauge symmetry 
to keep the gluon massless after mass diagonalization.   
}. 

We also give another characteristic feature of our model, the gauge invariance of the HLS, the exact symmetry (though spontaneously broken).
 The HLS forbids a possible mixing between the iso-triplet one-family walking technirho's, 
$\rho_\Pi^i, \rho_P^i$, which, were it not for the HLS, 
would mix each other by the explicit  breaking of the global $SU(8)_L\times SU(8)_R$ symmetry 
by the SM gauge interaction, thereby affecting the analyses of Ref.~\cite{Fukano:2015uga}. 
The $\rho^i_{\Pi}$ are produced by the Drell-Yan process, while the $\rho_P^i$ orthogonal to 
$\rho_{\Pi}^i$ are not produced by the Drell-Yan process and is totally irrelevant to 
the diboson processes in the absence of the mixing thanks to the HLS. 
We newly study the mixing effect between $\rho_\Pi^3$ and $\rho^0_P$ (iso-singlet) through the {\it transverse modes} of 
the $W/Z$ bosons (via $W/Z$ kinetic term mixing
after mass diagonalization), which is actually of higher order term of (obviously scale-invariant) ${\cal O} (p^4)$ under control of the HLS in the  HLS chiral perturbation theory \cite{Tanabashi:1993np,Harada:2003jx}. This effect was not considered in the previous studies dealing with only the longitudinal $W/Z$ modes to treat the $W/Z$ as the ``external fields'' (not dynamical). Considering such 
dynamical mixing and other related ${\cal O} (p^4)$ terms, additional two parameters $a$ and $z_8$ come into play in addition to the  previous two $F_\rho$ and $g_{\rho\pi\pi}$. 
We then demonstrate that
such higher order terms are negligible, as far as $a$ and $z_8$ 
are on the order of naive dimensional counting, $1\lesssim a\lesssim 10$, $z_3+z_8/2 \simeq 0$. 
For the same parameter region the possible mass splitting 
between $\rho_\Pi^3$ and $\rho^0_P$ is extremely small to be invisible at the present LHC diboson analyses.

Consequently, we show that  thanks to the power of the conformal barrier  and the HLS,  the essential features of  our previous results \cite{Fukano:2015hga} of the walking technirho for the ATLAS 2 TeV excesses 
remain unchanged, including the characteristic smallness of the decay width, 
after all the phenomenological analyses are newly performed under new setting and inputs.

The paper is organized as follows: 
In Sec.~\ref{HiddenSymmetries} we first demonstrate that the SM Higgs sector can formally be scale-invariant in terms 
of the nonlinear realization and the Higgs can always be viewed as a dilaton associated with the ``hidden" scale symmetry. 
This formulation is extended to a model having new vector bosons realized as gauge bosons of the hidden local symmetry, 
protected by the conformal barrier. 
Taking a generic massive spin 1 model as an example, 
in Sec.~\ref{ET} we discuss the incompatibility of the conformal barrier with the widely 
quoted ``equivalence theorem'' for the vector boson decays. 
In Sec.~\ref{power-of-HLS} 
we describe the power of hidden local symmetry of our s-HLS model 
for the walking technirho's in the one-family walking technicolor.
It is shown that mixing between the iso-triplet $\rho^i_{\Pi}$ and $\rho^i_P$ by the SM gauge interactions is strictly 
forbidden by the HLS as an exact gauge symmetry to all orders of the chiral perturbation theory 
(up to the HLS-invariant intrinsic parity odd (Wess-Zumino-Witten) term \cite{Fujiwara:1984mp,Harada:2003jx}). 
The order ${\cal O} (p^4)$ terms including the small mixing of $\rho^3_{\Pi}$ and $\rho^0_P$ are fully considered.  
In Sec.~\ref{pheno} we reanalyze the LHC diboson excesses 
under completely new setting and inputs based on the preliminaries set up in Sec.~\ref{power-of-HLS}.   
The results remain unchanged compared with the previous analyses. 
Section~\ref{summary} is devoted to summary and discussion. 
Appendix~\ref{diagonalization} provides the explicit way of diagonalization of the s-HLS model 
including possible mass and kinetic term mixing among the SM gauge and HLS gauge boson fields. 
The formulae for the partial decay widths of the 2 TeV technirhos  are given in Appendix~\ref{decay}. 
Appendix~\ref{breakdown_Signal} complements the analysis done in Sec.~\ref{pheno}.

\section{Hidden Scale Symmetry and Hidden Local Symmetry in the Standard-Model Higgs - Made Explicite via Nonlinear Realization}
\label{HiddenSymmetries}

In this section we demonstrate that 
a scale-invariance formally emerges in the SM Higgs sector and the theory can 
always be rewritten by the nonlinear realization of the scale symmetry with 
the Higgs identified with a dilaton ({\it Hidden Scale Symmetry}). 
 After that, this formulation is extended to the case incorporating new vector bosons 
as gauge bosons ({\it Hidden Local Symmetry} (HLS)). Important message of this section is that 
although Higgs as a dilaton can acquire mass only through the explicit breaking of the scale symmetry (pseudo-dilaton), the {\it vector boson mass 
as a gauge boson of HLS can be generated  without explicit breaking of the 
scale symmetry!!}

\subsection{Hidden Scale Symmetry - Higgs  as a dilaton in the ``conformal limit''}

The SM Higgs Lagrangian takes the form
\begin{equation}
{\cal L}_{\rm Higgs}= |\partial_\mu h|^2 -\mu^2 |h|^2 -\lambda|h|^4\,,
\label{Higgs}
\end{equation}
where 
\begin{equation}
h=
\left(\begin{array}{c}
\phi^+\\
\phi^0\end{array}  \right)=\frac{1}{\sqrt{2}}\left(\begin{array}{c}
i{\hat \pi}_1+{\hat \pi}_2\\
\hat{\sigma}-i {\hat \pi}_3\end{array}\right)
\label{hrepr}
\end{equation} 
is substituted back to the Lagrangian, which yields the $SU(2)_L\times SU(2)_R$ linear sigma model:
 \begin{equation}
 {\cal L}_{L\sigma}=\frac{1}{2} \left[ (\partial_\mu {\hat \sigma})^2 + (\partial_\mu {\hat \pi}_a)^2\right] -\frac{\mu^2}{2}  ({\hat \sigma}^2 +{\hat \pi}_a^2)
 -\frac{\lambda}{4}  ({\hat \sigma}^2 +{\hat \pi}_a^2)^2\,. 
 \end{equation}

Now define a $2\times2$
matrix
\begin{equation}
M=(i \tau_2 h^*, h) = \frac{1}{\sqrt{2}}\left({\hat \sigma}\cdot  1_{2\times 2} +2i {\hat \pi}\right)\, \quad \left({\hat \pi} \equiv {\hat \pi}_a \frac{\tau_a}{2}\right)\,,
\end{equation}
which transforms under $G=SU(2)_L\times SU(2)_R$ as:
\begin{equation}
M \rightarrow g_L \, M\, g_R^\dagger \,,\quad \left(g_{R,L} \in SU(2)_{R,L}\right)\,.
\end{equation}
Then the Lagrangian takes the form
\begin{equation}
 {\cal L}_{L\sigma}= \frac{1}{2} {\rm tr} \left( \partial_\mu M\partial^\mu M^\dagger \right)
 - \frac{\mu^2}{2} {\rm tr}\left(M M^\dagger\right)-\frac{\lambda}{4}  \left({\rm tr}\left(M M^\dagger\right)\right)^2 \,. 
 \label{Lag:sigma}
 \end{equation}

Note that only the Higgs mass term $\mu^2 {\rm tr}[M^\dag M]$ breaks 
the scale symmetry in Eq.(\ref{Lag:sigma}): 
under the scale transformation of an operator ${\cal O}(x)$ with the scale dimension $d_{\cal O}$, $\delta {\cal O}(x)=  (d_{\cal O} +x^\nu\partial_\nu) {\cal O}(x)$,
the action $S=\int d^4 x {\cal L} (x) $ is invariant if
\begin{equation}
\delta S=\int d^4 x (d_{\cal L}+x^\nu\partial_\nu) {\cal L} 
=\int d^4 x [(d_{\cal L}-4){\cal L}+\partial_\nu(x^\nu {\cal L})]
=\int d^4 x (d_{\cal L}-4){\cal L}=0\,,  \quad (d_{\cal L} =4)\,. 
\end{equation} 
Namely, only the operators with scale dimension $d_{\cal O}=4$ in the Lagrangian ${\cal L} =\sum_i {\cal O}_i$ can 
make  the theory be scale-invariant. 
Thus the mass term in Eq.(\ref{Lag:sigma}) has the scale-dimension two, while 
others do the scale dimension four, hence the scale symmetry is explicitly broken 
only by the mass term.

When the Higgs field gets the vacuum expectation value $v$, spontaneously breaking the 
chiral $SU(2)_L \times SU(2)_R$ symmetry down to the diagonal sum of them,  
$SU(2)_{L+R=V}$, the three NG bosons emerge in the chiral broken phase. 
 To make the symmetry of the model manifest in the broken phase, 
we parametrize the Higgs field $M$ in terms of the nonlinear realization of the 
chiral symmetry.   
Note first that any complex matrix $M$ can be decomposed into the Hermitian (always diagnonalizable) matrix $H$  and unitary matrix $U$ as $M=HU$ ( ``polar decomposition'' ): 
 \begin{equation}
 M = H\cdot U\,, \quad H=\frac{1}{\sqrt{2}} \left(\begin{array}{cc}
 \sigma & 0\\
 0  &\sigma
 \end{array}\right)
 \,, \quad U= \exp\left(\frac{2i \pi}{v}\right) \quad \left(v=\langle  \sigma \rangle\right) \,.
 \label{Polar}
 \end{equation}
 The chiral transformation of $M$ is inherited by $U$, while $H$ is a chiral singlet such that:
 \begin{equation}
 U \rightarrow g_L \, U\, g_R^\dagger\,,\quad H \rightarrow H\,,
 \end{equation}
 and $U \, U^\dagger=1$ implies $\langle U\rangle =\langle  \exp\left(\frac{2i \pi}{v}\right)\rangle=1 \ne 0$, namely the spontaneous breaking of the chiral symmetry is taken granted in the polar decomposition.
 Then the Lagrangian takes the form:
 \begin{eqnarray}
  {\cal L}_{L\sigma}
 &=& \frac{1}{2} \left(\partial_\mu \sigma\right)^2+ \frac{1}{4}{\sigma}^2\cdot {\rm tr} \left(\partial_\mu U \partial^\mu U^\dagger\right) -V(M)\nonumber\\
 V(M) &=& \frac{\mu^2}{2} {\sigma}^2  + \frac{\lambda}{4}  \sigma^4 
 =\frac{\lambda}{4} \left[\left(\sigma^2 -v^2\right)^2-v^4\right] \,,\quad \left(M_\sigma^2 =\frac{\partial^2}{\partial \sigma^2} V\Bigg|_{\sigma=v}= 
2 \lambda v^2 =- 2 \mu^2\right)\,. 
 \label{LagM}
 \end{eqnarray} 
 The potential $V(M)$ is independent of the NG bosons $\pi$ (phase modes) which disappear from the potential, entirely moved over 
  to the kinetic term. They actually get absorbed into the weak bosons, acting as the gauge parameters when the electroweak gauging is switched on through the covariant derivative in the kinetic term. On the other hand, the radial mode $\sigma$ is a chiral-singlet in contrast to $\hat \sigma$ which is chiral non-singlet, thereby the potential obviously is chiral-invariant.

In the strong coupling limit $\lambda \rightarrow \infty$ such that 
the potential gets decoupled, 
with leaving only $\sigma =v (={\rm const.})$, and
the linear sigma model becomes the nonlinear sigma model (NL$\sigma$):
\begin{equation}
   {\cal L}_{\rm Higgs}= {\cal L}_{L\sigma}  \mathop{\longrightarrow}^{\lambda \rightarrow \infty}  {\cal L}_{NL\sigma}  =
   \frac{v^2}{4}\cdot {\rm tr} \left(\partial_\mu U \partial^\mu U^\dagger\right)    \,,
   \label{NL}
    \end{equation}
where the  breaking of the scale invariance gets shifted to the kinetic term  $v^2\cdot {\rm Tr} \left(\partial_\mu U \partial^\mu U^\dagger\right)$,
which  no longer transforms as dimension 4 operator. 
Then the {\it ${\cal L}_{NL\sigma}$ is a good effective theory as a basis of the successful chiral perturbation theory, 
when the underlying theory is based on the strong 
dynamics like QCD having no scale symmetry}, 
perfectly consistent with each other. 
However, 
{\it if the underlying strong coupling theory possesses the scale symmetry, such as the 
walking technicolor~\cite{Yamawaki:1985zg},  
the corresponding effective theory turns out to be obtained  
not by taking the $\lambda \to \infty$ limit, 
but by sending $\lambda \to 0$, with $v=$ fixed}, as will be clarified below.

Actually, we can always parametrize $\sigma$ for arbitrary $\lambda$ as 
 \begin{equation} 
 \sigma =v \cdot \chi\,,\quad \chi=\exp\left(\frac{\phi}{F_\phi}
 \right)
 \,,
 \label{NLscale}
 \end{equation}
 where $F_\phi=v$ is the decay constant of the dilaton $\phi$, which  is  not necessarily the same as 
the decay constant of $F_\pi=v$ of $\pi$, $F_\phi\ne v$ in general case other than the SM Higgs.
 The scale transformations for these fields are 
 \begin{equation}
 \delta \sigma =(1 +x^\mu \partial_\mu) \sigma \,, \qquad  
\delta \chi=(1+x^\mu \partial_\mu) \chi\,, \qquad 
\delta \phi= F_\phi+x^\mu \partial_\mu\phi\,. 
 \end{equation}
Note that $\langle  \sigma\rangle= v \langle \chi \rangle = v\ne 0$ breaks spontaneously the scale symmetry, but not the chiral symmetry since 
$ \sigma$ ($\chi$ as well) is a chiral singlet.
This is a nonlinear realization of the scale symmetry: 
the $\phi$ is a dilaton, NG boson of the spontaneously broken scale symmetry. Note that although $\chi$ is a dimensionless field,
it transforms as that of dimension 1, while $\phi$ having dimension 1 transforms as the dimension 0, instead.

 Now the kinetic term in Eq.(\ref{LagM}) reads:
 \begin{eqnarray}
  {\cal L}_{\rm Kinetic}
 &=& \frac{F_\phi^2}{2} \left(\partial_\mu \chi \right)^2+ \frac{v^2}{4}{\chi}^2\cdot {\rm tr} \left(\partial_\mu U \partial^\mu U^\dagger\right) \nonumber\\
 &=&\chi^2 \cdot \left[ \frac{1}{2} \left(\partial_\mu \phi\right)^2  +\frac{v^2}{4}{\rm tr} \left(\partial_\mu U \partial^\mu U^\dagger\right)\right]\,,
 \label{Kinetic}
 \end{eqnarray}
which  yields {\it a scale-invariant action  
in spite of the nonzero vacuum expectation value $F_\phi=v\ne 0$}. 
The final form coincides with the {\it scale-invariant nonlinear sigma model} (s-NL$\sigma$)~\cite{Matsuzaki:2012gd,Matsuzaki:2015sya} as a basis of 
the scale-invariant chiral perturbation theory~\cite{Matsuzaki:2013eva}.

 On the other hand,  
{\it the potential is scale-violating but can be totally removed} 
by taking formally {\it the limit  $\lambda\rightarrow 0$ keeping $v=\sqrt{-\mu^2}/\lambda =$constant $\ne 0$} (``conformal limit''). 
 The potential  in Eq.(\ref{LagM}) reads
 \begin{equation}
 V(\chi)= \frac{\lambda}{4} v^4 \left[\left(\chi^2 -1\right)^2-1\right]  \,,
 \label{Potinchi}
 \end{equation}
with minimum at $\langle \chi\rangle =1$.
Obviously 
\begin{equation}
 V(\chi)\stackrel{\lambda \rightarrow 0}{\longrightarrow} 0\,.
  \end{equation}
The scale breaking part in the potential transforms (up to total divergence) as
\begin{equation}
\delta V(\chi) = -2  \frac{\lambda}{4} v^4 \left(\delta \chi^2 \right)= 
- \frac{\lambda}{2} v^4 \left(2 \chi^2 + x^\mu \partial_\mu \chi^2\right )= +\lambda v^4 \chi^2 
+ \textrm{total derivative} 
\,,
\label{Vtrans}
\end{equation}
which yields 
\begin{equation}
\partial^\mu {\bf D}_\mu=\theta^\mu_\mu= - \delta V(\chi)= - \lambda v^4 \chi^2\,,
\label{trace}
\end{equation}
where ${\bf D}_\mu$ is the dilatation current and $\theta^\mu_\mu$ is the trace of energy-momentum tensor. Then the Partially Conserved Dilatation Current (PCDC) reads
\begin{equation}
M_\phi^2 F_\phi^2 = - \langle 0|\partial^\mu {\bf D}_\mu|\phi\rangle F_\phi = - d_\theta \langle \theta^\mu_\mu\rangle  = 2\lambda v^4 \langle \chi^2\rangle= 2 \lambda v^4\,,
\label{PCDC}
\end{equation}
where $\theta^\mu_\mu$ has a scale dimension $d_{\theta} =2$ (as seen from Eq.(\ref{trace})). The result is consistent with the mass term of $\phi$:
$M_\phi^2= 2\lambda v^4/F_\phi^2$ (including the canonical value $M_\phi=2\lambda v^2,\, F_\phi=v$), as can be seen   by
expanding the potential Eq.(\ref{Potinchi}) as:
\begin{equation}
V(\chi) = \frac{1}{2} \left(\frac{2\lambda v^4}{F_\phi^2}\right) \phi^2+\cdots\,.
\end{equation}

  Thus the linear sigma model ${\cal L}_{L\sigma}$ goes over to the scale-invariant nonlinear sigma model in the limit $\lambda\rightarrow 0\,, v\ne 0$:
 \begin{eqnarray}
 {\cal L}_{\rm Higgs} 
= {\cal L}_{L\sigma}  
& \stackrel{\lambda \rightarrow 0}{\longrightarrow} & 
 {\cal L}_\textrm{s-NL$\sigma$}\nonumber \\ 
 &=&\chi^2 \cdot \left[ \frac{1}{2} \left(\partial_\mu \phi\right)^2  +\frac{v^2}{4}{\rm tr} \left(\partial_\mu U \partial^\mu U^\dagger\right)\right] 
\nonumber\\
  &=& \left( 1 +\frac{2\phi}{F_\phi} +\cdots\right)\cdot \left[ \frac{1}{2} \left(\partial_\mu \phi\right)^2  +\frac{v^2}{4}{\rm tr} \left(\partial_\mu U \partial^\mu U^\dagger\right)\right] 
\,, \quad \chi=\exp \left(\frac{\phi}{F_\phi}\right) 
 \label{SNL}
 \end{eqnarray}
which is a nonlinear realization of the chiral $SU(2)_L\times SU(2)_R$ but not just that: 
It can further be scale-invariant by the prefactor $\chi^2$  based on the nonlinear realization of the scale symmetry via the NG boson (dilaton) $\phi$. 
Then we are left with the scale-invariant Higgs Lagrangian (at the action level), 
having only the kinetic term in the limit $\lambda \rightarrow 0$, with the mass given by Eq.(\ref{PCDC}) for $\lambda\ll 1$. 
Now {\it the Higgs is nothing but a (pseudo-)dilaton $\phi$}.

The electroweak gauging is switched on through the covariant derivative:
 \begin{equation}
 \partial_\mu U \Rightarrow {\cal D}_\mu U= \partial_\mu U -i {\cal L}_\mu U + i U {\cal R}_\mu\,,
 \label{covariant}
 \end{equation}
 where ${\cal L}_\mu$ and ${\cal R}_\mu$ are the electroweak gauge boson fields. 
Then the gauged Lagrangian reads:
 \begin{equation}
{\cal L}^{\rm gauged}_\textrm{s-NL$\sigma$}
= \chi^2  \cdot \left[ \frac{1}{2} \left(\partial_\mu \phi\right)^2  +\frac{v^2}{4}{\rm tr} \left({\cal D}_\mu U {\cal D}^\mu U^\dagger\right)\right]\,,
\label{gaugedSNL}
\end{equation}
which is obviously still scale-invariant (at the action level), including the kinetic term of the electroweak gauge bosons. 
As can be seen from Eq.(\ref{Kinetic}), the coupling of $\phi$ to the SM particles are proportional to $1/F_\phi$ instead of $1/v$ of 
the SM Higgs coupling, and hence are even weaker coupling  than
the SM Higgs coupling by the factor $v/F_\phi (<1) $, if $F_\phi > v$.

Thus if the Higgs is a composite of an underlying {\it scale-invariant} strong coupling theory, 
such as the walking technicolor~\cite{Yamawaki:1985zg}, 
then the effective field theory is precisely given by Eq.(\ref{gaugedSNL}) plus kinetic terms of the electroweak gauge bosons 
(plus higher derivative terms as in the {\it scale-invariant  chiral perturbation theory} \cite{Matsuzaki:2013eva}),  
thanks to the {\it nonlinear
 realization of both the chiral and scale symmetries} via respective NG bosons, dilaton $\phi$ and the longitudinal modes $\pi$ of the weak bosons, 
with the chiral symmetry straightforwardly extendable so as to have a generic $N_F$ as $G=SU(N_F)_L \times SU(N_F)_R$. 
It was shown that $F_\phi \gg v=246\, {\rm GeV}$
 in the typical walking technicolor with $N_F=8$, 
in accord with the current LHC Higgs data \cite{Matsuzaki:2012gd,Matsuzaki:2012xx,Matsuzaki:2015sya}.

 The scale symmetry is not the exact symmetry of the quantum theory, even if the theory is scale invariant at the classical level, where
 there exists scale anomaly induced by the regularization, such as the intrinsic scale $\Lambda_{\rm QCD}$ in the QCD 
(This of course is also the case in the formally scale-invariant Higgs
 Lagrangian, Eq.(\ref{SNL})).  In the ordinary QCD
 there is no infrared scale symmetry and no small scale other than $\Lambda_{\rm QCD}$, i.e., 
$v ={\cal O} (\Lambda_{\rm QCD})$, while in the walking technicolor having approximate scale symmetry 
which is spontaneously broken by the strong coupling gauge interaction at the scale much smaller 
than the intrinsic scale $\Lambda_{\rm TC}$: $F_\phi \ll \Lambda_{\rm TC}$ (See, e.g., \cite{Matsuzaki:2015sya}). 
In the latter case the Higgs mass $M_\phi$ as the explicit breaking of the scale symmetry is given by the ``nonperturbative'' trace anomaly 
$\langle \theta^\mu_\mu\rangle$ (besides the usual anomaly related to the intrinsic scale), 
which is evaluated from the underlying theory such as the walking technicolor. 
Noting that $\langle \theta^\mu_\mu\rangle=- M_\phi^2 F_\phi^2/4$ via the PCDC, 
a new potential  generated by the trace anomaly takes the form 
(See, e.g., \cite{Matsuzaki:2015sya})~\footnote{
This potential is indeed obtained from the explicit computation of the effective potential of the scale-invariant underlying strong coupling gauge theory, 
the walking technicolor, at the conformal phase transition point in the gauged Nambu-Jona-Lasinio model~\cite{Miransky:1996pd}.  
See Eq.(65) of Ref.~\cite{Miransky:1996pd}.}: 
 \begin{equation}
 V(\phi)\Bigg|_{\rm anomaly}= \frac{M_\phi^2 F_\phi^2}{4} \chi^4 \left(\ln \chi -\frac{1}{4}\right)
 = -\frac{M_\phi^2F_\phi^2}{16}+ \frac{1}{2}M_\phi^2 \phi^2+ \frac{4M_\phi^2}{3F_\phi}\phi^3 + \frac{2 M_\phi^2}{F_\phi^2} \phi^4 +\cdots\,,
 \label{anomaly}
 \end{equation}
which indeed yields $\langle \delta V\rangle = M_\phi^2 F_\phi^2\langle \chi^4\rangle/4=M_\phi^2 F_\phi^2/4= -\langle \theta^\mu_\mu\rangle$ and  has a minimum at $\langle \chi\rangle=1$ as desired. 
If $F_\phi \gg v_{\rm EW}= 246$ GeV as in the walking technicolor with $N_F=8$, then {\it the self couplings of $\phi$ as well as the couplings to the $W,Z$ (including the SM fermions) in Eq.(\ref{gaugedSNL}) are all   weaker couplings than those of the SM Higgs}. Note that this form of the potential looks similar to  the Coleman-Weinberg type potential
which is of somewhat different origin, radiatively induced from the classically scale-invariant Higgs model (with $v=0$).

 \subsection{Hidden Local Symmetry - New Vector Boson as a Gauge Boson}

The scale-invariant version of the Higgs Lagrangian Eq.(\ref{SNL}) is a nonlinear realization of the spontaneous broken internal symmetry $G=SU(2)_L\times SU(2)_R$ down to $H=SU(2)_{L+R}=SU(2)_V$, based on the manifold $G/H$
as well as that of the spontaneously broken scale symmetry. 
The model actually has, besides the scale symmetry, a symmetry $ G_{\rm global} \times H_{\rm local}$ larger than
 $G$, with $H_{\rm local}$ being the hidden local symmetry (HLS) \cite{Bando:1984ej,Bando:1987ym}, which can be made explicit by dividing $U(x)$ into two parts:
\begin{equation}
 U(x)= \xi_L^\dagger(x) \cdot \xi_R(x)\,,
 \label{U:decomp}
\end{equation}  
where $\xi_{R,L}(x)$ transform under $G_{\rm global} \times H_{\rm local}$ as
\begin{equation}
\xi_{R,L}(x) \rightarrow h(x) \cdot \xi_{R,L}(x) \cdot {g^\prime}_{R,L}^\dagger\,,\quad 
U(x) \rightarrow {\hat g}_L U(x)  {g^\prime}_R^\dagger \quad \quad
 \left(h(x)\in H_{\rm local} \,, \,\,\, g^\prime_{R,L}\in G_{\rm global} \right)
\,. 
\end{equation}
The $H_{\rm local} $ is a gauge symmetry of group $H$ arising from the redundancy (gauge symmetry) how to divide
$U$ into two parts. 
Then we can introduce the HLS gauge boson fields $V_\mu(x)$ by the covariant derivatives acting on $\xi_{R,L}$ as 
\begin{equation}
D_\mu \xi_{R,L}(x) = \partial_\mu \xi_{R,L} (x)-i V_\mu(x) \xi_{R,L}(x) \,,
\label{HLScovariant}
\end{equation}
which transform in the same way as $\xi_{R,L}$. Then we have covariant objects transforming homogeneously under $H_{\rm local}$: 
  \begin{eqnarray}
 {\hat \alpha}_{\mu R,L}(x)&\equiv& \frac{1}{i}D_\mu \xi_{R,L}(x) \cdot \xi_{R,L}^\dagger(x) =\frac{1}{i}\partial_\mu \xi_{R,L}(x) \cdot \xi_{R,L}^\dagger(x) 
 - V_\mu(x)\,, \nonumber\\ 
 {\hat \alpha}_{\mu ||,\perp}(x)&\equiv & \frac{1}{2}\left({\hat \alpha}_{\mu R}(x) \pm  {\hat \alpha}_{\mu L}(x)\right) 
\nonumber \\  
&=& 
\Bigg\{ \begin{array}{c}
  \frac{1}{2i} \left(\partial_\mu \xi_{R}(x) \cdot \xi_{R}^\dagger(x)+ \partial_\mu \xi_{L}(x) \cdot \xi_{L}^\dagger(x) \right)- V_\mu(x)= {\alpha}_{\mu ||}(x) - V_\mu(x)\\  
 \frac{1}{2i} \left(\partial_\mu \xi_{R}(x) \cdot \xi_{R}^\dagger(x)- \partial_\mu \xi_{L}(x) \cdot \xi_{L}^\dagger(x)  \right) = {\alpha}_{\mu \perp}(x)  
\end{array}
\nonumber\,,\\
 \{  \hat{\alpha}_{\mu R,L}(x), \hat{\alpha}_{\mu ||, \perp}(x)\}
  &\rightarrow& h(x) \cdot \{ \hat{\alpha}_{\mu R,L}(x), \hat{\alpha}_{\mu ||,\perp}(x)\}\cdot h^\dagger(x)\,.  
      \end{eqnarray}
We thus have two independent invariants under the larger symmetry $G_{\rm global} \times H_{\rm local}$:
\begin{equation}
\,
{\cal L}_A=v^2\cdot {\rm tr} [{\hat \alpha}_{\mu \perp}^2(x)] 
\,,\quad 
 \,\quad
{\cal L}_V=v^2\cdot  {\rm tr} [{\hat \alpha}_{\mu ||}^2(x)] 
= v^2\cdot   {\rm tr} [ \left(V_\mu(x) - {\alpha}_{\mu ||}(x)\right)^2] 
\,. 
\end{equation} 

Hence the scale-invariant version of the Higgs Lagrangian Eq.(\ref{SNL}) can be extended to  a gauge-equivalent model,
the scale-invariant HLS model (s-HLS)~\cite{Kurachi:2014qma}: 
\begin{equation}
{\cal L}_{\rm s-HLS} = \chi^2 \cdot \left(\frac{1}{2} \left(\partial_\mu \phi\right)^2 + {\cal L}_A+ a {\cal L}_V\right)
+ {\cal L}_{\rm Kinetic} \left(V_\mu\right)\,,
\label{SHLS}
\end{equation} 
with $a$ being an arbitrary parameter, and  ${\cal L}_{\rm Kinetic} \left(V_\mu\right)$ is the kinetic term of the HLS gauge boson $V_\mu$
which is obviously scale-invariant. \footnote{
The s-HLS model was also discussed in a different context, the ordinary QCD in medium.\cite{Lee:2015qsa}
}

We now fix the gauge of HLS as $\xi_L^\dagger=\xi_R=\xi=e^{i \pi/v}$ such that $U=\xi^2$. 
Then  $H_{\rm local}$ and $H_{\rm global} (\subset G_{\rm global})$ 
get simultaneously broken spontaneously (Higgs mechanism), leaving the 
diagonal subgroup $H=H_{\rm local}+H_{\rm global}$, which is nothing but the subgroup of the original $G$ of $G/H$: $H\subset G$. 
According to the Higgs mechanism, the HLS gauge fields $V_\mu(x)$ acquire the masses $\frac{1}{2} a (g \, v)^2\,(V^a_\mu(x))^2$ 
through the invariant ${\cal L}_V$ after rescaling the kinetic term of $V_\mu$ by the HLS gauge coupling $g$ as 
$V_\mu(x) \rightarrow g\, V_\mu(x)$. {\it Obviously the vector boson mass terms are scale-invariant thanks to the nonlinear realization of the scale symmetry!}
When the kinetic term  is ignored in the low energy region $p^2\ll M_V^2= a (g\,v)^2$, the ${\cal L}_V$ term yields just 0, 
after the equation of motion $V_\mu=\alpha_{\mu ||}$ is used. 
Noting that  ${\cal L}_A=v^2 \cdot{\rm tr} 
[{\hat \alpha}_{\mu \perp}^2(x)]
=v^2  \cdot{\rm tr}[{\alpha}_{\mu \perp}^2(x)] 
=  \frac{v^2}{4}\cdot {\rm tr}[\partial_\mu U \partial^\mu U^\dagger] 
={\cal L}_{{\rm NL} \sigma}$ 
in Eq.(\ref{NL}) by a straightforward algebraic calculation,
Eq.(\ref{SHLS}) becomes identical to the scale-invariant version of the Higgs Lagrangian Eq.(\ref{SNL}). 
Hence {\it in the low energy $p^2\ll M_V^2= a (g\,v)^2$ where the massive vector boson $V_\mu$ gets decoupled, the  s-HLS Lagrangian Eq.(\ref{SHLS})
is in fact reduced back to precisely the original scale-invariant nonlinear sigma model Eq.(\ref{SNL}),}  which  (when including the non-zero potential Eq.(\ref{Potinchi})) is equivalent to the Higgs Lagrangian Eq.({\ref{Higgs})
 for $F_\phi=v\ne 0$.

The electroweak gauge interactions are introduced by extending the covariant derivatives in Eq.(\ref{HLScovariant}) 
in the same way as Eq.(\ref{covariant}), but this time by gauging $G_{\rm global}$, 
which is {\it independent of $H_{\rm local}$} in the HLS extension: 
\begin{equation}
D_\mu \xi_{R,L}(x)\Rightarrow {\hat D}_\mu \xi_{R,L}(x)\equiv  \partial_\mu \xi_{R,L} (x)-i V_\mu(x) \, \xi_{R,L}(x)  
+i \xi_{R,L}(x)\, {\cal R}_\mu(x) ( {\cal L}_\mu(x)) \,.
\end{equation}
As usual in the Higgs mechanism, the gauge bosons of ${\rm gauged-}H_{\rm global} (\subset 
{\rm gauged-}G_{\rm global}$) get mixed with the gauge bosons of HLS, leaving massless only the gauge bosons of the unbroken diagonal subgroup: $({\rm gauged-}H)=H_{\rm local} + ({\rm gauged-}H_{\rm global})$ after diagnolization.
 We then finally have a gauged s-HLS version of the Higgs Lagrangian (gauged-s-HLS):
 \begin{equation}
 {\cal L}^{\rm gauged}_{\rm s-HLS}=  \chi^2 \cdot 
 \left[
\frac{1}{2} \left(\partial_\mu \phi\right)^2  + {\hat {\cal L}}_A+ a {\hat {\cal L}}_V
  \right]  + {\cal L}_{\rm Kinetic} \left(V_\mu, W_\mu/B_\mu \right)\,, 
  \label{gaugedsHLS}
  \end{equation}
 with 
 \begin{equation}
  {\hat {\cal L}}_{A,V}= {\cal L}_{A,V} \left( D_\mu \xi_{R,L}(x) \Rightarrow {\hat D}_\mu \xi_{R,L}(x)\right) \,,
     \end{equation}
where $ {\cal L}_{\rm Kinetic} \left(V_\mu, W_\mu/B_\mu\right)$ stands for the kinetic terms of the HLS and SM gauge bosons.

It is straightforward to extend the internal symmetry group to $G_{\rm global}$ =$[SU(N_F)_L\times SU(N_F)_R]_{\rm global}$ and $H_{\rm local}= [SU(N_F)_V]_{\rm local}$. 
The Lagrangian then takes the form in general:
\begin{equation} 
 {\cal L}_{\rm s-HLS} 
 = \chi^2 \cdot  \left( 
 \frac{1}{2} \left(\partial_\mu \phi\right)^2
 + 
F_\pi^2 \left[ 
{\rm tr}[ \hat{\alpha}_{\mu \perp}^2 ] 
+ 
a \, {\rm tr}[ \hat{\alpha}_{\mu ||}^2 ] 
\right] 
\right) 
+ \cdots  
\,, \label{sHLS}
\end{equation}  
where $F_\pi$ is related to $F_\pi = v/\sqrt{N_F/2}$, and ``$+ \cdots$'' includes the kinetic term of the HLS and SM gauge bosons
and possible higher order terms in the derivative/loop expansion (``chiral perturbation theory''). 

{\it Thus all the mass of the gauge bosons (SM gauge bosons as well as HLS gauge bosons) acquired via Higgs mechanism are generated keeping the 
scale symmetry (spontaneously broken, realized in the nonlinear realization),
which is in sharp contrast to the dilaton $\phi$ which can be massive (pseudo-dilaton) only through the explicit-scale symmetry-violation.} 
This implies that when the underlying theory behind the Higgs has an approximate scale symmetry 
as in the walking technicolor, the composite vector bosons should have the masses in a 
scale-invariant way~\footnote{Other composite matter fields (non-NG boson fields) in the scale-invariant underlying theory can also have 
masses symmetric under the spontaneously broken symmetries via the nonlinear realization, 
e.g., mass term  (up to the Yukawa coupling) of the technibaryon may be included into the nonlinearly realized ``Higgs Lagrangian'' 
via  the obviously scale-invariant form 
$\bar \psi_L(x) \cdot M(x) \cdot \psi_R(x)  + {\rm h.c.}
= \frac{v}{\sqrt{2}} \cdot \bar \psi_L(x) \cdot  (\chi(x) \cdot U(x)) \cdot \psi_R(x)  + {\rm h.c.} 
= \frac{v}{\sqrt{2}} \cdot \chi(x) \cdot   ({\bar \Psi}_L(x) \Psi_R(x)) + {\rm h.c.}$, 
which is  also chiral-symmetric since 
the composite matter fields $\Psi_{R,L} $ transform as 
$\Psi_{R,L} \equiv \xi_{R,L} \psi_{R,L} \rightarrow h(x) {\Psi}_{R,L} $.
See Eqs.(\ref{Polar}) and (\ref{NLscale}). },  
in such a way that all the couplings of Higgs/dilaton $\phi$ to the gauge bosons (SM and HLS) are given through the overall prefactor $\chi^2$ 
in front in Eq.(\ref{sHLS}), 
hence the off-diagonal mass terms are rotated away by the mass diagonalization done independently of the Higgs/dilaton $\phi$ which lives only in the overall prefactor $\chi^2$. 
This is fairly insensitive to the tiny explicit-scale symmetry-violation responsible for the Higgs/dilaton mass $M_\phi$ via 
the PCDC in Eq.(\ref{PCDC}) or Eq.(\ref{anomaly}), 
arising from the potential term, instead of the ``kinetic term'' Eq.(\ref{sHLS}). 
This invalidates the popular ``equivalence theorem'', a phenomenon what we called  ``conformal barrier''\cite{Fukano:2015uga},  
as we discuss in details below.

\subsection{Conformal Barrier}

When one works on the typical one-family walking technicolor with $N_F=8$ 
as an example for the s-HLS, one finds that 
the SM gauging $G_{\rm global}$ includes the full $SU(3)_c \times SU(2)_W \times U(1)_Y \subset {\rm gauged-} G_{\rm global}$. 
The mass mixing in Eq.(\ref{sHLS}) then 
takes place between $\rho_{\theta_a}^0$ (color-octet, iso-singlet) 
and the QCD gluons, between $\rho_{\Pi^i}$ (color-singlet iso-triplet) and $W^i$, between $\rho_{P^0}$ (color-singlet iso-singlet) and the $U(1)_Y$ gauge boson $B$, as well as the usual $W^3-B$ mixing. (For the definition of the one-family walking-technirho fields, see Table~\ref{tab:TPTR} in the later section.)
After diagonalization, QCD gluons and photon remain massless, as they should. 
The  mass terms in Eq.(\ref{sHLS}) having the $\phi$ field in the overall conformal factor $\chi^2(x)$ yields 
 $\phi$ couplings to the diagonal pairs of the SM gauge bosons after diagonalization.    
Thus the new vector bosons do not decay to the weak bosons 
in association with the Higgs 
in the presence of the scale/conformal symmetry ({\it Conformal Barrier})~\cite{Fukano:2015uga}, i.e., 
\begin{equation} 
  V- W/Z - H \, {\rm coupling} \, = \, 0   \,, 
\end{equation} 
consequently the $V$ predominantly decays to the weak boson pairs $WW/WZ$. In other word, the ``equivalence theorem'' is invalidated by the scale/conformal symmetry.
The absence of $V\to WH/ZH$ signatures at the LHC Run-II thus could indirectly probe the existence of the (approximate) scale/conformal invariance of the system involving
$V$, $W, Z$ and $H$.

 The conformal/scale invariance should be approximate, hence the conformal barrier will 
be broken at higher order level of the perturbation theory. 
In fact, the scale symmetry will be broken at the one-loop level  by 
Yukawa interaction terms like $\phi \bar{f}f$ once one considers them. 
 Thus these breaking terms would potentially generate off-diagonal 
$V$-$W$-$\phi$ terms, 
so the conformal barrier might be badly melt down. 
However, the size of the breaking turns out to be negligibly small: 
among possible breaking terms, the maximal effect is expected to come from 
the top Yukawa coupling $\sim m_t/v$. Then the one-loop diagram constructed from 
the $\phi$-$t$-$t$, $V$-$t$-$t$ and $W$-$t$-$t$ vertices 
would yield the off-diagonal $V$-$W$-$\phi$ coupling, 
\begin{equation} 
 g_{V W \phi}^{\rm one-loop} \sim 
\frac{N_c}{(4\pi)^2}g_W g  \frac{m_t^2}{v} 
\,, 
\end{equation} 
which is compared to the ``equivalence theorem" coupling $g_{V W \phi} \sim g_W g v$ 
\begin{equation} 
 \frac{g_{V W \phi}^{\rm one-loop}}{g_{V W \phi}}
\sim \frac{N_c}{(4\pi)^2 } \left( \frac{m_t}{v} \right)^2 
\sim {\cal O}(10^{-2}) 
\,, 
\end{equation}
up to possible ${\cal O}(1)$ coefficients. 
Thus the one-loop induced off-diagonal coupling yields  
nonzero $V \to W \phi$ amplitude suppressed by a factor of ${\cal O}(10^{-2})$, 
leading to the extremely small LHC cross section suppressed by a factor of $\sim 10^{-4}$,   
compared 
with the `equivalence theorem'' coupling which if existed at all would give the same branching ratio as the diboson decays. 
Thus it 
will be too small to be detected at the LHC experiments. 
In this sense, the conformal barrier is still a powerful constraint for the vector boson.

One way out to avoid the conformal barrier may be to introduce multi Higgs fields which 
give the masses to new vector bosons as well as the weak bosons. 
The mixing among the Higgs bosons would make the mixing structures different 
for the $V$-$W$ and $V$-$W$-$\phi$. 
Models having such a vector boson - Higgs boson sector correspond to those studied in Refs.~\cite{Abe:2015jra,Abe:2015uaa}. 
However, some of those Higgs bosons would phenomenologically be heavy to be integrated out, 
such that, except the lightest 125 GeV Higgs,  
all the Higgs fields in the linear realization can be cast into the nonlinear forms 
keeping only the NG boson fields (nonlinear realization). 
The aforementioned models will then be effectively described as a model having the lightest Higgs 
and multi NG bosons eaten by weak and new vector bosons (or some of them would be real electroweak pions 
such as technipions).  
Then, the conformal barrier would be operative even for such those multi Higgs models.

\section{Incompatibility with the ``equivalence theorem''}
\label{ET}

In this section 
we show that 
the conformal barrier is actually incompatible with 
the so-called ``equivalence-theorem" result for the $V \to WW/WZ$ and $V \to WH/ZH$ 
decays, i.e., $\Gamma(V \to WW/WZ) \simeq \Gamma(V \to WH/ZH)$. 
It turns out that the conformal barrier is achieved  only 
by taking a special limit for the vector boson parameters.

To demonstrate this point clearly,  
we shall employ a generic model, 
called heavy-vector triplet (HVT) model~\cite{Pappadopulo:2014qza},     
which is quoted by the ATLAS and CMS groups for new vector boson searches as a benchmark.  
The model Lagrangian reads~\cite{Pappadopulo:2014qza} 
\begin{eqnarray} 
 {\cal L}_V 
&=&  
- \frac{1}{2} {\rm tr}[ V_{\mu\nu}^2] + m_V^2 {\rm tr}[V_\mu^2]  
\nonumber \\ 
&& 
+ g_V c_{_{H}} \, \left( i H^\dag V^\mu D_\mu H + {\rm h.c.} \right)
\nonumber \\ 
&& 
+ 2 g_V^2 c_{_{VVHH}} {\rm tr}[ V_\mu^2 ] H^\dag H 
\nonumber \\ 
&& 
+
{\cal L}_{\rm Higgs} 
+  \cdots 
\,, \label{LV} 
\end{eqnarray}
where we have put the standard-model Higgs terms ${\cal L}_{\rm Higgs}$ including 
the kinetic term $|D_\mu H|^2$ and the usual Higgs potential. 
In Eq.(\ref{LV}) we have defined 
\begin{eqnarray} 
V_{\mu\nu} &=& D_\mu V_\nu - D_\nu V_\mu 
\,, \nonumber \\ 
D_\mu V_\nu &=& \partial_\mu V_\nu - i g_W [W_\mu, V_\nu] 
\,, 
\end{eqnarray} 
with the $g_W$ being the weak gauge couping.    
We have not displayed terms which do not include the Higgs $H$ along with the new vector boson field $V$.

  When the Higgs field $H$ gets the vacuum expectation value $v$ $(\simeq$ 246 GeV), 
the new vector boson $V$ starts to mix with the weak boson $W$ through the $c_{_{H}}$ term in Eq.(\ref{LV}).  
We parameterize the $H$ as 
\begin{equation} 
H= \frac{v}{\sqrt{2}} \left(1 + \frac{\phi}{v} \right) 
\left( 
\begin{array}{c} 
0 \\ 
1 
\end{array} 
\right) 
\end{equation} 
plus the eaten NG boson terms (set to zero in the unitary gauge of the electroweak gauge interactions) . 
Momentarily, we ignore the hypercharge gauge for simplicity (without loss of generality for the following discussions).     
Then the mass matrix for ${\bf V}_\mu$ = ($V_\mu$, $W_\mu$)$^T$ is read off:   
\begin{equation}  
{\cal M}^2 
= 
 \left( 
\begin{array}{cc} 
m_V^2 + g_V^2 c_{_{VVHH}} v^2 &  \frac{1}{4} g_W g_V c_{_{H}} v^2 \\ 
  \frac{1}{4} g_W g_V c_{_{H}} v^2 & \frac{1}{4} g_W^2 v^2 
\end{array} 
\right) 
\,. \label{mass-matrix}
\end{equation}  
In addition, one has the Higgs ($\phi$) couplings to $V$ and $W$, 
\begin{equation} 
{\cal G}_{VW\phi} 
= 
 \left( 
\begin{array}{cc} 
g_V^2 c_{_{VVHH}} v^2 &  \frac{1}{4} g_W g_V c_{_{H}} v^2 \\ 
  \frac{1}{4} g_W g_V c_{_{H}} v^2 & \frac{1}{4} g_W^2 v^2 
\end{array} 
\right) 
\,. \label{VHW}
\end{equation} 
In the Lagrangian the ${\cal M}^2 $ and ${\cal G}_{VW\phi} $ terms look like 
\begin{equation} 
 {\cal L}_V = \frac{1}{2} 
{\bf V}^T_\mu \cdot {\cal M}^2 \cdot {\bf V}^\mu 
+ 
\frac{\phi}{v} \cdot {\bf V}_\mu^T \cdot {\cal G}_{VW\phi} \cdot {\bf V}^\mu 
+ \cdots 
\,.   
\end{equation} 
Note that the mass matrix ${\cal M}^2$ and the couplings to the Higgs $\phi$ differ 
only by the $m_V^2$ term.

Assuming a heavy vector limit, 
\begin{equation} 
x = \frac{g_W v}{g_V v} \ll 1  
\,, \label{heavy-vector-limit}
\end{equation}
and expanding terms in powers of $x$, 
we diagonalize the mass matrix Eq.(\ref{mass-matrix}) by an orthogonal rotation to get the 
mass eigenvalues for the mass eigenstates $\tilde{\bf V} = (\tilde{V}, \tilde{W})^T$: 
\begin{eqnarray} 
m_{\tilde V}^2 &=& \frac{g_V^2 v^2}{4} \left[ \frac{4 m_V^2}{g_V^2 v^2} + 4 c_{_{VVHH}} + \frac{c_{_{H}}}{\frac{4 m_V^2}{g_V^2 v^2} + 4 c_{_{VVHH}}} x^2 + \cdots \right] 
\,, \nonumber \\ 
 m_{\tilde W}^2 &=& \frac{g_V^2 v^2}{4} \left[ \frac{\frac{4 m_V^2}{g_V^2 v^2} + 4 c_{_{VVHH}} - c_{_{H}}^2}{\frac{4 m_V^2}{g_V^2 v^2} + 4 c_{_{VVHH}}} x^2 + \cdots \right] 
\,. \label{V-mass}
\end{eqnarray}
The corresponding rotation matrix is 
\begin{equation} 
\left( 
\begin{array}{c} 
W \\ 
V 
\end{array} 
\right) 
= 
\left( 
\begin{array}{cc} 
  1- \frac{1}{2} \left( \frac{4 m_V^2}{g_V^2 v^2} + 4 c_{_{VVHH}} \right)^2 x^2 (1 + {\cal O}(x^2)) &  
- \frac{c_{_H}}{\frac{4 m_V^2}{g_V^2 v^2} + 4 c_{_{VVHH}}} x (1 + {\cal O}(x^2)) \\
\frac{c_{_H}}{\frac{4 m_V^2}{g_V^2 v^2} + 4 c_{_{VVHH}}} x (1 + {\cal O}(x^2)) & 
 1- \frac{1}{2} \left( \frac{4 m_V^2}{g_V^2 v^2} + 4 c_{_{VVHH}} \right)^2 x^2 (1 + {\cal O}(x^2)) 
\end{array} 
\right) 
\left( 
\begin{array}{c} 
\tilde{W} \\ 
\tilde{V} 
\end{array} 
\right) 
\,. \label{rot}
\end{equation} 
One also finds the couplings such as 
$\tilde{V}$-$\tilde{V}$-$\phi$, $\tilde{W}$-$\tilde{W}$-$\phi$, as well as the off diagonal 
coupling $\tilde{V}$-$\tilde{W}$-$\phi$. 
By using the rotation matrix in Eq.(\ref{rot}) 
the off diagonal coupling $\tilde{V}$-$\tilde{W}$-$\phi$ is found to be 
\begin{equation} 
 {\cal G}_{\tilde{V} \tilde{W} \phi} = 
 \frac{m_V^2}{4 m_{\tilde V}^2} c_{_{H}} g_V^2 v^2 x + \cdots 
 = \frac{m_V^2}{4 m_{\tilde V}^2} c_{_H} g_V g_W v^2 + \cdots  
\,. 
\end{equation}
Crucial to notice is that the presence of the nonzero off-diagonal coupling $\tilde{V}$-$\tilde{W}$-$\phi$ 
is essentially due to the $m_V^2$ term in Eq.(\ref{LV}): without the $m_V^2$ term 
two mixing matrices ${\cal M}^2$ and ${\cal G}_{VW \phi}$ would become identical to be diagonalized simultaneously, 
so the $\tilde{V}$-$\tilde{W}$-$\phi$ coupling would completely be rotated away:
\begin{equation}
  {\cal G}_{\tilde{V} \tilde{W} \phi} = 0  \quad (m_V^2=0,\, m_{\tilde{V}}^2\ne 0)\,.
  \end{equation}
Note that absence of the $m_V^2$ term does not affect the mass $m_{\tilde{V}}^2$ of the new boson 
as clearly seen from Eq.(\ref{V-mass}) (See also Eq.(\ref{V-mass:C}) below).

Examining terms in Eq.(\ref{LV}) in quadratic order of the vector fields ${\bf V}_\mu$ = ($V_\mu$, $W_\mu$)$^T$ with the scale dimensions $d_{\bf V}=1$ taken into account,  
one readily realizes that 
the $m_V^2$ term in the model Lagrangian Eq.(\ref{LV}) transforms as   $\delta {\bf V}_\mu^2 = (2 +  x^\mu\partial_\mu) {\bf V}_\mu^2 $, 
which obviously violates the scale invariance.

Eliminating the $m_V^2$ term, we see that the mass matrix  reads
\begin{equation} 
{\cal M}^2|_{m_V=0}  
= 
 \left( 
\begin{array}{cc} 
g_V^2 c_{_{VVHH}} v^2 &  \frac{1}{4} g_W g_V c_{_{H}} v^2 \\ 
  \frac{1}{4} g_W g_V c_{_{H}} v^2 & \frac{1}{4} g_W^2 v^2 
\end{array} 
\right) 
= {\cal G}_{VW\phi} 
\,, \label{mass-matrix:C}
\end{equation}  
This is the same matrix as the ${\cal G}_{VW\phi}$ in Eq.(\ref{VHW}), 
hence the off-diagonal $\tilde{V}$-$\tilde{W}$-$\phi$ coupling goes away after the diagonalization of the vector boson 
sector: 
\begin{eqnarray} 
{\cal L}_V \Bigg|_{m_V=0} 
&=& 
\frac{1}{2} 
{\bf V}^T_\mu \cdot {\cal M}^2_{m_V=0} \cdot {\bf V}^\mu 
+ 
\frac{\phi}{v} \cdot {\bf V}_\mu^T \cdot {\cal G}_{VW\phi} \cdot {\bf V}^\mu 
+ 
\cdots  
\nonumber \\ 
&=& 
\frac{1}{2} 
\left(1 + \frac{2 \phi}{v} \right) 
{\bf V}^T_\mu \cdot 
{\cal M}^2_{m_V=0} 
\cdot 
{\bf V}^\mu 
+ \cdots 
\,. \label{LV:C}
\end{eqnarray} 
In terms of the mass eigenstate fields $\tilde{\bf V}_\mu = (\tilde{V}_\mu, \tilde{W}_\mu)^T$, 
the Lagrangian Eq.(\ref{LV:C}) goes like 
\begin{eqnarray} 
 {\cal L}_V \Bigg|_{m_V=0} 
& =& 
 \frac{1}{2} 
\left(1 + \frac{2 \phi}{v} \right) 
\tilde{\bf V}^T_\mu \cdot 
 \left( 
\begin{array}{cc} 
m_{\tilde V}^2 &  0 \\ 
  0 & m_{\tilde{W}}^2 
\end{array} 
\right)   
\cdot 
\tilde{\bf V}^\mu  
+ \cdots 
\,, 
\end{eqnarray}
with the masses of the mass eigenstate vectors $(m_{\tilde{V}}, m_{\tilde W})$. 
The corresponding mass eigenvalues are now modified from Eq.(\ref{V-mass}) to be 
\begin{eqnarray} 
 m_{\tilde V}^2 &\to& m_{\tilde V}^2 = c_{_{VVHH}} g_V^2 v^2 \left[ 1 + {\cal O}(x^2) \right] 
\,, \nonumber \\ 
 m_{\tilde W}^2 &\to& m_{\tilde W}^2 = \frac{g_W^2 v^2}{4} \left[ 1 - \frac{c_{_H}^2}{4 c_{_{VVHH}}} + {\cal O}(x^2) \right] 
\,. \label{V-mass:C}    
\end{eqnarray}
Thus the conformal barrier is realized by taking a special limit $m_V=0$ on 
the generic parameter space of the HVT model.

Now that we have established how to realize the conformal barrier from 
the generic HVT model, 
we show that the conformal barrier is actually incompatible with 
the ``equivalence-theorem" result for the vector boson. 
To this end, we explicitly compute the partial decay widths 
$\Gamma(\tilde{V} \to \tilde{W}\tilde{W})$ and $\Gamma(\tilde{V} \to \tilde{W} \phi)$ in the original HVT model Eq.(\ref{LV}) 
to get 
\begin{eqnarray} 
 \Gamma(\tilde{V} \to \tilde{W}\tilde{W}) 
&\simeq& 
 \Gamma(\tilde{V} \to \tilde{W}_L \tilde{W}_L)
 \simeq 
 \frac{g_{\tilde{V} \tilde{W}_L \tilde{W}_L}^2}{48 \pi} m_{\tilde V} 
 \,, \nonumber \\ 
  \Gamma(\tilde{V} \to \tilde{W}\phi) 
&\simeq &
\frac{1}{48 \pi} \frac{g_{\tilde{V}\tilde{W}\phi}^2}{4 m_{\tilde W}^2} m_{\tilde V} 
\,, \label{widths}
\end{eqnarray}  
where the limit $x=g_W/g_V \ll 1$ in Eq.(\ref{heavy-vector-limit}) 
was taken and the relevant couplings have come from 
the interaction Lagrangian parts: 
\begin{eqnarray} 
{\cal L}_{VW_LW_L} &=& 
 g_{\tilde{V} \tilde{W}_L \tilde{W}_L} \epsilon^{abc} \tilde{V}_\mu^a \partial^\mu \pi_W^b \pi_W^c 
\, \nonumber \\ 
{\cal L}_{VW\phi} &=& 
g_{\tilde{V}\tilde{W} \phi} \phi \tilde{V}_\mu^a \tilde{W}^{\mu a}  
\,, 
\end{eqnarray}
with the longitudinal component of $\tilde{W}$ being $\tilde{W}_L^\mu = \partial^\mu \pi_W/m_{\tilde W}$ 
and 
\begin{eqnarray} 
 g_{\tilde{V} \tilde{W}_L \tilde{W}_L } 
 &\simeq & \frac{ m_{\tilde V}^2 c_{_H}  g_V }{4 m_{\tilde V}^2 - c_{_H}^2 g_V^2 v^2} 
 \,, \nonumber \\ 
 g_{\tilde{V}\tilde{W} \phi}
 &\simeq & 
 \frac{m_V^2}{4 m_{\tilde V}^2} c_{_H} g_V g_W v
 \,. \label{couplings}
\end{eqnarray}
In addition to the limit in Eq.(\ref{heavy-vector-limit}), 
one now considers the mass parameter $m_V$ to be   
\begin{equation} 
 m_V \gg g_V v (\gg g_W v) 
 \,, \label{ET:limit}
\end{equation}
 such that the vector boson mass $m_{\tilde V}$ is almost saturated by the bare mass $m_V$. 
Then the couplings in Eq.(\ref{couplings}) approximately look like 
\begin{eqnarray} 
  g_{\tilde{V} \tilde{W}_L \tilde{W}_L } 
 &\approx & \frac{c_{_H}  g_V }{4} 
 \,, \nonumber \\ 
 g_{\tilde{V}\tilde{W} \phi}
 &\approx& 
 \frac{1}{4} c_{_H} g_V g_W v \simeq g_{\tilde{V} \tilde{W}_L \tilde{W}_L} (2 m_{\tilde W}) 
\,. \label{tree:coupling}
\end{eqnarray}
Hence one reaches the ``equivalence theorem", 
\begin{equation} 
\Gamma(\tilde{V} \to \tilde{W}\tilde{W}) 
\approx 
\Gamma(\tilde{V} \to \tilde{W}\phi) 
\approx 
\frac{g_{\tilde{V} \tilde{W}_L \tilde{W}_L}^2}{48 \pi} m_{\tilde V} 
\,. 
\end{equation}
Thus the limit Eq.(\ref{ET:limit}) to realize the ``equivalence-theorem" result   
is incompatible with the limit where the conformal/scale invariance is present, $m_V \to 0$.

\section{Power of Scale-invariant Hidden Local Symmetry} 
\label{power-of-HLS}

Hereafter we shall employ an explicit model in which the conformal barrier is 
realized and discuss the phenomenological consequences from 
the conformal barrier and the HLS.

The s-HLS Lagrangian in Eq.(\ref{sHLS}) is the effective theory  
realizing the (approximate) scale/conformal invariance and chiral symmetry of 
the underlying theory, the walking technicolor~\cite{Yamawaki:1985zg}. 
One phenomenologically interesting candidate for the walking technicolor is 
the one-family model having the technifermion flavor of the number 8 ($N_F=8$). 
In the model, the LHC Higgs is identified with the technidilaton ($\phi$), 
a composite pseudo NG boson for the (approximate) conformal/scale symmetry,  
and the new vector bosons are the technirhos ($V$).

The way of constructing the s-HLS for the one-family model 
will be just a straightforward extension of the procedure described in Sec.~\ref{HiddenSymmetries} for the simplest case of $N_F=2$. 
The features characteristic to $N_F=8$ will be seen when one considers 
the way of 
embedding the SM gauge fields and technirho fields into the $8 \times 8$ matrix 
form (See the discussions below).

\subsection{Preliminaries: the s-HLS for $N_F=8$}

The s-HLS action reflecting the underlying one-family model is constructed 
by nonlinear realization based on the coset space 
$G/H=SU(8)_L \times SU(8)_R/SU(8)_V$  
with the basic variable $U=e^{2 i\pi/F_\pi}$.  
We write the $U$:   
$ 
U(x)=e^{ i \frac{2\pi(x)}{F_\pi}} =\xi_L^\dagger(x) \cdot \xi_R(x)
$, in the same way as in Eq.(\ref{U:decomp}) for the simplest $N_F=2$ case. 
The variables  $\xi_{L/R}$ transform as  
$ 
\xi_{L/R} \rightarrow h(x)\,  \xi_{L/R} \, g_{L/R}^\dagger 
$, 
with $h(x) \in H_{\rm local}=SU(8)_{V}$ and $g_{L/R} \in G_{\rm global}=
SU(8)_L \times SU(8)_R$.  
The theory then has a larger symmetry $G_{\rm global} \times H_{\rm local}$, 
$H_{\rm local}$ being the HLS, where 
the redundant symmetry (HLS), $H_{\rm local}$ symmetry, 
is a spontaneously broken and exact gauge symmetry (Higgs mechanism),
with the vector mesons as the HLS gauge bosons: 
the HLS gauge bosons of $H_{\rm local}$ become massive  
by this spontaneous breakdown (Higgs mechanism) and is quantized as a unitary quantum theory, 
just like the SM electroweak gauge theory which is exact, i.e.,
without  explicit breaking of the gauge symmetry (though spontaneously broken).

We parametrize the $\xi_{L,R}$ as 
\begin{equation} 
 \xi_{L,R}(x) = e^{\frac{i {\cal P}(x)}{F_{\cal P}}} e^{\mp  \frac{i \pi(x)}{F_\pi}}  
 \,, 
 \qquad 
 ({\pi(x)=\pi^A(x) X^A}\,, \qquad {{\cal P}(x) = {\cal P}^A(x) X^A} )
\,, 
\end{equation} 
Here the broken generators are $X^A$ ($A=1,\cdots 63$) and  
the fictitious NG bosons ${\cal P}^A(x)$ 
along with the decay constant $F_{\cal P}$, 
related to the parameter $a$ in Eq.(\ref{sHLS}), as 
\begin{equation}  
a= \frac{F_{\cal P}^2}{F_\pi^2}
\,. 
\end{equation} 
The ${\cal P}^A(x)$ are to be absorbed into the hidden local gauge degree of freedom. 
When the hidden local gauge is fixed (e.g. unitary gauge ${\cal P}(x)=0$) as $\xi_L^\dagger(x) =\xi_R(x)=\xi(x)=e^{ i \frac{\pi(x)}{F_\pi}} $,  
$H_{\rm local}$ and $H_{\rm global}(\subset G_{\rm global})$ are both spontaneously broken down to a single $H$ which is a diagonal sum of both of them. 
Then $G_{\rm global}$ is reduced back to the original chiral symmetry $G(\neq G_{\rm global})$ in the model based on $G/H$. 
$\xi$ transforms as $\xi \rightarrow h(g, \pi)\, \xi\, g^\dagger_{L,R}$, with $h(g, \pi)$ being the $\pi(x)$-dependent (global) $H$ transformation of $G/H$.

As to the scale transformation property, 
the $\xi_{L,R}$ have no scale dimension so that they transform under the scale symmetry as 
\begin{equation} 
  \delta \xi_{L,R}(x) = x_\nu \partial^\nu \xi_{L,R}(x) 
  \,. 
\end{equation}

 Now switch on the SM gauge boson fields ($G_\mu, W_\mu, B_\mu$) 
by gauging the $G_{\rm global}$ 
whose charges ($g_c, g_W, g_Y$) are independent of the 
charge $g$ of the independent gauge symmetry, 
the $H_{\rm local}$. 
Then the $G_{\rm global}$ is explicitly broken by this gauging, 
so being the original $G$ 
in the unitary gauge, 
since $G_{\rm global}$ is explicitly broken. 
However, the HLS as well as the SM gauge symmetry is exact. 
The covariant derivatives acting on the $\xi_{L,R}$ are then written, in a way 
similar to Eq.(\ref{HLScovariant}) for $N_F=2$ case,  
to be  
\begin{equation}  
D_\mu \xi_{L,R} (x) = 
\partial_\mu \xi_{L,R}(x) - i V_\mu(x) \xi_{L,R}(x) + i \xi_{L,R} {\cal L}_\mu(x)({\cal R}_\mu(x))
\, , 
\end{equation} 
with the HLS gauge field $V_\mu$  and the external gauge fields ${\cal L}_\mu$ and ${\cal R}_\mu$, 
in which the $SU(3)_c \times SU(2)_W \times U(1)_Y)$ 
gauges are fully embedded, independently of the HLS. 
In addition, the technirho field strengths are defined as 
\begin{equation} 
  V_{\mu\nu} = \partial_\mu V_\nu - \partial_\nu V_\mu - i [V_\mu, V_\nu]
  \,,  
\end{equation}
which transform under the HLS homogeneously 
\begin{equation}
 V_{\mu\nu} \to h(x) \cdot V_{\mu\nu} \cdot h^\dag(x) 
 \,. 
\end{equation}

Finally, 
we introduce the nonlinear base for the scale symmetry $\chi(x)$ parametrized by 
the technidiaton field $\phi$, which is the same as in Eq.(\ref{SNL}), 
as 
\begin{equation} 
 \chi(\phi) = e^{\phi/F_\phi} 
 \,, 
\end{equation} 
with the decay constant $F_\phi$ (not necessarily identical to $F_\pi$). 
The $\chi$ transforms under the scale symmetry just like fields having the scale dimension 1, 
\begin{equation} 
 \delta \chi(x) = (1 + x_\nu \partial^\nu) \chi(x) 
 \,, 
\end{equation}
while the technidilaton field does nonlinearly, 
\begin{equation} 
 \delta \phi(x) = F_\phi + (1 + x_\nu \partial^\nu ) \phi(x) 
\,.   
\end{equation}

The s-HLS Lagrangian is thus written at the leading order of derivatives  
to be 
\begin{equation} 
 {\cal L}_{(2)} =  \chi^2(x) \cdot \left( 
\frac{1}{2} (\partial_\mu \phi)^2 
+
F_\pi^2 {\rm tr}[ \hat{\alpha}_{\mu \perp}^2] + a 
F_\pi^2 {\rm tr}[\hat{\alpha}_{\mu ||}^2]  \right) 
  - \frac{1}{2g^2} {\rm tr}[V_{\mu\nu}^2] 
  \,, \label{Lag:p2}
\end{equation}
where the gauge coupling $g$ is counted as ${\cal O}(p)$, $\phi(x)$ as ${\cal O}(p^0)$,
 and  
\begin{equation}
  \hat{\alpha}_{\mu \perp,||} = \frac{ D_\mu \xi_R \cdot \xi_R^\dag \mp D_\mu \xi_L \cdot \xi_L^\dag}{2i} 
\,, \label{trans:alpha}
\end{equation}
which transform  as 
\begin{equation} 
\hat{\alpha}_{\mu \perp,||} \rightarrow h(x)\cdot \hat{\alpha}_{\mu \perp,||}\cdot h^\dag(x) 
\,. 
\end{equation} 
Note again that without the kinetic term of the HLS gauge fields $V_\mu(x)$ (namely by integrating out the $V_\mu$), 
the Lagrangian is reduced 
to the nonlinear sigma model based on $G/H$ in the unitary gauge ${\cal P}(x)=0$ ($\xi_L^\dagger(x) =\xi_R(x)=\xi(x)=e^{ i \frac{\pi(x)}{F_\pi}} $).

The conformal/scale symmetry of the s-HLS is explicitly broken by 
the technidilaton mass in the potential of the 
form:
\begin{equation}
V(\chi) = \frac{F_\phi^2 m_\phi^2}{4} \chi^4 \left(\log \chi -\frac{1}{4} \right)\,,
\end{equation}
 which corresponds to the trace anomaly of the underlying walking technicolor: 
$\langle \theta^\mu_\mu\rangle = \langle \partial^\mu {\bf D}_\mu\rangle 
=-\delta V(\phi) = -\frac{m_\phi^2F_\phi^2}{4} \langle \chi^4\rangle 
= -\frac{m_\phi^2F_\phi^2}{4}$ 
in accord with the PCDC, 
with $m_\phi^2 F_\phi^2\simeq (v_{\rm EW}/2)^2\cdot (5 v_{\rm EW})^2 \cdot (8/N_F) (4/N_C)$ ($v_{\rm EW}=246 \,{\rm GeV}$) 
in the ladder calculations (See e.g., \cite{Matsuzaki:2015sya}). 
It was shown~\cite{Matsuzaki:2012gd,Matsuzaki:2012xx} that the technidilaton for the walking technicolor with $N_F=8$ and $N_C=4$ 
($F_\phi \simeq 5 v_{\rm EW}$) is best fit to the current data of the 125 GeV Higgs. 
The effect of the conformal/scale symmetry breaking arises only at ${\cal O}(p^6)$ or higher orders 
of the derivative expansion, since the ${\cal O}(p^4)$ terms are 
already scale-invariant without involving the technidilaton field $\chi = e^{\phi/F_\phi}$. 
Thus, additional Higgs (= $\phi$) potential terms are not generated at the ${\cal O}(p^4)$.

\subsection{Embedding gauge and pion fields into the s-HLS}

\begin{table}[t] 
\begin{tabular}{c|c|c|c|c}
\hline 
\hspace{15pt} technipion \hspace{15pt} & 
\hspace{15pt} technirho \hspace{15pt} & 
\hspace{15pt} constituent \hspace{15pt} & 
\hspace{15pt} color \hspace{15pt} & 
\hspace{15pt} isospin \hspace{15pt} 
\\  
\hline \hline 
$\theta_a^{i}$ & $\rho_{\theta_a}^i$ &  
$\frac{1}{\sqrt{2}}\bar{Q} \lambda_a \tau^{i} Q$ & octet & triplet \\  
$\theta_a^0$ & $\rho_{\theta_a}^0$ & 
$ \frac{1}{2\sqrt{2}} \bar{Q} \lambda_a Q$ & octet & singlet \\ 
$T_c^{i}$ $(\bar{T}_c^{i})$ & $\rho_{T_c}^i$ ($\bar{\rho}_{T_c}^i$) & 
$\frac{1}{\sqrt{2}} \bar{Q}_c \tau^{i} L$ (h.c.) & triplet & triplet  \\ 
$T_c^0$ ($\bar{T}_c^0$) & $\rho_{T_c}^0$ ($\bar{\rho}_{T_c}^0$) & 
$ \frac{1}{2 \sqrt{2}} \bar{Q}_c L $ (h.c.)   & triplet & singlet \\ 
\hline 
$P^i $ & $\rho_{P}^i$ & 
$\frac{1}{2 \sqrt{3}} (\bar{Q} \tau^i Q - 3 \bar{L} \tau^i L)$ & singlet & triplet \\  
$P^0$ & $\rho_{P}^0$ &   
$\frac{1}{4 \sqrt{3}} (\bar{Q} Q - 3 \bar{L}  L)$  & singlet & singlet \\  
\hline 
$\Pi^i $ & $\rho_{\Pi}^i$ & 
$\frac{1}{2} (\bar{Q} \tau^i Q + \bar{L} \tau^i L)$ & singlet & triplet \\   
\hline 
\end{tabular} 
\caption{ The technipions, technirhos and 
their associated constituent techni-quarks $Q_c=(U,D)_c$ and leptons $L=(N,E)$. 
Here $\lambda_a$ ($a=1,\cdots,8$) are the Gell-Mann matrices, 
$\tau^i$ $SU(2)$ generators defined as 
$\tau^i=\sigma^i/2$ ($i=1,2,3$) with the Pauli matrices $\sigma^i$, and 
the label $c$ stands for the QCD-three colors, $c=r,g,b$. 
}
\label{tab:TPTR} 
\end{table}

The 63 chiral NG boson (technipion) fields are classified by the isospin and QCD color charges, 
which are listed in Table~\ref{tab:TPTR}, 
where the notation follows the original literature~\cite{Farhi:1979zx}. 
The three of them ($\Pi^i$) become unphysical to be eaten by $W$ and $Z$ bosons in the same way as in 
the usual Higgs mechanism, while the other sixty Nambu-Goldstone bosons become pseudos, techni-pions, to be massive in several ways. 
They are embedded in the adjoint representation of $SU(8)$ group as~\cite{Jia:2012kd,Kurachi:2014qma}: 
\begin{eqnarray} 
\sum_{A=1}^{63} \pi^A(x) X^A 
&=& 
\sum_{i=1}^3 \Pi^i(x) X_{\Pi}^i  
+ 
\sum_{i=1}^3 P^i(x) X^i_{P} + P^0 (x) X_{P} 
\nonumber \\  
&& 
+ \sum_{i=1}^3 \sum_{a=1}^8 \theta^i_a (x) X_{\theta a}^i 
+ 
\sum_{a=1}^8 \theta_a^0 (x) X_{\theta a}  
\nonumber \\ 
&& 
+ 
\sum_{c=r,g,b} \sum_{i=1}^3 \left[ T_c^{i}(x) X_{T c}^{i} + \bar{T}_c^{i}(x) X_{\bar{T} c}^{i} \right] 
+
\sum_{c=r,g,b} \left[ T_c^0 (x) X_{T c}  + \bar{T}_c^0 (x) X_{\bar{T} c} \right] 
\,, 
\end{eqnarray} 
where ($\tau^i=\sigma^i/2$)
\begin{eqnarray} 
 X_{\Pi}^i 
 &=& 
\frac{1}{2} \left( 
\begin{array}{c|c} 
  \tau^i \otimes {\bf 1}_{3\times 3}  &  \\ 
  \hline 
   & \tau^i 
\end{array}
\right)  \,, 
\qquad 
X^i_P 
= 
\frac{1}{2 \sqrt{3}}
\left( 
\begin{array}{c|c} 
  \tau^i \otimes {\bf 1}_{3\times 3}  &  \\ 
  \hline 
   & -3 \cdot \tau^i 
\end{array}
\right)  \,, \qquad 
X_P
= 
\frac{1}{4 \sqrt{3}}
\left( 
\begin{array}{c|c} 
  {\bf 1}_{6 \times 6}  &  \\ 
  \hline 
   & -3\cdot {\bf 1}_{2 \times 2} 
\end{array}
\right) 
\,, \nonumber \\ 
 X_{\theta a}^i 
 &=& 
\frac{1}{\sqrt{2}}  \left( 
\begin{array}{c|c} 
  \tau^i \otimes \lambda_a  &  \\ 
  \hline 
   & 0 
\end{array}
\right)  \,, 
\qquad 
X_{\theta_a} 
= 
\frac{1}{2 \sqrt{2}}
\left( 
\begin{array}{c|c} 
  {\bf 1}_{2\times 2} \otimes \lambda_a  &  \\ 
  \hline 
   & 0  
\end{array}
\right)  \,, \nonumber \\ 
X^{i}_{T c} 
&=&  
\frac{1-i}{2}
\left( 
\begin{array}{c|c} 
 & \tau^i \otimes {\bf e}_c    \\ 
  \hline 
\tau^i \otimes {\bf e}_c^\dag  &  
\end{array}
\right) 
\,, \qquad 
X^{i}_{\bar{T} c}  
= 
\left( X^{i}_{T c}  \right)^\dagger
\,, \nonumber \\ 
X_{T c}  
&=&  
\frac{1-i}{4}
\left( 
\begin{array}{c|c} 
 & {\bf 1}_{2 \times 2} \otimes {\bf e}_c    \\ 
  \hline 
{\bf 1}_{2 \times 2} \otimes {\bf e}_c^\dag  &  
\end{array}
\right) 
\,, \qquad 
X_{\bar{T} c}  
= 
\left( X_{T c}  \right)^\dagger
\,, 
\end{eqnarray}
with 
${\bf e}_c$ being a three-dimensional unit vector in color space  
and the generators normalized as ${\rm Tr}[X^A X^B]=\delta^{AB}/2$ (except for $X_{T_c}^{i,0}$ and $\bar{X}_{T_c}^{i,0}$ which respectively satisfy the normalization 
${\rm tr}[X_{T_c}^{i} \bar{X}_{T_c}^{j}]=\delta^{ij}/2$ and 
${\rm tr}[X_{T_c}^{0} \bar{X}_{T_c}^{0}]=1/2$).  
Among the above, $\Pi^i$ become longitudinal degrees of freedom of the SM $W^{\pm}$ and $Z$ bosons. 
It is convenient to express $\pi$ in a blocked $8 \times 8$ matrix form as 
\begin{equation} 
 \pi^A X^A = 
 \Bigg(
 \begin{array}{c|c} 
(\pi_{QQ})_{6 \times 6} & (\pi_{QL})_{2 \times 6} \\ 
\hline  
 (\pi_{LQ})_{6 \times 2} & (\pi_{LL})_{2 \times 2} 
 \end{array} 
 \Bigg) 
 \,, \label{pi:para}
\end{equation}
where 
\begin{eqnarray} 
\pi_{QQ} &=& \left[ \sqrt{2} \theta  + \frac{1}{\sqrt{2}} \theta^0   \right] 
+ \left(  \frac{1}{2} \Pi  + \frac{1}{2\sqrt{3}} P  + \frac{1}{4 \sqrt{3}} P^0 \right) \otimes {\bf 1}_{3 \times 3}
\,, \nonumber \\ 
\pi_{QL} &=&   T + \frac{1}{2} T^0  
\,, \nonumber \\ 
\pi_{LQ} &=& \pi_{QL}^\dag = 
\bar{T}  + \frac{1}{2} \bar{T}^0  
\,, \nonumber \\ 
\pi_{LL} &=& 
 \left( \frac{1}{2} \Pi  - \frac{\sqrt{3}}{2} P  - \frac{\sqrt{3}}{4} P^0   \right) 
 \,, \nonumber \\
\theta &=& \theta_a^i \tau^i \frac{\lambda_a}{2} 
\,, \qquad 
\theta^0 = \theta_a^0  \cdot {\bf 1}_{2 \times 2} \cdot \frac{\lambda_a}{2} 
\,, \nonumber \\ 
T &=& T_c^i {\bf e}_c \tau^i 
\,, \qquad 
T^0 = T_c^0 {\bf e}_c 
\,, \nonumber \\ 
P &=&  P^i \tau^i 
\,, \qquad 
P^0 = P^0 \cdot {\bf 1}_{2 \times 2} 
\,, \nonumber \\ 
\Pi &=& \Pi^i \tau^i 
\,.   \nonumber 
\end{eqnarray}

All the 60 technipions not eaten by the $W$ and $Z$ bosons (except for the $\Pi$'s) 
in the one-family model acquire  
masses   due to the explicit breaking of the $[SU(8)_L\times SU(8)_R]_{\rm global}$ chiral symmetry by the electroweak and QCD as well as the extended technicolor couplings,
which are enormously enhanced to  the order of ${\cal O}(\textrm{a few TeV})$ by the large anomalous dimension $\gamma_m \simeq 1$
as a salient  feature of the walking technicolor~\cite{Kurachi:2014xla}. 
Hence they should be larger, or as large as  the technirho mass. 
Thereby, we will ignore the 2 TeV $\rho_\Pi$ couplings to technipions.

The 63 technirho fields are classified in terms of the technifermion fields by the SM charges as listed in Table~\ref{tab:TPTR}   
in a way similar to $\pi$. (The notation of the generators is taken to be the same as that of $\pi$, though the generators for the technirho fields 
are not broken generators.) 
They are parametrized as   
\begin{eqnarray} 
\sum_{A=1}^{63} \rho^A(x) X^A 
&=& 
\sum_{i=1}^3 \rho_{\Pi}^i  (x) X_{\Pi}^i  
+ 
\sum_{i=1}^3  \rho_{P}^i (x) X^i_{P} +  \rho_{P}^0  (x) X_{P} 
\nonumber \\  
&& 
+ \sum_{i=1}^3 \sum_{a=1}^8  \rho_{\theta a}^i  (x) X_{\theta a}^i 
+ 
\sum_{a=1}^8 \rho_{\theta_a}^0 (x) X_{\theta a}  
\nonumber \\ 
&& 
+ 
\sum_{c=r,g,b} \sum_{i=1}^3 \left[ \rho_{T c}^{i}(x) X_{T c}^{i} + \bar{\rho}_{T c}^{i}(x) X_{\bar{T} c}^{i} \right] 
+
\sum_{c=r,g,b} \left[ \rho_{T c}^0(x) X_{T c}  + \bar{\rho}_{T_c}^0(x) X_{\bar{T} c} \right] 
\,. \label{rho-parametrize}
\end{eqnarray}  
They are embedded in a  $8 \times 8$ block-diagonal form, $V_\mu=V^A_\mu X^A$, as   
\begin{equation} 
 \rho^\mu  = \frac{V^{\mu A} X^A}{g} = 
 \Bigg(
 \begin{array}{c|c} 
(\rho^\mu_{QQ})_{6 \times 6} & (\rho^\mu_{QL})_{2 \times 6} \\ 
\hline  
 (\rho_{LQ}^\mu)_{6 \times 2} & (\rho^\mu_{LL})_{2 \times 2} 
 \end{array} 
 \Bigg) 
 \,, \label{rho:para}
\end{equation}
with 
\begin{eqnarray} 
\rho^\mu_{QQ} &=& \left[ \sqrt{2} \rho^{\mu}_\theta + \frac{1}{\sqrt{2}} \rho^{\mu 0}_\theta   \right]  
+ \left(  \frac{1}{2} \rho^{\mu}_{\Pi}  + \frac{1}{2\sqrt{3}} \rho^{\mu}_P  + \frac{1}{4 \sqrt{3}} \rho^{0 \mu}_P  \right) \otimes {\bf 1}_{3\times 3} 
\,, \nonumber \\ 
\rho_{QL}^\mu &=& \rho^\mu_T + \frac{1}{2} \rho^{0\mu}_T
\,, \nonumber \\ 
\rho^\mu_{LQ} &=& (\rho^\mu_{QL})^\dag = 
\bar{\rho}^\mu_T  + \frac{1}{2} \bar{\rho}^{0\mu}_T  
\,, \nonumber \\ 
\rho^\mu_{LL} &=& 
 \left( \frac{1}{2} \rho^\mu_\Pi  - \frac{\sqrt{3}}{2} \rho^\mu_P - \frac{\sqrt{3}}{4} \rho^{0\mu}_P \right) 
 \,,\nonumber \\
 \rho_{\theta}^\mu &=& \rho_{\theta a}^{i\mu} \tau^i \frac{\lambda_a}{2} 
\,, \qquad 
\rho^{0\mu}_\theta = \rho_{\theta a}^{0\mu}  \cdot {\bf 1}_{2 \times 2} \cdot \frac{\lambda_a}{2} 
\,, \nonumber \\ 
\rho^\mu_T &=& \rho_{Tc}^{i\mu} {\bf e}_c \tau^i 
\,, \qquad 
\rho^{0\mu}_T = \rho_{Tc}^{0\mu} {\bf e}_c 
\,, \nonumber \\ 
\rho_P^\mu &=&  \rho_P^{i\mu} \tau^i 
\,, \qquad 
\rho_P^{0\mu} = \rho_P^{0\mu} \cdot {\bf 1}_{2 \times 2} 
\,, \nonumber \\ 
\rho^\mu_\Pi &=& \rho^{i\mu}_\Pi \tau^i 
\,. \nonumber 
\end{eqnarray}

Here we used the same basis of $SU(8)_V$ matrix as that of $\pi$. 
This base allows us to definitely separate the isospin-triplet technirhos into two classes: 
one can be produced by the Drell-Yan (DY) process through the mixing with $W$ and $Z$ bosons, 
while the other cannot.  
The former corresponds to $\rho_\Pi^i$ in Table~\ref{tab:TPTR} and 
the latter is $\rho_P^i$ which can be seen from the orthogonality of ${\rm tr}[X_{\Pi}^i X_P^j]=0$. 
We shall hereafter call this base the {\it DY-base}.    
We will be back to this point in the next subsection.

The external gauge fields ${\cal L}_\mu$ and ${\cal R}_\mu$ involve the $SU(3)_c$, $SU(2)_W$ and $U(1)_Y$ gauge fields 
$(G_\mu, W_\mu, B_\mu)$ in the SM as follows: 
\begin{eqnarray} 
{\cal L}_\mu 
&=& 
2 g_W W_\mu^i X^i_{\Pi} + \frac{2}{\sqrt{3}} g_Y B_\mu X_P  + \sqrt{2} g_s  G_\mu^a X_{\theta_a}  
\,, \nonumber \\ 
{\cal R}_\mu 
&=&
2 g_Y B_\mu \left( X_{\Pi}^3 + \frac{1}{\sqrt{3}} X_P \right) + \sqrt{2} g_s G_\mu^a X_{\theta_a} 
\,. 
\end{eqnarray}
Through the standard diagonalization procedure, the left and right gauge fields are 
expressed in terms of the mass eigenstates $(W^\pm, Z, \gamma, g)$ as 
\begin{eqnarray} 
{\cal L}_\mu 
&=& 
g_s G_\mu^a \Lambda^a + e Q_{\rm em} A_\mu + \frac{e}{sc} \left( I_3 - s^2 Q_{\rm em} \right) Z_\mu + 
\frac{e}{\sqrt{2}s} \left( W_\mu^+ I^+ + {\rm h.c.} \right)
\,, \nonumber \\ 
{\cal R}_\mu 
&=&
g_s G_\mu^a \Lambda^a + e Q_{\rm em} A_\mu  - \frac{es}{c} Q_{\rm em} Z_\mu  
\,, 
\end{eqnarray}
where $s$ ($c^2=1-s^2$) denotes the standard weak mixing angle defined by $g_W=e/s$ and $g_Y=e/c$, and  
\begin{eqnarray} 
\Lambda_a &=& \sqrt{2} \, X_{\theta_a} \,, \qquad  
I_3 = 2 \, X_{\Pi}^3 \,  , \qquad 
Q_{\rm em} = I_3 + Y \,, \nonumber \\ 
Y&=& \frac{2}{\sqrt{3}} X_P  \,, \qquad 
I_+ = 2(X_{\Pi}^1 + i X_{\Pi}^2)   \,, \qquad 
I_- = (I_+)^\dag 
\,. 
\end{eqnarray}
It is convenient to define the vector and axial-vector gauge fields ${\cal V}_\mu$ and ${\cal A}_\mu$ as
\begin{equation} 
 {\cal V}_\mu = \frac{{\cal R}_\mu + {\cal L}_\mu }{2} 
 \,,\qquad
 {\cal A}_\mu = \frac{{\cal R}_\mu - {\cal L}_\mu}{2} 
\,, 
\end{equation}  
so that they are expressed in a blocked-$8 \times 8$ matrix form:  
\begin{eqnarray} 
 {\cal V}^\mu &=& 
 \Bigg( 
 \begin{array}{c|c} 
 ({\cal V}^\mu_{QQ})_{6 \times 6} & {\bf 0}_{2 \times 6} \\ 
 {\bf 0}_{6 \times 2} & ({\cal V}^\mu_{LL})_{2 \times 2}
\end{array} 
 \Bigg) 
 \,,\label{v:para} \\ 
  {\cal A}^\mu &=& 
 \Bigg( 
 \begin{array}{c|c} 
 ({\cal A}^\mu_{QQ})_{6 \times 6} & {\bf 0}_{2 \times 6} \\ 
 {\bf 0}_{6 \times 2} & ({\cal A}^\mu_{LL})_{2 \times 2}
\end{array} 
 \Bigg) 
 \,, \label{v:a:para}
\end{eqnarray}
where 
\begin{eqnarray} 
 {\cal V}^\mu_{QQ} &=& 
{\bf 1}_{2 \times 2} \cdot g_s G^\mu_a \frac{\lambda_a}{2}   
+ 
\left[ 
 e Q_{\rm em}^q A^\mu + \frac{e}{2sc} z_V^q Z^\mu + \frac{e}{2\sqrt{2}s} \left( \tau^+ W^{\mu +} + {\rm h.c.} \right)
\right] \cdot {\bf 1}_{3 \times 3} 
\,, \nonumber \\ 
 {\cal V}^\mu_{LL} &=& 
\left[ 
 e Q_{\rm em}^l A^\mu + \frac{e}{2sc} z_V^l Z^\mu + \frac{e}{2\sqrt{2}s} \left( \tau^+ W^{\mu +} + {\rm h.c.} \right)
\right]
\,, \nonumber \\ 
{\cal A}^\mu_{QQ} &=& 
- \left(
 \frac{e}{2sc} \tau^3 Z^\mu + \frac{e}{2 \sqrt{2}s}\left( \tau^+ W^{\mu +} + {\rm h.c.} \right) 
\right) \cdot {\bf 1}_{3 \times 3} 
\,, \nonumber \\
{\cal A}^\mu_{LL} &=& 
- \left(
 \frac{e}{2sc} \tau^3 Z^\mu + \frac{e}{2 \sqrt{2}s}\left( \tau^+ W^{\mu +} + {\rm h.c.} \right) 
\right) 
\,, \nonumber \\
Q_{\rm em}^{q} &=& 
\left(
\begin{array}{cc}
 2/3 & 0 \\ 
 0 & -1/3 
\end{array}
\right) 
\,, \qquad 
Q_{\rm em}^{l} = 
\left(
\begin{array}{cc}
 0 & 0 \\ 
 0 & -1 
\end{array}
\right) 
\,, \nonumber \\ 
z_V^{q,l} &=& \tau^3 - 2 s^2 Q_{\rm em}^{q,l} 
\,, \qquad 
\tau^+ = 
\left( 
 \begin{array}{cc}
 0 & 1 \\ 
 0 & 0 
\end{array}
\right) 
\,, \qquad 
\tau^- = (\tau^+)^\dag  
\,. \nonumber 
\end{eqnarray}

\subsection{Walking Technirho Couplings}

Now that all the relevant fields are explicitly incorporated into the s-HLS Lagrangian 
Eq.(\ref{Lag:p2}), it is straightforward to extract the relevant couplings between 
technirhos and SM gauge fields. 
In the present study we shall focus on the technirhos which allow to couple to the dibosons $W$ and $Z$.

Among 63 technirhos classified as in Table~\ref{tab:TPTR},  
it turns out that only the $\rho_\Pi^i$ and $\rho_P^0$ are relevant to the study of diboson signatures  
thanks to the HLS: actually, there exist other isospin triplet technirhos, $\rho_P^i$. 
However, the $\rho_P^i$ does not couple to the diboson due to the $SU(8)_V$ orthogonality:  
\begin{equation} 
\rho_P^i - W -W/Z \qquad  {\rm coupling} =0
\qquad 
{\rm since} 
\qquad  
{\rm tr}[X_\Pi^i X_P^j]=0
\, . 
\end{equation} 
One might further suspect the presence of mixing between isospin-triplet technirhos $\rho_\Pi^i$ and $\rho_P^i$ 
because the $SU(8)_V$ symmetry is merely approximate to be explicitly broken by 
the SM gauge interactions. 
However, it is not the case thanks to the gauge invariance of the HLS.  
We shall first demonstrate this point below.

\subsubsection{Power of the HLS}

The SM gauge bosons ($G_\mu, W_\mu, B_\mu$)  introduced by gauging the {\it $G_{\rm global}$ 
whose charges ($g_s, g_W, g_Y$) are independent of the charge of the HLS ($g$) since it is 
the independent gauge symmetry, $H_{\rm local}$}. 
Then the global chiral $G_{\rm global}=[SU(8)_L \times SU(8)_R]_{\rm global}$ is explicitly broken by this gauging, 
while the HLS as well as the SM gauge symmetry is intact. 
For instance, the HLS prohibits the $SU(8)_V$ breaking terms like 
\begin{equation} 
 {\rm tr}[\hat{\alpha}_{\mu ||} \hat{\alpha}^\mu_{||} X_{\theta_a}]
 \,, \qquad 
 {\rm tr}[\hat{\alpha}_{\mu ||} X_{\theta_a} \hat{\alpha}^\mu_{||} X_{\theta_a}]
\,, \label{breaking}
\end{equation}
which, if they are present, would yield the $\rho_\Pi^i$ - $\rho_P^i$ mixing at the lowest order of 
the derivative ${\cal O}(p^2)$. 
Obviously, terms such as above explicitly break the HLS (Recall the transformation law of $\hat{\alpha}_{\mu ||}$ in Eq.(\ref{trans:alpha})), 
hence these are not incorporated in the Lagrangian Eq.(\ref{Lag:p2}) which has to be gauge invariant under the HLS.

Actually, one can write down $G_{\rm global}$-breaking terms without breaking the HLS as done in Ref.~\cite{Bando:1985rf} for $N_F=3$ case, 
which do not take the invariant form made only by the 1-form $\hat{\alpha}_{\mu ||,\perp}$ covariantized through the SM gauging.  
Note also that 
the explicit breaking of the gauge symmetry, the HLS  
simply destroys the unitarity of the quantized theory, i.e., meaningless as a quantum theory. 
Unitarity of the quantum theory is manifestly proved in terms of the BRS symmetry which is a quantum version of the gauge symmetry.

If one employed vector boson models other than the HLS, 
vector bosons would be introduced as conventional massive-spin 1 fields of the $SU(8)_V$ flavor symmetry, $R_\mu$, 
transforming like 
$R_\mu \to g_V \cdot R_\mu  \cdot g_V^\dag$, where $g_V \in SU(8)_V$. 
Since the global $SU(8)_V$ 
is nothing but an approximate symmetry to be explicitly broken by the SM gauges, 
one naively introduces terms such as in Eq.(\ref{breaking}), 
$
{\rm tr}[R_\mu R^\mu  X_{\theta_a}] \,, \cdots  
\,, 
$
leading to the isospin-triplet $\rho_\Pi^i$ - $\rho_P^i$ mixing.

  \begin{figure}[t]
\begin{center}
   \includegraphics[width=6.5cm]{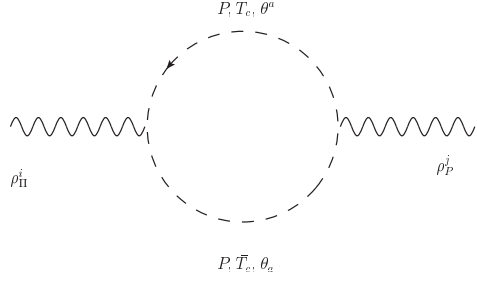} 
\caption{ 
The Feynman graphs contributing to the $\rho_\Pi^i$ - $\rho_P^j$ two-point function at one-loop level. 
\label{pi-loop-graphs}
}
\end{center} 
 \end{figure}

Beyond the leading order, one might think that loop corrections would generate the 
$\rho_\Pi^i$- $\rho_P^i$ mixing, for instance, by the technipion and SM gauge boson loops. 
However, it cannot happen, either, because of the HLS invariance: 
examining the s-HLS Lagrangian in Eq.(\ref{Lag:p2}) and expanding terms 
in powers of technipion fields, one readily finds the $\rho$-$\pi$-$\pi$ coupling term in the $8 \times 8$ matrix form, 
\begin{equation} 
 ag {\rm tr}[\rho_\mu [\partial^\mu \pi, \pi]] 
 \,.  \label{rhopipi:term}
\end{equation}  
This term arises from the $a F_\pi^2 {\rm tr}[\hat{\alpha}_{\mu ||}^2]$ in the HLS-invariant manner, 
hence the $\rho_\Pi^i$- $\rho_P^i$ mixing breaking the HLS cannot be generated even through the one-loop diagrams 
constructed from this part. 
To see this clearly, consider the $SU(8)$ algebraic form to appear in the one-loop diagram corresponding to the $\rho^A$- $\rho^{A'}$ two-point function  
made of the term in Eq.(\ref{rhopipi:term}). 
It should come with the $SU(8)$ group structure like 
\begin{equation} 
  f^{ABC}f^{A'BC} = C_2(G) \delta^{AA'} 
  \,, 
\end{equation}
where $X^A X^A = C_2 (G) \cdot 1_{8 \times 8}$ for the adjoint representation.  
Hence the $\rho^A$- $\rho^B$ two-point function should be diagonal, i.e., include no mixing between the technirhos.

One can easily check this also in terms of the DY base parametrized as in Eqs.(\ref{pi:para}) and (\ref{rho:para}): 
the $\rho$-$\pi$-$\pi$ couplings in the DY-base are~\footnote{ 
In Ref.~\cite{Kurachi:2014qma} 
the rho couplings to the isospin-triplet color-octet pions $\theta_a^i$ are missing, which have properly been incorporated in Eq.(\ref{rhopipi:coupling:DYbase}). 
} 
\begin{eqnarray} 
{\cal L}_{\rho_{\Pi}^i/\rho_P^i-\pi-\pi} 
&=& - ag \Bigg[ \frac{1}{4} \epsilon^{ijk} \partial_\mu \Pi^i \Pi^j \rho_\Pi^{k \mu} 
+ \frac{1}{4} \epsilon^{ijk} \partial_\mu P^i P^j \rho_\Pi^{k \mu}  
\nonumber \\ 
&& 
+ \frac{1}{4} \epsilon^{ijk} \left( \partial_\mu \bar{T}^i_c T^j_c - \bar{T}^i_c \partial_\mu T^j_c \right) \rho_\Pi^{k \mu} 
+ \frac{1}{4} \epsilon^{ijk} \partial_\mu \theta^i_a \theta^j_a \rho_\Pi^{k \mu} 
\nonumber \\ 
&& 
- \frac{1}{2 \sqrt{3}} \epsilon^{ijk} \partial_\mu P^i P^j \rho_P^{k \mu} 
- \frac{1}{4\sqrt{3}} \epsilon^{ijk} \left( \partial_\mu \bar{T}^i_c T^j_c - \bar{T}^i_c \partial_\mu T^j_c \right) \rho_P^{k \mu} 
\nonumber \\ 
&& 
+ \frac{1}{4\sqrt{3}} \epsilon^{ijk} \partial_\mu \theta^i_a \theta^j_a \rho_P^{k \mu} 
- \frac{1}{2\sqrt{3}} \left( \partial_\mu \bar{T}_c^i T^0_c + \partial_\mu \bar{T}^0_c T^i_c 
- \bar{T}^i_c \partial_\mu T^0_c - \bar{T}^0_c \partial_\mu T^i_c 
 \right) \rho_P^{i\mu} 
 \Bigg]
\,. \label{rhopipi:coupling:DYbase}
\end{eqnarray} 
 From these couplings one can easily compute the $\rho_\Pi^i$ - $\rho_P^j$ two-point functions at the one-loop level 
as depicted in Fig.~\ref{pi-loop-graphs}. 
The one-loop terms are then evaluated for each graph to be dramatically canceled as   
\begin{eqnarray}
&& 
\left[ \frac{1}{2} \cdot \left( - \frac{1}{2 \sqrt{3}} \right) \left(  \frac{1}{4}  \right) \right]_{PP{\rm-loop}} 
\nonumber \\ 
+ && 
\left[ 3 \cdot\left( - \frac{1}{4 \sqrt{3}} \right) \left(  \frac{1}{4}  \right) \right]_{T\bar{T}{\rm-loop}} 
\nonumber\\ 
+ && 
\left[ \frac{1}{2} \cdot 8 \cdot \left(  \frac{1}{4 \sqrt{3}} \right) \left(  \frac{1}{4}  \right) \right]_{\theta\theta{\rm-loop}} 
= 0 
\,. 
\end{eqnarray}
Thus the $\rho_\Pi^i$ - $\rho_P^j$ mixing cannot be generated even at the loop level.

As to the SM gauge loop corrections, 
it turns out that the HLS also protects the absence of the $\rho_\Pi^i$ - $\rho_P^j$ mixing 
at any order of the quantum theory. One might naively think that 
even when the $\rho_P^i$ does not couple to weak bosons at ${\cal O}(p^2)$ , 
the $\rho$ coupling to weak boson might arise when one considers higher derivative order of ${\cal O}(p^4)$
such as in Eq.(\ref{Lag:p4:text}) in the next subsection. 
However, it is not the case: 
examining the couplings one finds that it takes the form  
\begin{equation} 
 {\rm tr}[X_P^i [X_\Pi^j, X_\Pi^k]] 
\,, \label{comm}
\end{equation}
where the $X_P^i$ comes from the $\rho_P^i$ field and two $X_\Pi^i$ from 
the $W$ boson field. 
Noticing that 
\begin{equation} 
 [X_\Pi^i, X_\Pi^j] = \frac{1}{\sqrt{N_F/2}} \epsilon^{ijk} X_\Pi^k 
\,, 
\end{equation}
 with $N_F=8$, and the orthogonality ${\rm tr}[X_P^i X_\Pi^j]=0$, 
one readily arrives at no coupling between $\rho_P^i$ and $WW$, 
\begin{equation} 
 g_{\rho_P^i WW} =0
\,. 
\end{equation}
The coupling is still zero even when the hypercharge $(\propto X_P)$ is turned on since 
$[X_\Pi^i, X_P]=0$.

Actually, the absence of the $\rho_\Pi^i$ - $\rho_P^i$ mixing is {\it exact  
to all orders of derivative expansions of the chiral perturbation theory,     
as far as the intrinsic-parity~\footnote{ 
The intrinsic parity of a particle is defined to be even, if its parity is $(-1)^{\rm spin}$, 
and odd otherwise.  
} even sector is concerned}: 
beyond the one-loop (${\cal O}(p^4)$) order, one can find 
that the triple gauge vertices like $\rho_P^i$ - $W/B$- $W$ would arise 
at the ${\cal O}(p^6)$ from the operators, 
\begin{equation} 
 {\rm tr}[ V_{\mu\nu} [ \hat{\cal V}^{\nu \lambda}, \hat{\cal V}_{\lambda}^\mu] ] 
 \,, \qquad 
  {\rm tr}[ V_{\mu\nu} [ \hat{\cal A}^{\nu \lambda}, \hat{\cal A}_{\lambda}^\mu] ] 
\,, \label{Op6}
\end{equation}
where 
\begin{eqnarray} 
 \hat{\cal V}_{\mu\nu} &=& 
\frac{\hat{\cal R}_{\mu \nu} + \hat{\cal L}_{\mu\nu}}{2}  
\,, \nonumber \\ 
 \hat{\cal A}_{\mu\nu} &=& 
\frac{\hat{\cal R}_{\mu \nu} - \hat{\cal L}_{\mu\nu}}{2} 
\,, \nonumber \\ 
\hat{\cal R}_{\mu\nu} 
&=& \xi_R R_{\mu\nu} \xi_R^\dag 
\,, \qquad 
\hat{\cal L}_{\mu\nu} 
= \xi_L L_{\mu\nu} \xi_L^\dag 
\,, \nonumber \\ 
R(L)_{\mu\nu} 
&=& 
\partial_\mu R(L)_\nu - \partial_\nu R(L)_\mu 
- i [R(L)_\mu , R(L)_\nu] 
\,. 
\end{eqnarray}
In writing operators in Eq.(\ref{Op6}) 
the charge conjugation, parity and intrinsic-parity invariance has been taken into account. 
Note, however, that the terms in Eq.(\ref{Op6})  
take the same form as in Eq.(\ref{comm}) (or $[X_P^i [X_P, X_\Pi^j] ]$), hence 
do not yield the coupling between the $\rho_P^i$ and weak bosons by the same argument for 
the ${\cal O}(p^4)$ terms as above. 
Even if one further goes to higher order terms, one cannot see any operators 
which generate triple gauge vertices since terms beyond ${\cal O}(p^6)$ 
should involve at least four gauge fields by construction. 
Thus the $\rho_P^i$ - weak boson - weak boson coupling, hence the $\rho_P^i$ - $\rho_\Pi^i$ mixing 
 is completely forbidden by the HLS chiral perturbation theory, the HLS gauge invariance.

If one considers terms having the intrinsic-parity odd property, 
such as $\epsilon^{\mu\nu\sigma\lambda} \partial_\mu B_\nu W_\sigma^i \rho_{P \lambda}^i$, 
at the ${\cal O}(p^4)$ level,  
one could see the nonzero $\rho_P^i$-$W^i$-$B$ coupling because those terms 
do not take the commutator form for generators as in Eq.(\ref{comm}).  
However, the contribution to the $\rho_\Pi^i$ - $\rho_P^i$ mixing generated by 
such terms at the ${\cal O}(p^6)$ level will be 
highly suppressed since it is in total of order of the two-loop level, 
to be safely negligible, as noted in Ref.~\cite{Fukano:2015hga}.

Thus the HLS is powerful and 
gives the strong constraint for the vector boson phenomenology, such as 
no mixing between isospin triplet states like among $\rho_\Pi^i$ and $\rho_P^i$.

 \subsubsection{The $\rho_\Pi^i$ and $\rho_P^0$ couplings in the mass-eigenstate basis}

Now we are allowed to safely pick up only the $\rho_\Pi^i$ and $\rho_P^0$ thanks to the HLS invariance. 
In Ref.~\cite{Fukano:2015hga} the $\rho_\Pi^i$ couplings to the SM particles have been discussed. 
The $\rho_\Pi^i$ couplings to the SM fermions were evaluated by 
the vector meson dominance, through mixing between the $\rho_\Pi^i$ 
and the SM gauge bosons. 
As for the $\rho_P^0$, since the couplings to longitudinal modes of weak bosons ($\pi_W$ and $\pi_Z$)   
vanish due to the orthogonality, ${\rm tr}[X_P [X_\Pi^i, X_\Pi^j]] \propto {\rm tr}[X_P X_\Pi^i]=0$, 
the $\rho_P^0$ was excluded in studying the diboson signatures.   
As clearly seen in Appendix~\ref{diagonalization}, it turns out, however, that the $\rho_\Pi^i $ actually mixes with the $\rho_P^0$ 
through the hypercharge gauge when one takes into account transverse modes of the dynamical weak bosons, 
so that the $\rho_P^0$ couplings to diboson can arise through the mixing with the $\rho_\Pi^i$. 
In this subsection, we shall update the evaluation of the 2 TeV technirho couplings  
by taking into account such dynamical contributions of weak bosons. 
Possible higher derivative coupling terms will also be incorporated, some of which were not listed 
in Ref.~\cite{Fukano:2015hga} because the weak bosons were taken not to be dynamical. 
This will make the diboson analysis more accurate than Ref.~\cite{Fukano:2015hga} as will be seen in the next section.

To discuss the dynamical mixing between the $\rho_\Pi^i, \rho_P^0$ and the weak bosons, 
we first introduce the weak boson kinetic terms, 
\begin{equation} 
 - \frac{1}{2 g_W^2} {\rm tr}[W_{\mu\nu}^2] - \frac{1}{4g_Y^2} B_{\mu\nu}^2 
 \,, 
\end{equation} 
where $g_W$ and $g_Y$ are counted as ${\cal O}(p^2)$ and 
\begin{eqnarray} 
 W_{\mu\nu} &=& \partial_\mu W_\nu - \partial_\nu W_\mu - i [W_\mu, W_\nu] 
 \,, \nonumber \\ 
 B_{\mu\nu} &=& \partial_\mu B_\nu - \partial_\nu B_\mu 
 \,, 
\end{eqnarray}
and the trace is taken only over the $SU(2)$ isospin space.

Using the explicit expressions for the external and HLS fields 
given in the last section, 
one can easily find the mixing mass matrices for the charged and 
neutral gauge boson sectors. 
Actually, beyond the leading order ${\cal O}(p^2)$, one has 
the kinetic mixing terms at the ${\cal O}(p^4)$, so may incorporate them 
to diagonalize the gauge propagator matrix.

The full s-HLS Lagrangian up to ${\cal O}(p^4)$~\cite{Tanabashi:1993np,Harada:2003jx} relevant to the present study 
is thus written as 
\begin{eqnarray}
{\cal L}_{\rm s-HLS}
&=& \chi^2 F_\pi^2 \left(
 {\rm tr}[\hat{\alpha}_{\perp \mu}^2] 
 + 
 a 
 {\rm tr}[\hat{\alpha}_{|| \mu}^2]
 \right) 
 - \frac{1}{2g^2} {\rm tr}[V_{\mu\nu}^2] 
 - \frac{1}{2g_W^2} {\rm tr}[W_{\mu\nu}^2] 
 - \frac{1}{2g_Y^2} {\rm tr}[B_{\mu\nu}^2] 
+ {\cal L}_{4} 
\,, \label{Lag:start:text} \\ 
{\cal L}_4 
&=& z_3 {\rm tr}[\hat{\cal V}_{\mu\nu} V^{\mu\nu}] 
- i z_4 {\rm tr}[V_{\mu\nu} \hat{\alpha}_\perp^\mu \hat{\alpha}_\perp^\nu] 
+ i z_5 {\rm tr} [V_{\mu\nu} \hat{\alpha}_{||}^\mu \hat{\alpha}_{||}^\nu] 
\nonumber \\ 
&& 
+ i z_6 {\rm tr}[\hat{\cal V}_{\mu\nu} \hat{\alpha}_\perp^\mu \hat{\alpha}_\perp^\nu] 
+ i z_7 {\rm tr}[\hat{\cal V}_{\mu\nu} \hat{\alpha}_{||}^\mu \hat{\alpha}_{||}^\nu] 
- i z_8 {\rm tr}[\hat{\cal A}_{\mu\nu} (\hat{\alpha}_\perp^\mu \hat{\alpha}_{||}^\nu 
+ \hat{\alpha}_{||}^\mu \hat{\alpha}_{\perp}^\nu 
)]  
\,.  \label{Lag:p4:text}
\end{eqnarray} 
Among ${\cal O}(p^4)$ couplings in Eq.(\ref{Lag:p4:text}) 
only the $z_3$ and $z_4$ terms have been incorporated in Ref.~\cite{Fukano:2015hga}, 
which contribute to the couplings to the SM fermions ($F_\rho$) and the longitudinal modes of weak bosons 
$(g_{\rho\pi\pi})$ as will be seen below (Eqs.(\ref{Frho:def}) and (\ref{rhopipi:def})). 
Other ${\cal O}(p^4)$ couplings ($z_5, z_6, z_7, z_8$) are newly introduced here, 
which can enter the couplings to the transverse modes of weak bosons as the ${\cal O}(p^4)$ corrections. 
 As it turns out, however, only the $z_8$ terms contribute  by accident 
(See Eqs.(\ref{gWZrho})-(\ref{gWWrho})).

Let the mass and kinetic eigenstates be fields with tilde, 
\begin{equation} 
  \tilde{\rho}_\Pi^{\pm, 0}, \tilde{\rho}_P^0 
\,,  
\end{equation}
and the weak bosons be $\tilde{W}$ and $\tilde{Z}$, similarly. 
One can thus find the relevant couplings of the technirhos as shown below. 
The details on the diagonalization are given in Appendix~\ref{diagonalization}. 
Here we just list some formulae by which one can easily follow the coupling expressions. 
We evaluate the couplings by assuming 
\begin{equation} 
x=\frac{g_W}{g} \simeq \sqrt{a} \left( \frac{m_W}{M_{\tilde{\rho}_{\Pi, P}}} \right) 
\simeq 0.04 \sqrt{a}  \ll 1
\,, \label{x:def}
\end{equation}
where the second approximation follows from the explicit expressions of 
mass eigenvalues of the $W$ and the  technirho masses given below (See Eqs.(\ref{mass:rhoc:text}), (\ref{EV_rhoP0:text}) and (\ref{EV_rhoPi3:text}) ).  
Then the gauge eigenstates $(W^\pm_\mu, \rho_{\Pi \mu}^\pm)$ in the charged current sector 
and $(W_\mu^3, \rho_{\Pi \mu}^3, B_\mu, \rho_{P \mu}^0)$ in the neutral current sector 
are related to the mass eigenstates at ${\cal O}(p^4)$ as follows: 
\begin{eqnarray} 
\left( 
\begin{array}{c} 
W_\mu^\pm \\ 
\rho_{\Pi \mu}^\pm
\end{array}
\right) 
&=& 
\left[ 
\begin{array}{c|c}  
-1 & - (1 - g^2 z_3) x \\ 
\hline 
- x & 1 
\end{array} 
\right] 
\left( 
\begin{array}{c} 
\tilde{W}_\mu^\pm \\ 
\tilde{\rho}_{\Pi \mu}^\pm
\end{array}
\right) 
\,,  \label{wavefunc:CC:text} 
\\ 
\left( 
\begin{array}{c}
W_\mu^3 \\ 
\rho_{\Pi \mu}^3 \\ 
B_\mu \\ 
\rho_{P \mu}^0 
\end{array}
\right) 
&=& 
\left[ 
\begin{array}{c|c|c|c}  
c & s & -(1 - g^2 z_3)c_\rho x & (1 - g^2 z_3) s_\rho x \\  
\hline 
c (1 - t^2) x & 2 s x & c_\rho & - s_\rho \\ 
\hline 
- s & c & - (1 - g^2 z_3) (c_\rho + \frac{2}{\sqrt{3}} s_\rho) t x & (1 - g^2 z_3) (s_\rho - \frac{2}{\sqrt{3}} c_\rho) t x \\ 
\hline 
- \frac{2}{\sqrt{3}} c t^2 x & \frac{2}{\sqrt{3}} s x & s_\rho & c_\rho  
\end{array} 
\right] 
\left( 
\begin{array}{c} 
\tilde{Z}_\mu \\ 
\tilde{A}_\mu \\ 
\tilde{\rho}_{\Pi \mu}^3 \\ 
\tilde{\rho}_{P \mu}^0 
\end{array}
\right) 
\,,  \label{wavefunc:NC:text}
\end{eqnarray} 
up to corrections of ${\cal O}(x^2)$, 
where we have defined the mixing angles as 
\begin{eqnarray}
t &=& \frac{g_W}{g_Y} 
\,, \qquad 
c = \frac{1}{\sqrt{1+t^2}}
\,, \qquad 
s = \sqrt{1 - c^2} 
\,, \\  
c_\rho &=& \frac{1}{\sqrt{2}} \sqrt{ \frac{(3-t^2) + \sqrt{ (3-t^2)^2 + 48 t^4 }}{\sqrt{(3 -t^2)^2 + 48t^4}} }
\,, \qquad 
s_\rho = \sqrt{1 - c_\rho^2}
\,,  
\end{eqnarray}
in which $t = \tan\theta_W = \sin\theta_W/\cos\theta_W$ is identical to the weak mixing angle (up to ${\cal O}(x^2)$) 
and $c_\rho$ ($s_\rho$) denotes the mixing angle between the gauge-eigenstate $\rho_\Pi^3$ and $\rho_P^0$. 
Then the electromagnetic coupling $e$ is expressed as $e= g_W s = g_W \sin\theta_W + {\cal O}(x^2)$. 
   The angle parameter $t$ can numerically be fixed by the experimental values of the $Z$ boson mass, the 
electromagnetic coupling $\alpha_{\rm EM}$ (at the scale of $Z$ mass) and the Fermi constant $G_F$ as 
\begin{equation} 
 t \simeq 0.55 
\,. 
\end{equation}
This allows us to estimate also the mixing angle $c_\rho$ ($s_\rho$) between the $\rho_\Pi^i$ and $\rho_P^0$: 
\begin{equation} 
 c_\rho = \frac{1}{\sqrt{2}} \sqrt{ \frac{(3-t^2) + \sqrt{ (3-t^2)^2 + 48 t^4 }}{\sqrt{(3 -t^2)^2 + 48t^4}} }
 \simeq 0.95 
 \,, \qquad 
 s_\rho = \sqrt{1 - c_\rho^2} 
 \simeq 0.33 
\,,  \label{srho:val}
\end{equation} 
which implies that the gauge-eigenstate $\rho_\Pi^i$ and $\rho_P^0$ hardly mix each other, 
such that the $\tilde{\rho}_P^0$ has almost vanishing couplings to weak bosons, in accord with 
the analysis done in Ref.~\cite{Fukano:2015hga}.      
The mass eigenvalues 
corresponding to the mass eigenstates are evaluated up to corrections of ${\cal O}(x^2)$ 
to be  
\begin{eqnarray} 
 m_{\bar{W}}^2 
 &=&  
\frac{g_W^2 v_{\rm EW}^2}{4} 
\left (1 + {\cal O}(x^2) \right) 
= m_W^2 \left (1 + {\cal O}(x^2) \right) 
\,, 
\\ 
M_{\bar{\rho}_\Pi^\pm}^2 
&=& 
 \frac{a g^2 v_{\rm EW}^2}{4} 
\left (1 + {\cal O}(x^2) \right) 
= M_\rho^2 (1 + {\cal O}(x^2) ) 
\,,  \label{mass:rhoc:text} 
\\ 
 m_{\bar{Z}}^2 
 &=&  
\frac{g_W^2 v_{\rm EW}^2}{4c^2} 
\left (1 + {\cal O}(x^2) \right) 
= m_Z^2 \left (1 + {\cal O}(x^2) \right) 
\,, \label{mass:Z} \\ 
m_{\bar{A}}^2 
&=& 0 
\,, \\ 
M_{\bar{\rho}_\Pi^3}^2 
&=& 
 \frac{a g^2 v_{\rm EW}^2}{4} 
\left (1 + {\cal O}(x^2) \right) 
= M_\rho^2 (1 + {\cal O}(x^2) ) 
\,, \label{EV_rhoPi3:text}\\
M_{\bar{\rho}_P^0}^2 
&=& 
 \frac{a g^2 v_{\rm EW}^2}{4} 
\left (1 + {\cal O}(x^2) \right) 
= M_\rho^2 (1 + {\cal O}(x^2) ) 
\,, \label{EV_rhoP0:text}
\end{eqnarray} 
where we have defined 
\begin{equation} 
m_W^2 = \frac{g_W^2 v_{\rm EW}^2}{4} 
\,, \qquad 
M_\rho^2 = \frac{a g^2 v_{\rm EW}^2}{4}
\,.   
\end{equation} 
Note that the mass difference between $\tilde{\rho}_{_\Pi}^3$ and $\tilde{\rho}^0_{_P}$ is  
very small:  
\begin{eqnarray} 
\Delta M &=& \frac{M^2_{\tilde{\rho}^0_{_P}} - M^2_{\tilde{\rho}_{_\Pi}^3} }{M_{\tilde{\rho}^0_{_P}} + M_{\tilde{\rho}_{_\Pi}^3} } 
\simeq 
\frac{M^2_{\tilde{\rho}^0_{_P}} - M^2_{\tilde{\rho}_{_\Pi}^3} }{2 M_\rho} 
\nonumber \\ 
&\simeq& \frac{M_{\rho}}{2} \times {\cal O}(x^2) 
\,, 
\label{mass:diff}
\end{eqnarray}  
where use has been made of 
Eqs.(\ref{EV_rhoPi3:text}), (\ref{EV_rhoP0:text}) and (\ref{x:val}).

The $\tilde{\rho}_\Pi^{\pm, 3}$ and $\tilde{\rho}_P^0 $ couplings to the SM fermions 
arise from the weak bosons coupled to fermions 
via the propagator mixing between the weak bosons and $\tilde{\rho}_\Pi^{\pm, 3}$ and $\tilde{\rho}_P^0 $. 
 Using Eqs.(\ref{wavefunc:CC:text}) and (\ref{wavefunc:NC:text}) one can easily find the couplings:  
\begin{eqnarray} 
{\cal L}_{\tilde{\rho}_\Pi^3 ff} 
&=&  \bar{\psi} \gamma^\mu \left[ A_{V}^\psi + A_{A}^\psi \gamma_5 \right] \psi \, \tilde{\rho}_{\Pi \mu}^3  
\,, \nonumber \\ 
{\cal L}_{\tilde{\rho}_P ff} 
&=& 
\bar{\psi} \gamma^\mu \left[ B_{V}^\psi + B_{A}^\psi \gamma_5 \right]  \psi \, \tilde{\rho}_{P \mu}^0   
\,,  \nonumber \\
 {\cal L}_{\tilde{\rho}_\Pi^{\pm} ff} 
 &=&    
C_L^\psi 
 \left( \bar{\psi}_{u_L} \gamma^\mu \psi_{d_L} \right) \tilde{\rho}_{\Pi \mu}^+ + {\rm h.c.}  
\,, 
\end{eqnarray}
where $\psi_{{u(d)}_L}$ denotes the left-handed SM fermion fields having the isospin charge $\tau_3^\psi=1/2(-1/2)$ 
and 
\begin{eqnarray} 
 A_{V}^\psi &=& - \frac{e^2}{2 s^2} \left( \frac{\sqrt{N_D}}{2} \right) \left( \frac{F_\rho}{M_\rho} \right) 
 \left\{ 
 \left[ 
 c_\rho - \left( c_\rho + \frac{2}{\sqrt{3}} s_\rho \right) t^2 \right] \tau_3^\psi + 2 t^2 \left(
 c_\rho + \frac{2}{\sqrt{3}} s_\rho 
 \right) Q_{\rm em}^\psi
 \right\} 
 \,, \nonumber \\ 
  A_{A}^\psi &=&  \frac{e^2}{2 s^2} \left( \frac{\sqrt{N_D}}{2} \right) \left( \frac{F_\rho}{M_\rho} \right) 
\left[  c_\rho - \left( c_\rho + \frac{2}{\sqrt{3}} s_\rho \right) t^2  
 \right]\tau_3^\psi
 \,, \nonumber \\ 
  B_{V}^\psi &=& \frac{e^2}{2 s^2} \left( \frac{\sqrt{N_D}}{2} \right) \left( \frac{F_\rho}{M_\rho} \right) 
 \left\{ 
 \left[ 
 s_\rho - \left( s_\rho - \frac{2}{\sqrt{3}} c_\rho \right) t^2 \right] \tau_3^\psi + 2 t^2 \left(
 s_\rho - \frac{2}{\sqrt{3}} c_\rho 
 \right) Q_{\rm em}^\psi
 \right\} 
 \,, \nonumber \\ 
  B_{A}^\psi &=& - \frac{e^2}{2 s^2} \left( \frac{\sqrt{N_D}}{2} \right) \left( \frac{F_\rho}{M_\rho} \right) 
 \left[ 
 s_\rho - \left( s_\rho - \frac{2}{\sqrt{3}} c_\rho \right) t^2 \right] \tau_3^\psi  
 \,, \nonumber \\ 
 C_L^\psi 
 &=& 
 - \frac{e^2}{\sqrt{2}s^2} \left( \frac{\sqrt{N_D}}{2} \right) \left( \frac{F_\rho}{M_\rho} \right) 
 \,, \label{couplings:fermions}
\end{eqnarray}
with 
\begin{eqnarray} 
Q_{\rm em}^q
&=& 
\left( 
\begin{array}{cc} 
 2/3 & 0 \\ 
 0 & -1/3 
\end{array}
 \right)
 \,, \qquad 
 Q_{\rm em}^l
=
\left( 
\begin{array}{cc} 
 0 & 0 \\ 
 0 & -1 
\end{array}
 \right)
\,, \\
F_\rho &=& 
\sqrt{a} F_\pi (1 - g^2 z_3)
\,. \label{Frho:def}
\end{eqnarray} 
The prefactor $\sqrt{N_D}$ in Eq.(\ref{couplings:fermions}) 
stands for the number of electroweak doublets formed by technifermions, which is 4 in the case of the one-family model.   
Note that the $N_D$ dependence is canceled out in the combination $(\sqrt{N_D} F_\rho)$: 
$\sqrt{N_D} F_\rho \propto \sqrt{N_D} \sqrt{a} F_\pi = \sqrt{a} v_{\rm EW}$ where $v_{\rm EW} \simeq 246$ GeV. 
In the limit where the $\rho_\Pi^3$ - $\rho_P^0$ mixing is turned off, i.e., $s_\rho =0$, and the rho mass scale is much larger than 
weak boson masses ($M_\rho \gg m_{W/Z}$), 
the $\tilde{\rho}_\Pi^{\pm, 3}$ couplings to fermions in Eq.(\ref{couplings:fermions}) 
precisely become the same as those (without the symbol of tilde)  
given in Ref.~\cite{Fukano:2015hga}, as it should.

The couplings to the weak bosons $\tilde{W}\tilde{W}/\tilde{W}\tilde{Z}$ in the mass-basis arise from 
the non-Abelian vertex terms in ${\rm tr}[V_{\mu\nu}^2]$ and ${\rm tr}[W_{\mu\nu}^2]$ in Eq.(\ref{Lag:start:text}) 
as well as the ${\cal O}(p^4)$ terms in Eq.(\ref{Lag:p4:text}) 
to be 
\begin{eqnarray} 
 {\cal L}_{\rho \tilde{W}\tilde{W}/\tilde{W}\tilde{Z}} 
 &=& i \Bigg[ 
 g_{\tilde{\rho}_\Pi \tilde{W}\tilde{Z}} 
 \left( 
\partial_\mu \tilde{\rho}_{\Pi \nu}^- - \partial_\nu \tilde{\rho}_{\Pi \mu}^- 
\right) \tilde{Z}^\mu \tilde{W}^{+ \nu} 
+ 
 g_{\tilde{W} \tilde{Z}\tilde{\rho}_\Pi} 
 \left( 
\partial_\mu \tilde{W}_{\nu}^+ - \partial_\nu \tilde{W}_{\mu}^+ 
\right) \tilde{Z}^\mu \tilde{\rho}^{- \nu}_\Pi   
\nonumber \\ 
&& 
+  
 g_{\tilde{Z} \tilde{W}\tilde{\rho}_\Pi} 
 \left( 
\partial_\mu \tilde{Z}_{\nu} - \partial_\nu \tilde{Z}_{\mu} 
\right) \tilde{W}^{\mu +} \tilde{\rho}^{- \nu}_\Pi 
\nonumber \\ 
&& 
+ 
\frac{1}{2} 
g_{\tilde{\rho}_\Pi \tilde{W}\tilde{W}} 
    \left( 
\partial_\mu \tilde{\rho}_{\Pi \nu}^3 - \partial_\nu \tilde{\rho}_{\Pi \mu}^3 
\right) \tilde{W}^\mu \tilde{W}^{+ \nu} 
+ 
\frac{1}{2} 
g_{\tilde{\rho}_P \tilde{W}\tilde{W}} 
    \left( 
\partial_\mu \tilde{\rho}_{P \nu}^0 - \partial_\nu \tilde{\rho}_{P \mu}^0 
\right) \tilde{W}^\mu \tilde{W}^{+ \nu} 
\nonumber \\ 
&& +  
 g_{\tilde{W} \tilde{W}\tilde{\rho}_\Pi} 
 \left( 
\partial_\mu \tilde{W}_{\nu}^+ - \partial_\nu \tilde{W}_{\mu}^+ 
\right) \tilde{W}^{\mu -} \tilde{\rho}^{3 \nu}_\Pi 
+  
 g_{\tilde{W} \tilde{W}\tilde{\rho}_P} 
 \left( 
\partial_\mu \tilde{W}_{\nu}^+ - \partial_\nu \tilde{W}_{\mu}^+ 
\right) \tilde{W}^{\mu -} \tilde{\rho}^{0 \nu}_P
 \Bigg]
\nonumber \\ 
&& 
+ {\rm h.c.}
\,. 
\end{eqnarray}
To the nontrivial-leading order of expansion in $x = g_W/g = \sqrt{a} (m_W/M_\rho)\ll 1$, we thus have 
\begin{eqnarray} 
   g_{\tilde{\rho}_\Pi \tilde{W}\tilde{Z}} 
&=& 
\frac{e}{\sqrt{N_D} sc} x \left(  1 + \frac{1}{2}g^2 z_4 \right)
=  
\frac{1}{c} \frac{2}{\sqrt{N_D}} \left( \frac{m_W}{M_{\rho}}\right)^2 
g_{\rho\pi\pi}
\,, \qquad 
g_{\rho\pi\pi} 
= \frac{1}{2} ag \left(1 + \frac{1}{2} g^2 z_4  \right) 
\,,  \label{rhopipi:def} \\ 
g_{\tilde{\rho}_\Pi \tilde{W}\tilde{W}} 
&=& 
- \frac{2}{\sqrt{N_D}} c_\rho \, \left( \frac{m_W}{M_\rho}\right)^2  g_{\rho\pi\pi} 
\,, \\ 
g_{\tilde{\rho}_P \tilde{W}\tilde{W}} 
&=& 
\frac{2}{\sqrt{N_D}} s_\rho \, \left( \frac{m_W}{M_\rho}\right)^2 g_{\rho\pi\pi} 
\,, \label{grhoPWW} \\ 
 g_{\tilde{W} \tilde{Z}\tilde{\rho}_\Pi} 
&=& 
- \frac{e}{ \sqrt{N_D} sc} x \left[ 1 - g^2 \left( z_3 + \frac{1}{2} z_8   \right) \right] 
= 
- \frac{e}{ \sqrt{N_D} sc} \sqrt{a}\left( \frac{m_W}{M_\rho}  \right) \left[ 1 - g^2 \left( z_3 + \frac{1}{2} z_8   \right) \right] 
\,, \label{gWZrho} \\ 
 g_{\tilde{Z} \tilde{W}\tilde{\rho}_\Pi} 
&=& 
 \frac{e}{\sqrt{N_D} sc} x \left[ 1 - g^2 \left( z_3 + \frac{1}{2} z_8   \right) \right] 
= 
 \frac{e}{\sqrt{N_D} sc} \sqrt{a} \left( \frac{m_W}{M_\rho} \right) \left[ 1 - g^2 \left( z_3 + \frac{1}{2} z_8   \right) \right] 
\,, \\ 
 g_{\tilde{W} \tilde{W}\tilde{\rho}_\Pi} 
&=& 
- \frac{e}{\sqrt{N_D} s} x \, c_\rho \, \left[ 1 - g^2 \left( z_3 + \frac{1}{2} z_8   \right) \right] 
= 
- \frac{e}{\sqrt{N_D} s} \sqrt{a} \left( \frac{m_W}{M_\rho} \right) \, c_\rho \, \left[ 1 - g^2 \left( z_3 + \frac{1}{2} z_8   \right) \right] 
\,, \\ 
 g_{\tilde{W} \tilde{W}\tilde{\rho}_P} 
&=& 
 \frac{e}{\sqrt{N_D} s} x \, s_\rho \, \left[ 1 - g^2 \left( z_3 + \frac{1}{2} z_8   \right) \right] 
= 
 \frac{e}{\sqrt{N_D} s} \sqrt{a} \left(  \frac{m_W}{M_\rho} \right) \, s_\rho \, \left[ 1 - g^2 \left( z_3 + \frac{1}{2} z_8   \right) \right] 
\,. \label{gWWrho}
\end{eqnarray}
The longitudinal modes of the weak bosons only contribute to the couplings proportional to $g_{\rho\pi\pi}$ 
in Eqs.(\ref{rhopipi:def}) - (\ref{grhoPWW}). 
One can easily check this by replacing the weak boson fields $W_\mu, Z_\mu$ with 
the eaten pion fields as $W_\mu = \partial_\mu \Pi_W/m_W$ and $Z_\mu = \partial_\mu \Pi_Z/m_Z$, 
to see that only the couplings  in Eqs.(\ref{rhopipi:def}) - (\ref{grhoPWW}) survive. 
Then the couplings to the longitudinal modes of the weak bosons are precisely the same as 
those given in Ref.~\cite{Fukano:2015hga}, and the suppression factor $(m_W/M_\rho)^2$ cancels when 
one considers to multiply by the longitudinal polarization vector of weak bosons in amplitudes 
evaluated at the onshell of weak bosons, potentially yielding the dominant contribution to the rho widths, 
inn accord with Ref.~\cite{Fukano:2015hga}.

As to the couplings to the transverse modes of 
weak bosons in  Eqs.(\ref{gWZrho}) - (\ref{gWWrho}), 
one should note that the parameter $a$ dependence explicitly enters there. 
We will come back to this point when the LHC 
phenomenology is addressed in the next section.

Crucial to note is also the flavor dependence of the couplings to weak bosons, $N_D=N_F/2=4$, 
which makes the total width much smaller than the naive scale-up version of QCD with $N_F=2$, 
to be lower than 100 GeV, in accord with the ATLAS diboson data (For the detail, see discussions in the next section). 
When the partial decay width to the longitudinal parts of $WW/WZ$, 
$\tilde{\rho}_{\Pi} \to W_LW_L/W_LZ_L$, is evaluated  
one can see the narrowness more explicitly: 
\begin{equation} 
  \Gamma(\tilde{\rho}_\Pi \to W_LW_L/W_LZ_L) \simeq \left( \frac{1}{N_D} \right) \cdot  \frac{g_{\rho\pi\pi}^2}{48 \pi} \cdot c_\rho^2 \cdot M_{\tilde{\rho}_\Pi} 
  \,, 
\end{equation}
 where the heavy rho mass limit $M_{\tilde{\rho}} \gg m_{W/Z}$ has been taken (The exact expression for the partial decay widths are shown in Appendix~\ref{decay}). 
Thus the $\tilde{\rho}_\Pi$ width generically gets smaller as the number of the flavor $(N_D=N_F/2)$ increases.

Note also that the custodial partner of the $\tilde{\rho}_\Pi^\pm$  
involves both $\tilde{\rho}_\Pi^3$ and $\tilde{\rho}_P$ due to $c_\rho^2 + s_\rho^2 =1$, 
in such a way that 
\begin{eqnarray} 
c^2 \, g_{\tilde{\rho}_\Pi \tilde{W}\tilde{Z}}^2 
&=&  
g_{\tilde{\rho}_\Pi \tilde{W}\tilde{W}}^2 
+ 
g_{\tilde{\rho}_P \tilde{W}\tilde{W}}^2   
\,, \\ 
c^2 \,  g_{\tilde{W} \tilde{Z}\tilde{\rho}_\Pi} ^2 
&=& 
 g_{\tilde{W} \tilde{W}\tilde{\rho}_\Pi}^2
+ 
 g_{\tilde{W} \tilde{W}\tilde{\rho}_P}^2
\,, \\ 
c^2 \,  g_{\tilde{Z} \tilde{W}\tilde{\rho}_\Pi} ^2 
&=& 
 g_{\tilde{W} \tilde{W}\tilde{\rho}_\Pi}^2
+ 
 g_{\tilde{W} \tilde{W}\tilde{\rho}_P}^2
\,. \label{custodial}
\end{eqnarray}

Again, the technirhos do not couple to the Higgs (technidilaton) plus weak gauge bosons, 
due to the conformal barrier: 
\begin{eqnarray}
 g_{\tilde{\rho}_\Pi^{\pm,3}/\tilde{\rho}_P^0 - \tilde{W}/\tilde{Z}-\phi} = 0 
 \,. 
\end{eqnarray}

\section{LHC phenomenology of walking technirho at 2 TeV} 
\label{pheno}

In this section, we explore the 2 TeV walking technirho phenomenology at the LHC.  
We first set the LHC-Run I limits on the technirho couplings explicitly derived in the last section. 
Then we analyze the walking technirho signals in the diboson channel to see if the signals can 
explain the recently reported ATLAS diboson excesses~\cite{Aad:2015owa}.  
This section will update the previous analysis in Ref.~\cite{Fukano:2015hga} including 
the mixing effect from the dynamical weak gauge bosons (transverse components of $W$ and $Z$) as noted in the last section.

First of all, we summarize the parameters of the 2 TeV walking technirhos 
$\tilde{\rho}_{\Pi}^{\pm, 3}$ and $\tilde{\rho}_{P}^0$
relevant to the LHC phenomenology and discuss the way to fix those parameters.

Looking at the couplings to the SM fermions and weak bosons 
listed in the previous section, 
one can find that they are controlled by 
the six parameters, 
\begin{equation} 
F_\pi, \qquad 
a, \qquad 
g, \qquad 
z_3, \qquad 
z_4, \qquad 
z_8 
\,. 
\end{equation} 
Among these parameters, 
the technipion decay constant $F_\pi$ is related to the electroweak scale 
$v_{\rm EW}\simeq 246$ GeV for the one-family model with four weak-doublets, $N_D=N_F/2=4$, as
\begin{equation} 
F_\pi = v_{\rm EW}/\sqrt{N_D}\simeq 123 \,  {\rm GeV} 
\,,  \label{Fpi}
\end{equation} 
through the $W/Z$ mass formula obtained by examining the $F_\pi^2$ term in Eq.(\ref{Lag:start:text}).
In place of the original parameters $z_3$ and $z_4$, we shall use 
$F_\rho$ and $g_{\rho\pi\pi}$ defined in Eqs.(\ref{Frho:def}) and (\ref{rhopipi:def}). 
The HLS gauge coupling $g$ is determined through the rho mass formulae in Eqs.(\ref{mass:rhoc:text}), (\ref{EV_rhoP0:text}) and 
(\ref{EV_rhoPi3:text}), once 
the rho mass is set to 2 TeV and the parameter $a$ is chosen: 
\begin{equation} 
 g = \frac{M_\rho}{\sqrt{a} F_\pi} \simeq \frac{16}{\sqrt{a}}
 \,. 
\end{equation} 
Note from Eqs.(\ref{rhopipi:def}) - (\ref{grhoPWW}) 
that as far as the longitudinal-mode contributions of weak bosons (Eqs.(\ref{rhopipi:def}) - (\ref{grhoPWW}) and Eq.(\ref{couplings:fermions})) are concerned, 
all the rho couplings are completely free from the parameter $a$, as addressed in Ref.~\cite{Fukano:2015hga}.  
The $a$-dependence thus only comes in the transverse-mode contributions of weak bosons in Eqs.(\ref{gWZrho}) - (\ref{gWWrho}).   
To be consistent with our perturbative analysis in expansion with respect to $x=\sqrt{a} (m_W/M_\rho) \simeq 0.04 \sqrt{a} \ll 1$ 
(See Eq.(\ref{x:def})), 
we may choose the value of $a$ to be moderately small, say, 
\begin{equation} 
1 \lesssim a \lesssim 10
\,, \qquad 
 s.t., \qquad 
0.04 \lesssim x=\sqrt{a} \left( \frac{m_W}{M_\rho} \right) \lesssim 0.13 
\,.  \label{x:val}
\end{equation}
Then the HLS gauge coupling $g$ is determined to be in a range~\footnote{
The large value of $g$
implies large corrections from higher order in the HLS chiral perturbation theory, 
with the expansion parameter $\xi= N_F\cdot g^2/(4\pi)^2\simeq N_F \cdot p^2 /(4\pi F_\pi)^2|_{p=M_\rho} \simeq 16 \gg 1\, (N_F=8)$, 
in comparison with the real-life QCD ($N_F=3$), $\xi \simeq 1$.
Origin of the large corrections will be discussed in the Summary and discussion. }, 
\begin{equation} 
5.1 \lesssim g \lesssim 16 
\,. 
\end{equation}

As to the remaining ${\cal O}(p^4)$ coupling $z_8$ included in the transverse $\tilde{\rho}_{\Pi, P}$-$W$-$W/Z$ couplings 
in Eqs.(\ref{gWZrho}) - (\ref{gWWrho}),  we may take a special choice~\footnote{ 
The order of magnitude for ${\cal O}(p^4)$ parameters can be estimated 
by the naive dimensional analysis to be $z_i={\cal O}(N_C/(4\pi)^2)={\cal O}(10^{-2})$ for $N_C=3,4$. 
Hence the cancellation in Eq.(\ref{choice}) possibly takes place, though 
the nonperturbative estimate has not been done on the size of $z_8$ even in the QCD case.       
}, 
\begin{equation} 
 z_3 + \frac{1}{2} z_8 \simeq 0 
 \,, \label{choice}
\end{equation} 
such that the technirho couplings to the transverse modes of $W$ and $Z$ 
in Eqs.(\ref{gWZrho}) - (\ref{gWWrho})  
are highly suppressed by a factor of $x= \sqrt{a} (m_W/M_\rho)$. 
Then the rho couplings to $W$ and $Z$ are almost saturated by 
the longitudinal modes of $W$ and $Z$, i.e. $g_{\rho\pi\pi}$, 
in accordance with the analysis in Ref.~\cite{Fukano:2015hga}.

Note also that 
the mass difference between $\tilde{\rho}_\Pi^3$ and $\tilde{\rho}_P^0$ is very small 
for the value of $x$ in Eq.(\ref{x:val}) (See Eq.(\ref{mass:diff})),   
\begin{equation}
\Delta M 
\simeq 
(1.6 - 16) \,{\rm GeV} 
\,, \qquad 
{\rm for} 
\qquad 
0.04 \lesssim x \lesssim 0.13
\,.  
\label{mass:diff:val}
\end{equation}

Thus, with these inputs, we are left only with two parameters 
\begin{equation}
\left( F_\pi, \, a, \, g, \, z_3, \, z_4, \, z_8 \right)
\qquad 
\longrightarrow 
\qquad 
\left(F_\rho, \, g_{\rho\pi\pi} \right), 
 \end{equation} 
which control the $\rho_\Pi$ couplings to the SM fermions 
(Eq.(\ref{couplings:fermions})) and the longitudinal modes of weak gauge bosons (Eqs.(\ref{rhopipi:def}) - (\ref{grhoPWW})), respectively.

\subsection{Constraining walking technirho couplings: $g_{\rho\pi\pi}$ and $F_\rho$}

The LHC limits obtained from searches for $W'/Z'$ candidates~\cite{Aad:2014aqa,Khachatryan:2015sja,Khachatryan:2014xja,Aad:2014pha,Aad:2014xka,Khachatryan:2014gha,Aad:2015ufa,Khachatryan:2014tva,ATLAS:2014wra,Aad:2014cka,Khachatryan:2014fba,Khachatryan:2014hpa} constrain the 2 TeV technirho couplings 
to fermions ($F_\rho$) and weak bosons ($g_{\rho \pi\pi}$). 
We compute the partial decay rates and the DY cross sections by using the analytic formulae given in Appendix~\ref{decay}. 
 The DY cross section is evaluated by using the narrow width approximation (NWA), 
\begin{eqnarray} 
   \sigma_{\rm DY} (pp \to \rho) \Bigg|_{\rm NWA}  
 &= &
 \frac{16 \pi^2}{3 s} 
\, \sum_{q={\rm quarks}} \frac{\Gamma(\rho \to q\bar{q})}{M_\rho} \, 
\int_{-Y_B}^{Y_B} d\eta \, f_{q/p}\left(\frac{M_{\rho}}{\sqrt{s}} e^\eta, M_{\rho}^2 \right) 
f_{\bar{q}/p}\left( \frac{M_{\rho}}{\sqrt{s}} e^{-\eta},M_{\rho}^2 \right)  \,,
\end{eqnarray}
where $Y_B = -\frac{1}{2}\ln(M_\rho^2/s)$ and $f_{q/p}$ denotes the parton distribution function~\cite{Stump:2003yu}.   
  We constrain the size of the total widths to be $\lesssim$ 100 GeV 
in light of the ATLAS data on the diboson-tagged dijet mass distribution~\cite{Aad:2015owa}, 
so that the relevant couplings $F_\rho$ and $g_{\rho\pi\pi}$ are constrained. 
The constraint plots are displayed in Fig.~\ref{constraining-Frho-grhopipi-updated}. 
As to the dijet limit,  
we have quoted the upper limit set on generic narrow resonances (with the width being 0.1 percent of the mass)
decaying to the $qq$-jet reported from 
the CMS group~\cite{Khachatryan:2015sja}, while the ATLAS bound~\cite{Aad:2014aqa} includes all the jet candidates.

In addition to limits shown in Fig.~\ref{constraining-Frho-grhopipi-updated}, 
there are other limits from $W'/Z' \to $ Higgs plus weak bosons 
reported from the ATLAS and CMS~\cite{Aad:2015yza,Khachatryan:2015bma}, 
in which the analyses are based on the Higgs decay to $bb$ or $WW$. 
However, the $\rho_\Pi$s in the present study 
do not decay to the Higgs candidate (technidilaton), 
due to the conformal barrier.

   \begin{figure}[ht]
\begin{center}
   \includegraphics[width=7.0cm]{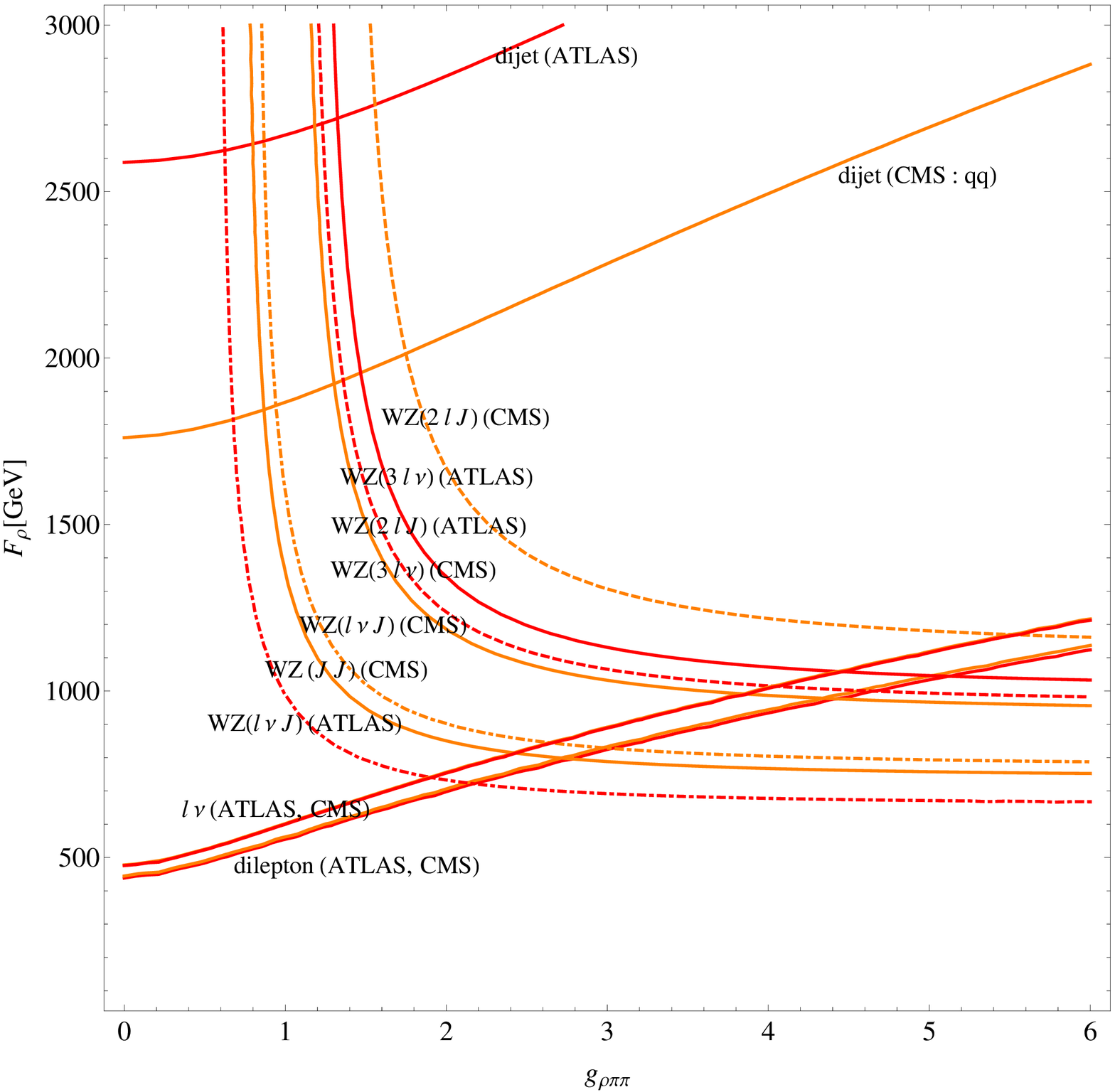}
\hspace{10pt}    
\includegraphics[width=7.0cm]{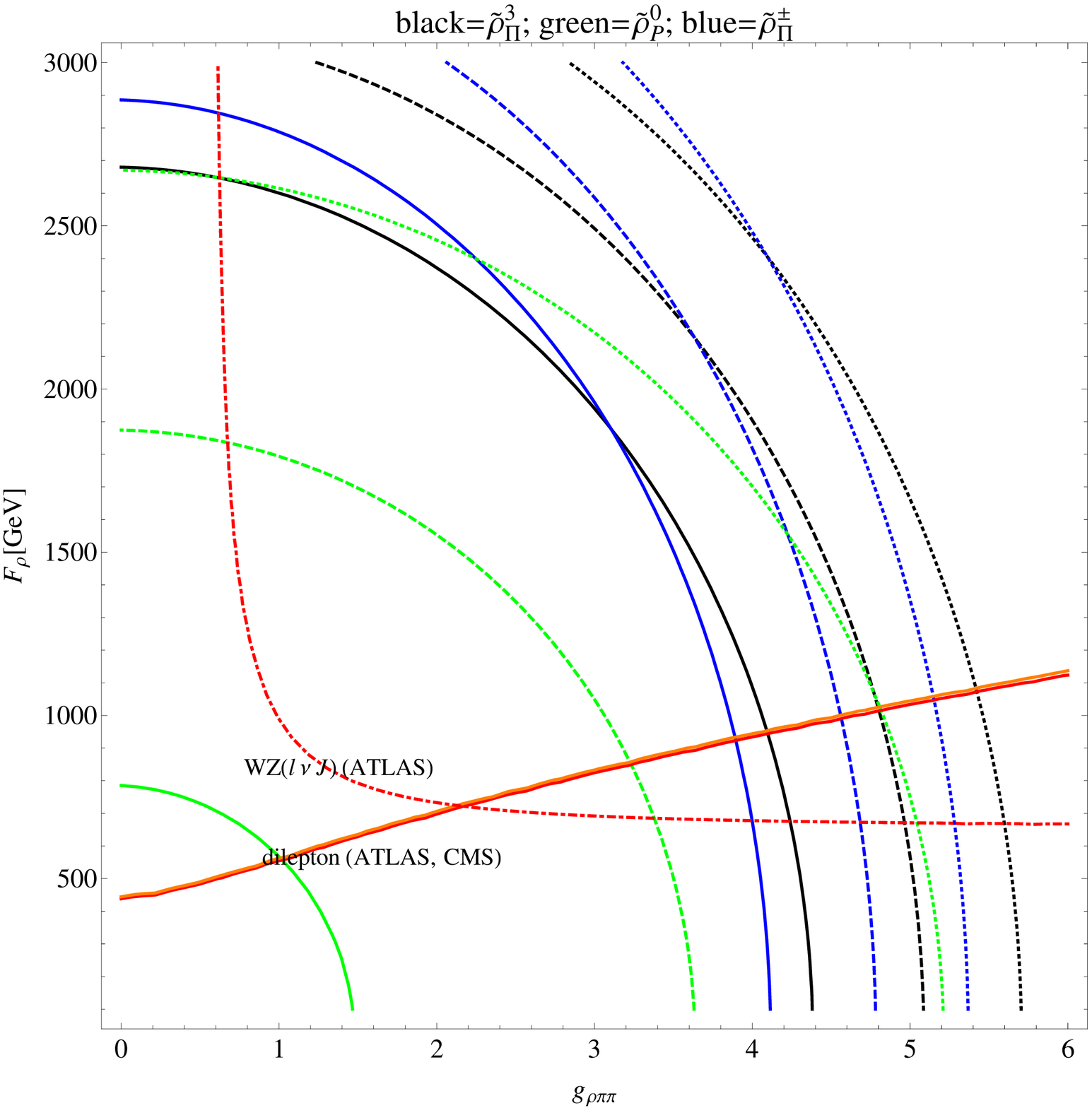}
\caption{ 
Left panel: The 95\% C.L.limits on the $(g_{\rho\pi\pi}, F_\rho)$ plane 
from the ATLAS and CMS data in the LHC-Run I~\cite{Aad:2014aqa,Khachatryan:2015sja,Khachatryan:2014xja,Aad:2014pha,Aad:2014xka,Khachatryan:2014gha,Aad:2015ufa,Khachatryan:2014tva,ATLAS:2014wra,Aad:2014cka,Khachatryan:2014fba,Khachatryan:2014hpa}.  
Right panel: 
The total widths as a function of $g_{\rho \pi\pi}$ and $F_\rho$. 
The solid-, dashed- and dot-dashed curves running along the vertical axes respectively denote  
$\Gamma=60(1), 80(5)$ and 100(10) GeV for the $\tilde{\rho}_\Pi^{\pm,3}$ ($\tilde{\rho}_P^0$),   
together with the most stringent constraints from dilepton and diboson channels extracted from the left panel.  
\label{constraining-Frho-grhopipi-updated} 
}
\end{center} 
 \end{figure}

Recently the ATLAS Collaboration released preliminary results on the combined limits on the $W'$ and Randall-Sundrum 
gravitons from various diboson search channels~\cite{ATLAS:diboson_comb}. The combination leads to a tighter constraint
over the entire mass range between 300 and 2500~GeV, except the 2 TeV mass region where the excess was observed 
in the diboson-tagged dijet channel~\cite{Aad:2015owa}. Given this, the constraints shown in Fig.~\ref{constraining-Frho-grhopipi-updated}
are not altered by the combined results. 

From Fig.~\ref{constraining-Frho-grhopipi-updated}, we see that the current LHC limits require 
\begin{equation} 
 F_\rho \lesssim 650\,{\rm GeV} 
\,.  \label{upperlimit}
\end{equation}
For reference values of $F_\rho$
we shall therefore take 
\begin{equation} 
 F_\rho[{\rm GeV}] = 250, 500, 650 
\,.  
\end{equation}
As for the $g_{\rho\pi\pi}$ coupling, we may take 
\begin{equation}
 g_{\rho\pi\pi} 
 = 3, 4, 5 
 \,, 
\end{equation}
in which the middle number can be supported by a holographic estimate~\cite{MY} 
and the last one can be deduced from the large $N_C$ scaling of $g_{\rho\pi\pi}$, 
\begin{equation}
g_{\rho\pi\pi} \sim \sqrt{\frac{3}{N_C}} [g_{\rho\pi\pi}]_{\rm QCD} 
\simeq 5.2 
\,, 
\qquad 
{\rm for} \qquad 
[g_{\rho\pi\pi}]_{\rm QCD} \simeq 6 
\,, 
\end{equation}
with $N_C=4$, to be consistent with the technidilaton coupling property versus the LHC Higgs~\cite{Matsuzaki:2012gd,Matsuzaki:2012xx,Matsuzaki:2015sya}.

The decay and production properties of the $\tilde{\rho}_\Pi^{\pm, 3}$ and $\tilde{\rho}_P^0$ are summarized for $F_\rho = (250, 500, 650)$ GeV  
as follows: 
\begin{itemize} 
\item The partial decay widths to diboson:  
\begin{eqnarray}  
\begin{array}{ccccc}
& & g_{\rho\pi\pi}=3 & g_{\rho\pi\pi}=4 &  g_{\rho\pi\pi}=5 \\ 
 \Gamma(\tilde{\rho}_\Pi^3 \to \tilde{W}\tilde{W}) [{\rm GeV}] 
& = 
& (29.2, 29.2, 29.2), &(50.3, 50.3, 50.3), &(77.3, 77.3, 77.3) \\ 
  \Gamma(\tilde{\rho}_\Pi^3 \to \tilde{W}_L \tilde{W}_L) [{\rm GeV}] 
& = 
&(26.6, 26.6, 26.6), &(47.2, 47.2, 47.2), & (73.8, 73.8, 73.8) \\  
  \Gamma(\tilde{\rho}_\Pi^\pm \to \tilde{W}^\pm \tilde{Z}) [{\rm GeV}] 
&= 
&(33.1, 33.1, 33.1), &(56.8, 56.8, 56.8), & (87.1, 87.1, 87.1) \\ 
  \Gamma(\tilde{\rho}_\Pi^\pm \to \tilde{W}_L^\pm \tilde{Z}_L) [{\rm GeV}] 
& = 
& (29.7, 29.7, 29.7), &(52.9, 52.9, 52.9), & (82.6, 82.6, 82.6) \\  
 \Gamma(\tilde{\rho}_P \to \tilde{W}\tilde{W}) [{\rm GeV}] 
&= 
& (3.48, 3.48, 3.48), &(6.00, 6.00, 6.00), &(9.22, 9.22, 9.22) \\  
  \Gamma(\tilde{\rho}_P \to \tilde{W}_L \tilde{W}_L) [{\rm GeV}] 
& = 
& (3.17, 3.17, 3.17), &(5.63, 5.63, 5.63), &(8.80, 8.80, 8.80) 
\,.   
\end{array}
\end{eqnarray}
\item The total widths: 
\begin{eqnarray} 
\begin{array}{ccccc} 
& & g_{\rho\pi\pi}=3 & g_{\rho\pi\pi}=4 &  g_{\rho\pi\pi}=5 \\ 
\Gamma_{\tilde{\rho}_\Pi^3}^{\rm tot} [{\rm GeV}]
&= 
& (29.7, 31.2, 32.6), &(50.8, 52.4, 53.8), & (77.8, 79.4, 80.8) \\  
\Gamma_{\tilde{\rho}_\Pi^\pm}^{\rm tot} [{\rm GeV}]
&= 
&(33.5, 34.8, 36.0), &(57.2, 58.5, 59.8), & (87.5, 88.9, 90.1) \\   
\Gamma_{\tilde{\rho}_P}^{\rm tot} [{\rm GeV}]
&=
&(3.56, 3.82, 4.06), & (6.08, 6.34, 6.58), & (9.3, 9.56, 9.80)    
\, .  
\end{array} 
\label{totwidth}
\end{eqnarray}
\item The relevant branching ratios:  
\begin{eqnarray} 
\begin{array}{ccccc} 
& & g_{\rho\pi\pi}=3 & g_{\rho\pi\pi}=4 &  g_{\rho\pi\pi}=5 \\ 
{\rm Br}[\tilde{\rho}_\Pi^3 \to \tilde{W}\tilde{W}] [\%]
&= 
&(98.2, 93.4, 89.4), & (99.0, 96.8, 93.6), & (99.3, 97.4, 95.7) \\ 
{\rm Br}[\tilde{\rho}_\Pi^\pm \to \tilde{W}^\pm \tilde{Z}] [\%]
&=
&(98.7, 95.0, 91.8), & (99.2, 97.6, 95.0), & (99.5, 98.0, 96.7) \\ 
{\rm Br}[\tilde{\rho}_P \to \tilde{W}\tilde{W}] [\%]
&=
&(97.6, 91.0, 85.6), & (98.6, 95.5, 91.1), & (99.1, 96.4, 94.0) 
\, ,    
\end{array} 
\end{eqnarray} 
for $j=u,d,s,c,b$, 
\begin{eqnarray} 
\begin{array}{ccccc} 
& & g_{\rho\pi\pi}=3 & g_{\rho\pi\pi}=4 &  g_{\rho\pi\pi}=5 \\ 
{\rm Br}[\tilde{\rho}_\Pi^3 \to jj] [\%]
&=
&(1.2, 4.6, 7.5), & (0.71, 2.8, 4.5), &(0.46, 1.8, 3.0) \\  
{\rm Br}[\tilde{\rho}_\Pi^\pm \to jj] [\%]
&=
&(0.66, 2.5, 4.1), & (0.38, 1.5, 2.5), & (0.25, 0.99, 1.6) \\  
{\rm Br}[\tilde{\rho}_P \to jj] [\%]
&=
&(1.6, 6.0, 9.5), & (0.94, 3.6, 5.9), & (0.61, 2.4, 3.9) 
\,,  
\end{array} 
\end{eqnarray}
and for $l=e, \mu$, 
\begin{eqnarray} 
\begin{array}{ccccc} 
& & g_{\rho\pi\pi}=3 & g_{\rho\pi\pi}=4 &  g_{\rho\pi\pi}=5 \\ 
{\rm Br}[\tilde{\rho}_\Pi^3 \to ll] [\%]
&=
&(0.30, 1.2, 1.9), & (0.18, 0.69, 1.1), & (0.12, 0.46, 0.76) \\  
{\rm Br}[\tilde{\rho}_\Pi^\pm \to l\nu] [\%]
&=
&(0.22, 0.84, 1.4),  &(0.13, 0.50, 0.83), &(0.084, 0.33, 0.55)  \\  
{\rm Br}[\tilde{\rho}_P \to ll] [\%]
&=
&(0.23, 0.87, 1.4), & (0.14, 0.52, 0.85), & (0.089, 0.35, 0.57) 
\,.    
\end{array}
\end{eqnarray}
\item The DY production cross sections at $\sqrt{s} = 8$ TeV decaying to $WW/WZ$:   
\begin{eqnarray} 
\begin{array}{ccccc} 
& & g_{\rho\pi\pi}=3 & g_{\rho\pi\pi}=4 &  g_{\rho\pi\pi}=5 \\ 
 \sigma_{\rm DY} (pp \to \tilde{\rho}_{\Pi}^3 \to \tilde{W}\tilde{W}) [{\rm fb}]
 &=
&(0.69, 2.6, 4.2), & (0.69, 2.7, 4.4), &  (0.69, 2.7, 4.5) \\   
  \sigma_{\rm DY} (pp \to \tilde{\rho}_{\Pi}^\pm \to \tilde{W}^\pm \tilde{Z}) [{\rm fb}]
 &
=& (1.3, 5.1, 8.3), & (1.3, 5.1, 8.5), &(1.3, 5.1, 8.5) \\  
  \sigma_{\rm DY} (pp \to \tilde{\rho}_P^0 \to \tilde{W}\tilde{W}) [{\rm fb}]
 &=
&(0.094, 0.35, 0.56), & (0.094, 0.36, 0.59), & (0.095, 0.45, 0.61) 
\,. 
\end{array}   
\label{cross:number:text} 
\end{eqnarray}
\end{itemize} 
Note that thanks to the presence of the custodial symmetry reflected by the relations in Eq.(\ref{custodial}),  
the decay rates for the charged $\tilde{\rho}_\Pi^\pm$ 
are almost the same as those for the $\tilde{\rho}_{\Pi}^3$ combined with the $\tilde{\rho}_P^0$ (via $c_\rho^2 + s_\rho^2 =1$). 
As to the cross sections, 
the custodial (isospin) symmetry is also well reflected as 
$\sigma_{\rm DY}(pp \to \tilde{\rho}_{\Pi}^{\pm}) \simeq 2 \sigma_{\rm DY}(pp \to \tilde{\rho}_\Pi^3 + \tilde{\rho}_P^0)$.

Let us compare the present analysis with the previous one in Ref.~\cite{Fukano:2015hga} where the mixing between 
${\rho}_{\Pi}^3$ and $\rho_P^0$ via the dynamical hypercharge gauge boson 
and other dynamical SM gauge contributions, such as the transverse modes of $W$ and $Z$ 
were not taken into account. 
Actually, as seen from Eq.(\ref{cross:number:text}), 
the numerical numbers of the decay rates and cross sections for $\tilde{\rho}_\Pi^{\pm, 3}$
are almost the same, since the $\rho_\Pi^3$ - $\rho_P^0$ mixing is tiny (which is supplied by $s_\rho \simeq 0.3$ in Eq.(\ref{srho:val})) 
and the $W/Z$ transverse-mode contributions are set to be quite small in accord with our perturbation analysis 
based on the assumption that $x = \sqrt{a}(m_W/M_\rho) \ll 1$ (Eq.(\ref{x:val})). 
Hence the LHC limits shown in Fig.~\ref{constraining-Frho-grhopipi-updated} look like almost the same as 
in Ref.~\cite{Fukano:2015hga}, 
so do the diboson signatures.

Considering the LHC detection of the 
DY coupling of the technirho, $(F_\rho/M_\rho)$, 
contributions from not only the walking technicolor sector, but also an extended technicolor (ETC)  
sector should be included there.  
(ETC communicates between the technifermion and the SM fermion sectors, 
being necessary to account for the SM-fermion mass generation.)  
The ETC generates an effective four-fermion interaction 
among the technifermions of the form 
$(\bar{F} \gamma_\mu F)^2$.  
This effective interaction affects $F_\rho$ to shift the 
value obtained from the walking technicolor sector alone.  
In that sense, one can say that 
the constraint on $F_\rho$ from the current LHC data in Fig.~\ref{constraining-Frho-grhopipi-updated} 
would imply the desired amount of the ETC contributions to the $F_\rho$. 
To see the desired ETC value more quantitatively,  
one may take the DY coupling estimated only from the walking technicolor sector 
to be $F_\rho^{\rm TC} \simeq 250$ GeV, 
which is supported from the result of nonperturbative 
calculations~\cite{Harada:2003dc,Appelquist:2014zsa}. 
Then the remaining amount may be supplied from the ETC, 
\begin{equation} 
F_\rho^{\rm ETC} = 0 - 400 \, {\rm GeV}  
\,, 
\qquad 
\textrm{for the total} \,\,
F_\rho = ( F_{\rho}^{\rm TC} + F_{\rho}^{\rm ETC}) = 250 - 650 \, {\rm GeV} 
\,. 
\end{equation}  
 This can be thought of as an indirect constraint on modeling of the ETC  
derived from the current LHC data. 
Looking back the present analysis, 
one may say that a role similar to    
such ETC effects has been played by  
the parameter $z_3$ in Eq.(\ref{Frho:def}) which shifts 
the $F_\rho$ at ${\cal O}(p^2)$ 
just like $F_\rho|_{{\cal O}(p^2)} \to F_\rho|_{{\cal O}(p^4)} = F_\rho^{\rm TC} + F_\rho^{\rm ETC}$.

One might suspect that the walking technirho with somewhat large DY coupling $F_\rho/M_\rho \sim (250 - 650\,{\rm GeV})/(2000\,{\rm GeV}) = 0.1 - 0.3$,   
leads to  
the sizable contribution to the $S$ parameter~\cite{Peskin:1990zt+}. 
One can in fact estimate the size of the $S$ coming only from the technirho contribution 
within the s-HLS model, including the above-mentioned $(\bar{F} \gamma_\mu F)^2$-type ETC contributions, to find  
\begin{equation} 
S|_{\rho} \simeq 4 \pi N_D \left( \frac{F_\rho}{M_{\rho}} \right)^2 
\simeq 0.8 - 5.3   
\, ,
\qquad 
{\rm for} \qquad F_\rho = 250 - 650\, {\rm GeV} 
\, ,
\end{equation}   
where $N_D =4$ for the  
one-family model. 
However, the techni-axialvector (techni-$a_1$) may destructively contribute to the $S|_\rho$ term, even including the above ETC effects. 
It has been suggested from several approaches~\cite{Harada:2003dc,Haba:2010hu,Matsuzaki:2012xx,Appelquist:2014zsa} 
that the masses of the techni-$\rho$ and -$a_1$ mesons are degenerate, $M_{\rho} \simeq M_{a_{1}}$, 
due to the characteristic walking feature. 
Taking into account this, one may add the techni-$a_1$ meson contribution to the $S$ as 
\begin{equation} 
S = S|_\rho + S|_{a_1} = 
4 \pi N_D \left( \frac{F_\rho}{M_{\rho}} \right)^2  \left[
1 - \left( \frac{F_{a_1}}{F_{\rho}} \right)^2 
\right] 
\,. \label{S-cancel}
\end{equation}        
Thus, if one has $F_{a_1} \simeq F_{\rho}$,  
the $S$ parameter can be vanishingly small, as it happens in a different context~\cite{Casalbuoni:1995yb}.    
This can actually occur in the one-family model of the walking technicolor, in 
view of a holographic dual (without ETC sector)~\cite{MY}.

The contribution to the $S$ parameter may actually come not only from the walking technicolor sector, 
but also from the 
ETC sector through the possible four-fermion operator $(\bar{F} \gamma_\mu f_{\rm SM})^2$ other than the above $(\bar{F} \gamma_\mu F)^2$, 
in a way analogous to ``delocalization" effect proposed in other models of electroweak symmetry breaking~\cite{Cacciapaglia:2004rb}. 
The possible cancellation in the $S$ parameter may therefore take place including such extra terms.

One should also note that  
the $g_{\rho\pi\pi}$ coupling generically gets modified 
due to the mixing between the pions and axialvector mesons such as the techni-$a_1$ 
mesons, 
as it happens in the case of the generalized HLS model~\cite{Bando:1987ym}, 
in application to QCD. 
Actually, in place of the techni-$a_1$ mesons,   
the $z_4$ term in Eq.(\ref{rhopipi:def}) mimics    
such a role of the shift effects, 
in that the $g_{\rho \pi\pi}$ coupling gets 
modified from the ${\cal O}(p^2)$ form ($1/2 ag$) 
by the $z_4$ term of ${\cal O}(p^4)$. 
In this respect, as well as the cancellation to the $S$ parameter in Eq.(\ref{S-cancel}),
it is of importance to discover the 2 TeV techni-$a_1$ mesons at the LHC Run-II, 
which will be definitely a smoking-gun of the one-family model of the walking technicolor 
to be pursued elsewhere.

\subsection{LHC diboson signatures at 2 TeV}

In this subsection, 
without the NWA, 
we perform the Monte Carlo simulation of the full hadronic decaying diboson analysis 
for the $2\,\TeV$ technirho mesons 
at the LHC.

We use the \texttt{FeynRules}~\cite{Alloul:2013bka,Christensen:2008py} to implement the coupling between 
technirho mesons and the SM particles 
into the \texttt{UFO} format~\cite{Degrande:2011ua},
then the events at the parton level 
($pp \to \tilde{\rho}_\Pi^\pm/\tilde{\rho}_\Pi^3/\tilde{\rho}^0_{_P} \to VV$, $V\to qq'$ 
where $V=W^\pm,Z^0$)
are generated by \texttt{Madgraph5\_aMC@NLO-2.3.0}~\cite{Alwall:2014hca}.
Here we set the factorization scale ($\mu_F$) and renormalization scale ($\mu_R$) to 
the $2\,\TeV$, $\mu_F = \mu_R = 2\,\TeV$.
The hadronization and parton showering are performed 
by using the \texttt{Pythia8.186}~\cite{Sjostrand:2007gs}.
We use the \texttt{CTEQ6L1}~\cite{Stump:2003yu} as a parton distribution function 
and the A2/AU2 tune~\cite{ATLAS:2011krm} for \texttt{Pythia8}.
The jets are reconstructed through the Cambridge-Aachen algorithm with the radius parameter
$R=1.2$ (CA12) by the \texttt{FastJet-3.0.6}~\cite{Cacciari:2011ma}.
To make a direct comparison with the ATLAS analysis~\cite{Aad:2015owa},
the CA12 jets are processed through a splitting and filtering algorithm described 
BDRS-A in Ref.~\cite{BDRS-A}. 
We  call this CA12 jets groomed jets hereafter. 
In addition, to account for the migration due to the ATLAS detector resolution, 
the groomed jet momentum/energy and mass values are smeared by Gaussian distributions 
with the mean of $0$ 
and the standard deviations of 5\% (as in Ref.~\cite{Aad:2015owa}) 
and 8\% (taken from $600<p_T<1000$~GeV bin in Table~2 of Ref.~\cite{BDRS-A}), respectively. 
We generate $10000$ signal events for the 
$M_{\tilde{\rho}_{_\Pi}^\pm} = M_{\tilde{\rho}_{_\Pi}^3} = 2\,\TeV$ 
and then we scale it to the DY cross section, 
$\sigma_{\text{DY}}(pp \to \tilde{\rho}_\Pi^\pm/\tilde{\rho}_\Pi^3/\tilde{\rho}^0_{_P} \to VV \to jj)$, 
times luminosity, ${\cal L}=20.3\,\text{fb}^{-1}$, for $\sqrt{s}=8\,\TeV$.

We apply the following event selections for groomed jets:
\begin{description}
\item[Cut 1:] 
(\# of groomed jets) $\geq 2$,  $p_{T_1} > 540\,\GeV$, 
$|\eta_{1,2}| < 2$ and $|y_1 - y_2| < 1.2$,

\item[Cut 2:]
$(p_{T_1} - p_{T_2})/(p_{T_1} + p_{T_2}) < 0.15$,

\item[Cut 3 (i):]
$\sqrt{y} \geq 0.45$, $n_{\text{trk}} < 30$,

\item[Cut 3 (ii):]
$|m_j - m_V| < 13.0\,\GeV$ where $m_W = 82.4\,\GeV$ and $m_Z = 92.8\,\GeV$,
\end{description}
where 
$\sqrt{y} = \min(\hat{p}_{T_{j1}},\hat{p}_{T_{j2}}) \Delta \hat{R}_{12}/\hat{m}_{12}$ 
is the momentum balance, 
$\hat{p}_{T_j}$ is the transverse momentum of the two subjets,
$\Delta \hat{R}_{12}$ is the distance between two subjets 
and $\hat{m}_{12}$ is the mass of the parent groomed jets. 
$n_{\text{trk}}$ is the number of the charged particle tracks which is associated with 
the original, ungroomed jets. 

In the previous analysis in \cite{Fukano:2015hga},
it was assumed that  
the $\tilde{\rho}_\Pi^{\pm,3}$ has the same $n_{\rm trk}$ distribution 
as the $W'$ signal used in ATLAS (Fig.~1b of Ref.~\cite{Aad:2015owa}). 
The charged particle multiplicity and kinematic properties are studied in the present analysis
and it is confirmed that the two signals have consistent charged particle distributions. 
It is also checked that the combined $n_{\text{trk}}$  cut efficiency and scaling factor 
($0.90\pm0.08$ in Ref.~\cite{Aad:2015owa} to account for the efficiency difference 
in data and the $W'$ simulation)
is properly reproduced by the cut on the charged particle track multiplicity. 
In the present analysis, we thus apply the $n_{\text{trk}}$ requirement 
to the simulated technirho samples without any scaling factors.

The Cut 1 and 2 correspond to the event topology requirements in Ref.~\cite{Aad:2015owa}
and 
the Cut 3 does to the boson tagging requirements. 
We find from above event selection Cut 3 (ii) that 
the $\tilde{\rho}_\Pi^\pm/\tilde{\rho}_\Pi^3/\tilde{\rho}^0_{_P} \to WW/WZ$  events 
may contaminate in the $ZZ$ selection 
since 
$WW,WZ$ and $ZZ$ selections are distinguished only by the jet mass window of the groomed jets.
In order to see the contamination in the $ZZ$ selection, 
it is convenient to use the efficiencies of the jet mass window cuts in Cut 3 (ii).
For this purpose, 
we divide Cut 3 (ii) into the six categories~\cite{Allanach:2015hba}: 
\begin{description}
\item[Category A:] 
$m_W -13.0 \leq m_{j_1,j_2} \leq m_Z - 13.0$,
\item[Category B:]
$m_W -13.0  \leq \min (m_{j_1},m_{j_2}) \leq m_Z -13.0$ 
and 
$m_Z -13.0  \leq \max (m_{j_1},m_{j_2}) \leq m_W +13.0$,
\item[Category C:]
$m_Z -13.0 \leq m_{j_1,j_2} \leq m_W + 13.0$,
\item[Category D:]
$m_W -13.0  \leq \min (m_{j_1},m_{j_2}) \leq m_Z -13.0$ 
and 
$m_W+13.0  \leq \max (m_{j_1},m_{j_2}) \leq m_Z +13.0$,
\item[Category E:]
$m_Z -13.0  \leq \min (m_{j_1},m_{j_2}) \leq m_W +13.0$ 
and 
$m_W +13.0  \leq \max (m_{j_1},m_{j_2}) \leq m_Z +13.0$,
\item[Category F:]
$m_W +13.0 \leq m_{j_1,j_2} \leq m_Z + 13.0$.
\end{description}
The number of events after each cut of of signal events from 
$\tilde{\rho}_{_\Pi}^\pm$, $\tilde{\rho}_{_\Pi}^3 +\tilde{\rho}^0_{_P}$ 
are listed  in Appendix \ref{breakdown_Signal}. 
The numbers of signal events at $1.85\,\TeV \leq m_{jj}\leq 2.15\,\TeV $, 
corresponding to three $m_{jj}$ bins centered at the $m_{jj}=2\,\TeV$ bin,   
are also shown in Appendix~\ref{breakdown_Signal}.

\begin{table}[ht]
\centering
\begin{tabular}[t]{|c|c|c|c|}
\multicolumn{4}{c}{The ATLAS results~\cite{Aad:2015owa}}
\\ \hline
\parbox[c][4ex][c]{0ex}{}
$m_{jj}$ bin [GeV] & $WW$ selection & $WZ$ selection & $ZZ$ selection  
\\ \hline \hline
\parbox[c][4ex][c]{0ex}{}
$1850 \leq m_{jj} \leq1950$ 
& $1.33 \pm 2.04$ & $1.88^{+ 2.28}_{-2.29}$ & $3.81\pm2.25$
\\
\parbox[c][4ex][c]{0ex}{}
$1950 \leq m_{jj} \leq 2050$ 
& $5.16^{+2.66}_{-2.67}$ & $5.92^{+ 2.85}_{-2.86}$ & $2.24^{+1.74}_{-1.75}$
\\
\parbox[c][4ex][c]{0ex}{}
$2050 \leq m_{jj} \leq 2150$ 
& $0.7^{+1.43}_{-1.44}$ & $0.59^{+ 1.44}_{-1.45}$ & $0.5\pm1.01$
\\ \hline \hline
\parbox[c][4ex][c]{0ex}{}
$1850 \leq m_{jj} \leq 2150$ 
& $7.19^{+3.62}_{-3.64}$ & $8.39^{+3.89}_{-3.92}$ & $6.55^{+3.01}_{-3.02}$
\\ \hline
\end{tabular}
\caption{The number of events obtained by subtracting the predicted events of the background-only fit 
from the events observed in the $WW$, $WZ$, and $ZZ$ selections 
in each dijet mass bin.}\label{ATLAS_results}
\end{table}

In Fig.~\ref{results_technirhos}, 
we show the signal events in dijet mass distributions/100 GeV bin 
with $WW$ (top panel), $WZ$ (middle panel), $ZZ$ (bottom panel) selections 
at $\sqrt{s}=8$ TeV and the integrated luminosity ${\cal L}=20.3 \, {\rm fb}^{-1}$ 
where each signal is the sum of all signal events from 
$\tilde{\rho}_{_\Pi}^\pm$, $\tilde{\rho}_{_\Pi}^3$ and $\tilde{\rho}^0_{_P}$. 
In Fig.~\ref{results_technirhos}, 
the data given in Table.~\ref{ATLAS_results} 
in the $WW$, $WZ$ and $ZZ$ selections in the dijet mass with the mass window 
$1850\,\GeV \leq m_{jj} \leq 2150\,\GeV$.
are also shown for comparison. 
One should notice it hard to distinguish $\tilde{\rho}^0_{_P}$ from $\tilde{\rho}_{_\Pi}^3$ 
at the present level of the resolution of the detector (100 GeV bin) 
since the mass difference 
$\Delta M = M_{\tilde{\rho}_\Pi^3} - M_{\tilde{\rho}_P^0}\simeq (1.6 - 16)\,{\rm GeV}$ 
is very small as in Eq.(\ref{mass:diff:val}). 
In addition, the total decay width of $\tilde{\rho}^0_{_P}$, $\Gamma_{\tilde{\rho}^0_{_P}}$,  
is much smaller than $\Gamma_{\tilde{\rho}_{_\Pi}^3}$ 
(See Eq.(\ref{totwidth}) or Fig.~\ref{constraining-Frho-grhopipi-updated}).  
Therefore, the diboson events/100 GeV bin generated from the $\tilde{\rho}_P^0$ 
contaminate in the 2 TeV events from the $\tilde{\rho}_\Pi^{\pm, 3}$. 
In Fig.~\ref{results_technirhos} 
the numbers of events in the histograms thus include 
the contribution from the $\tilde{\rho}_P^0$.


 From Fig.~\ref{results_technirhos} we thus see that 
the 2 TeV walking technirhos ($\tilde{\rho}_{\Pi}^{\pm, 3}, \tilde{\rho}_P^0$) 
account for the observed excesses in the $WW$, $WZ$ and $ZZ$ selections, 
with the DY coupling $F_\rho$ in a range,  
\begin{equation} 
 F_\rho = 450 - 650 \, {\rm GeV} 
\qquad  
 \left( 
 F_\rho/M_\rho = 0.23 - 0.33 
 \right)
\,,  
\end{equation}  
and the coupling to diboson $g_{\rho \pi\pi}$, 
\begin{equation} 
g_{\rho \pi\pi} = 3 ,4, 5 
\,,  
\end{equation} 
which controls the total width to be $\lesssim 100 $ GeV. 
Thus, if at the ongoing Run II 
the excesses grow in the diboson channels, 
but not in the $VH$ channel, 
then it can be a strong hint of the presence of the 2 TeV walking technirhos 
with the conformal barrier.

\begin{figure}[h]
\begin{center}
\begin{tabular}{lll}
\includegraphics[scale=0.25]{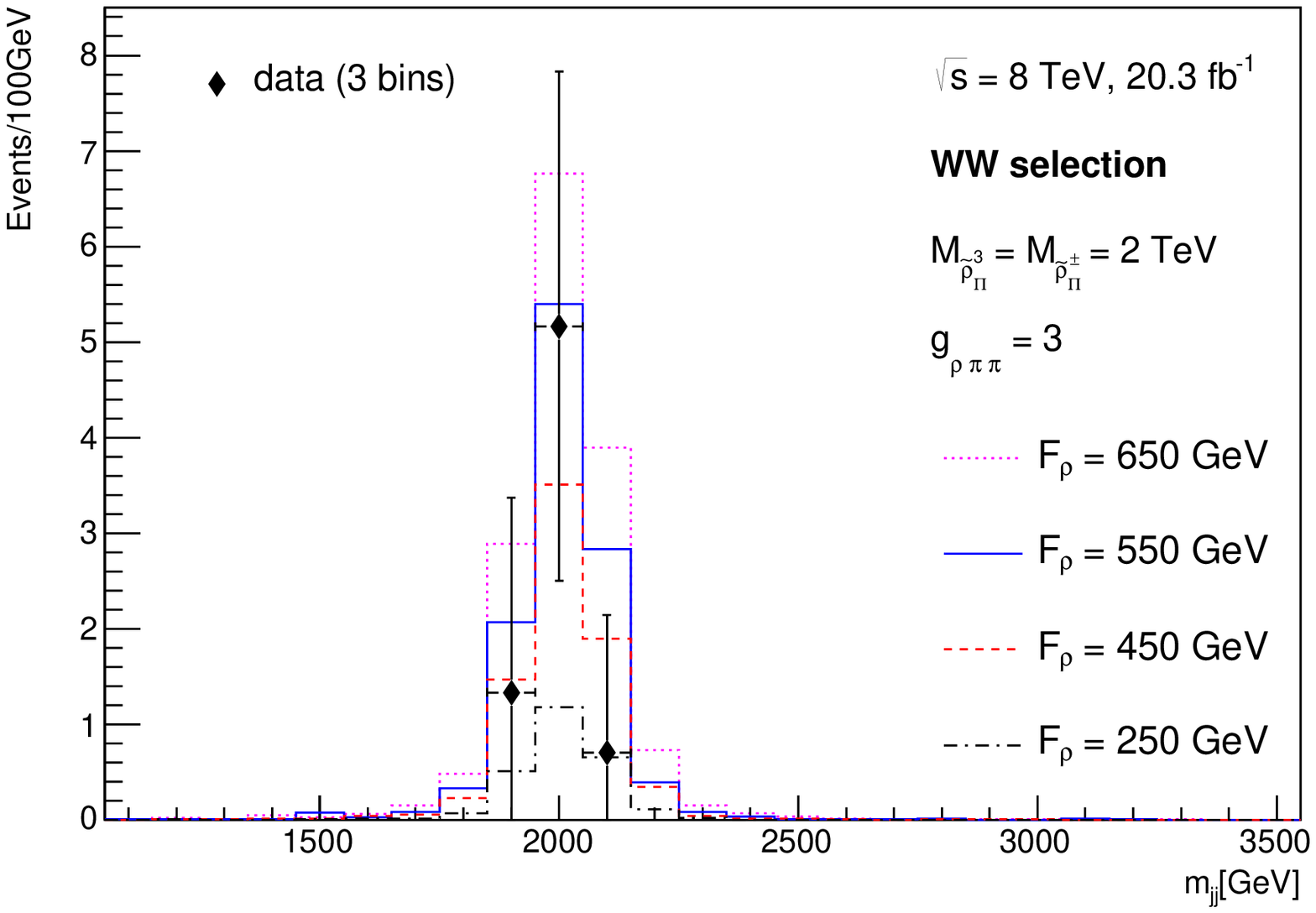} 
&
\includegraphics[scale=0.25]{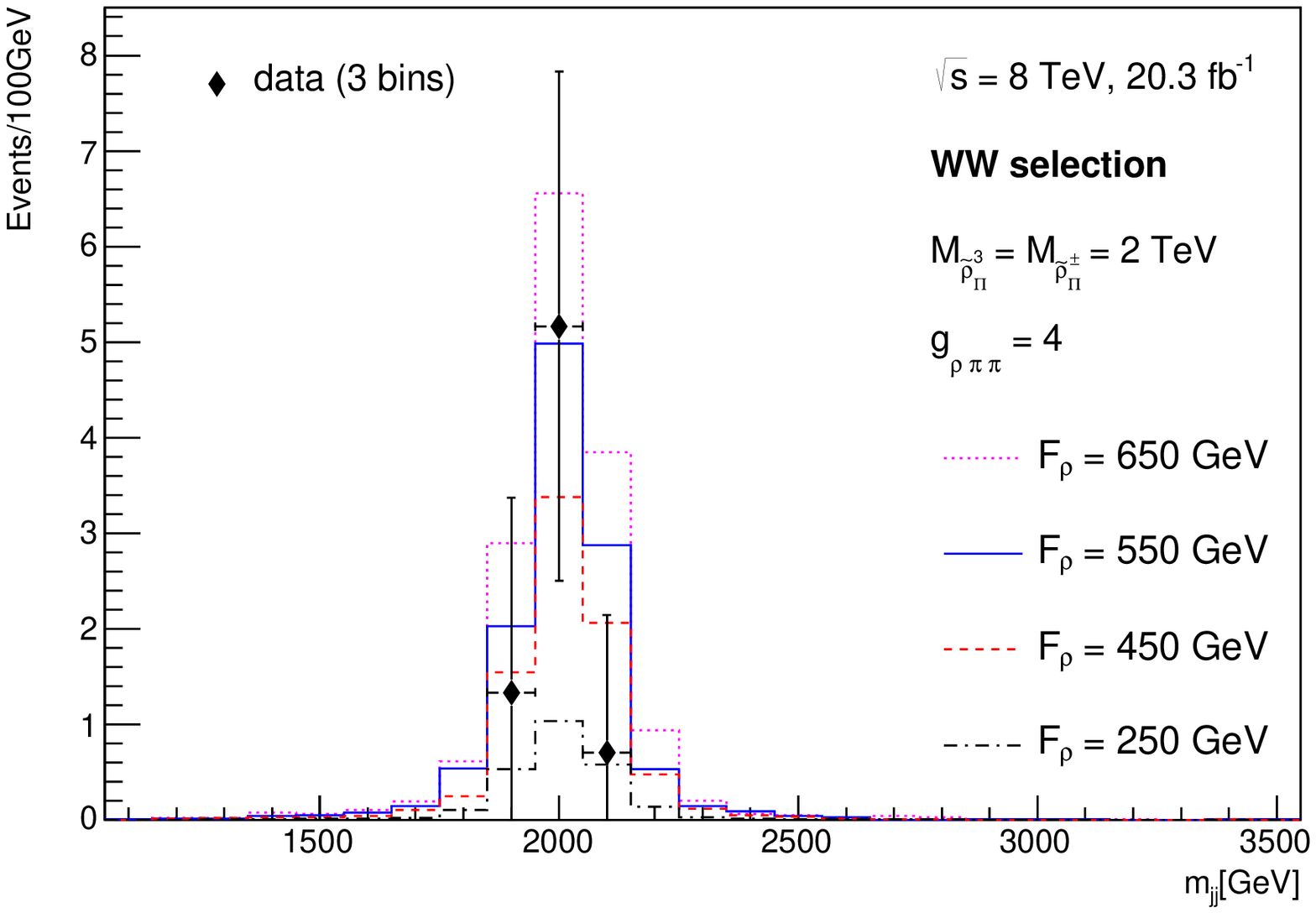} 
&
\includegraphics[scale=0.25]{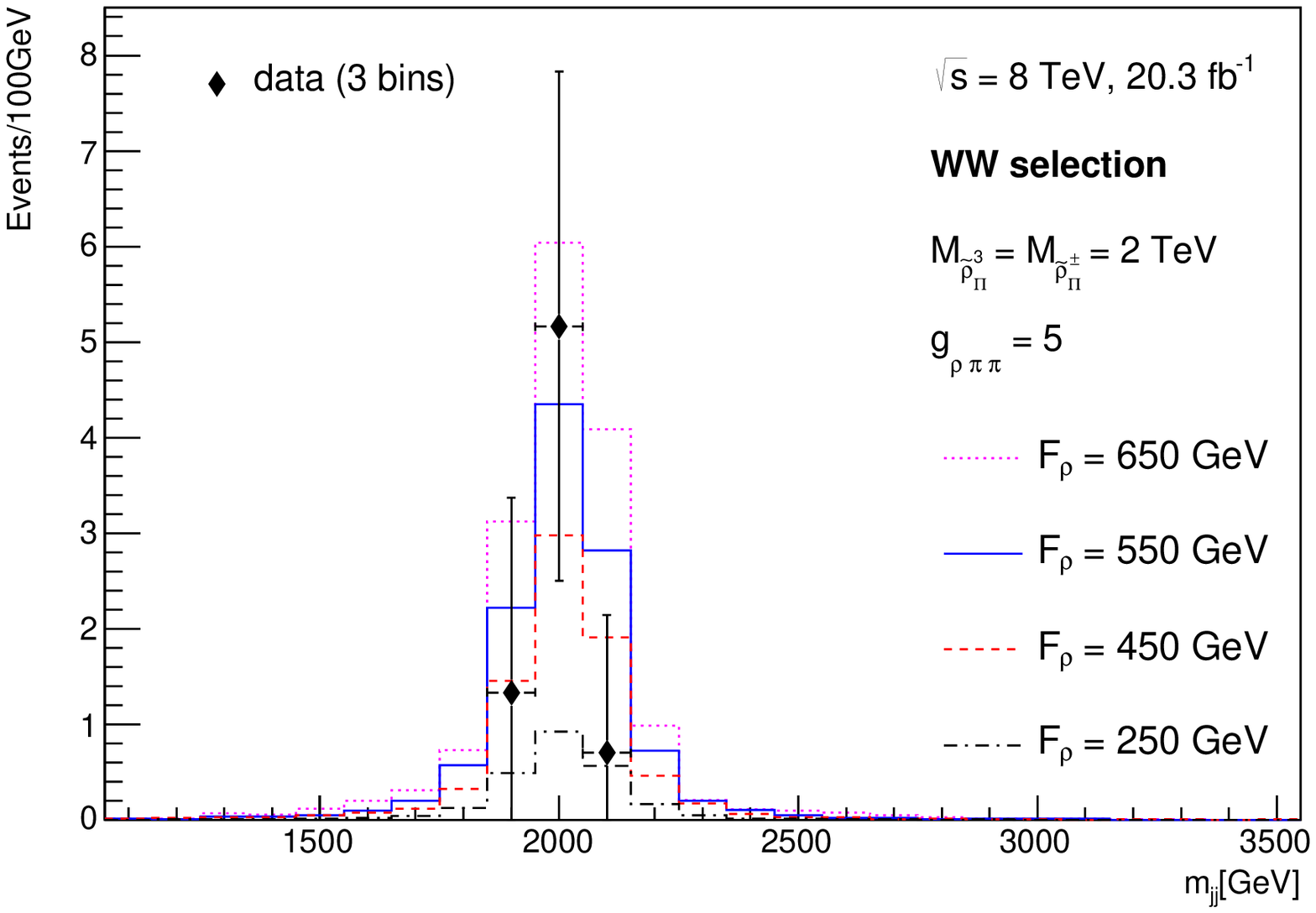} 
\\
\includegraphics[scale=0.25]{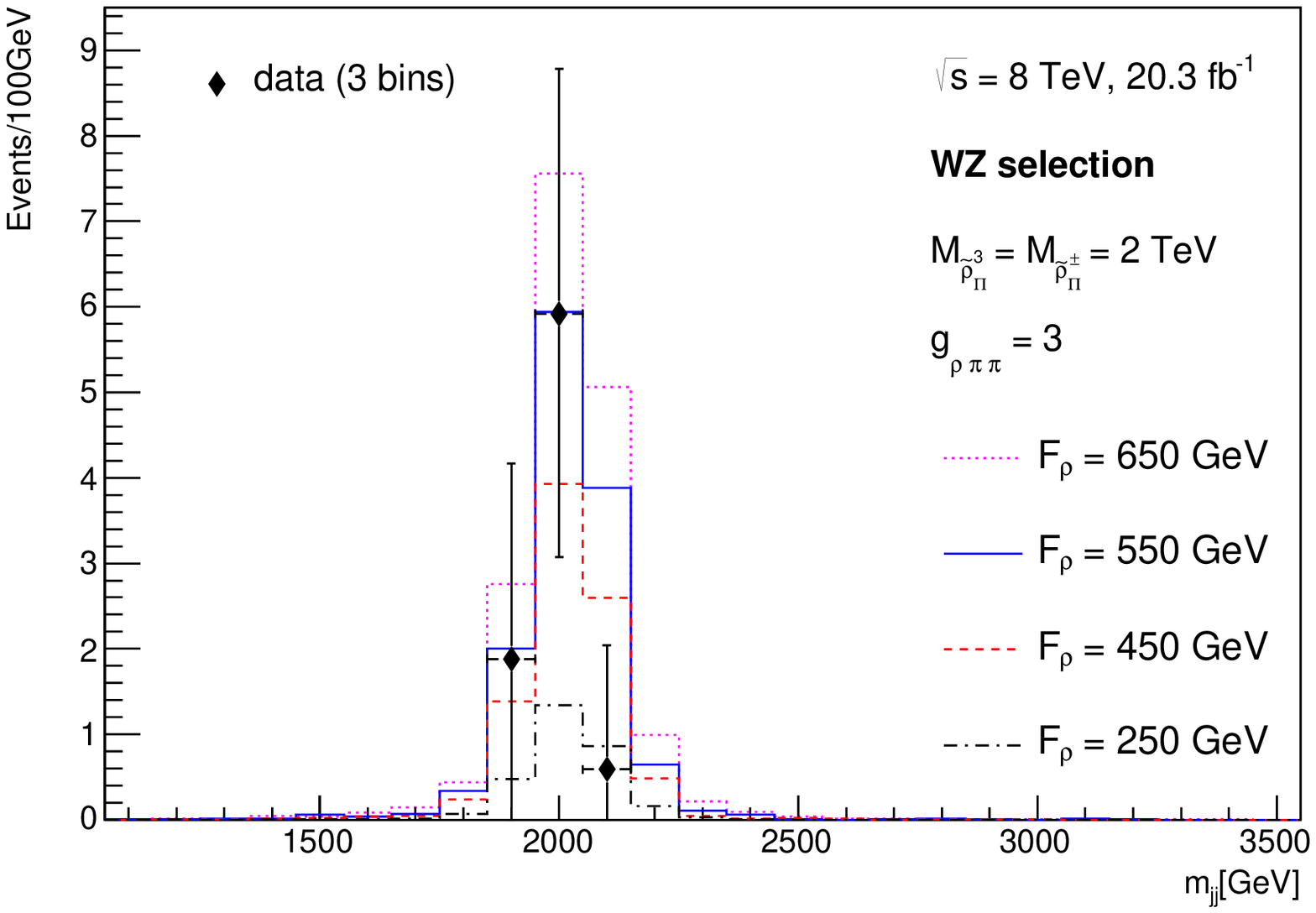} 
&
\includegraphics[scale=0.25]{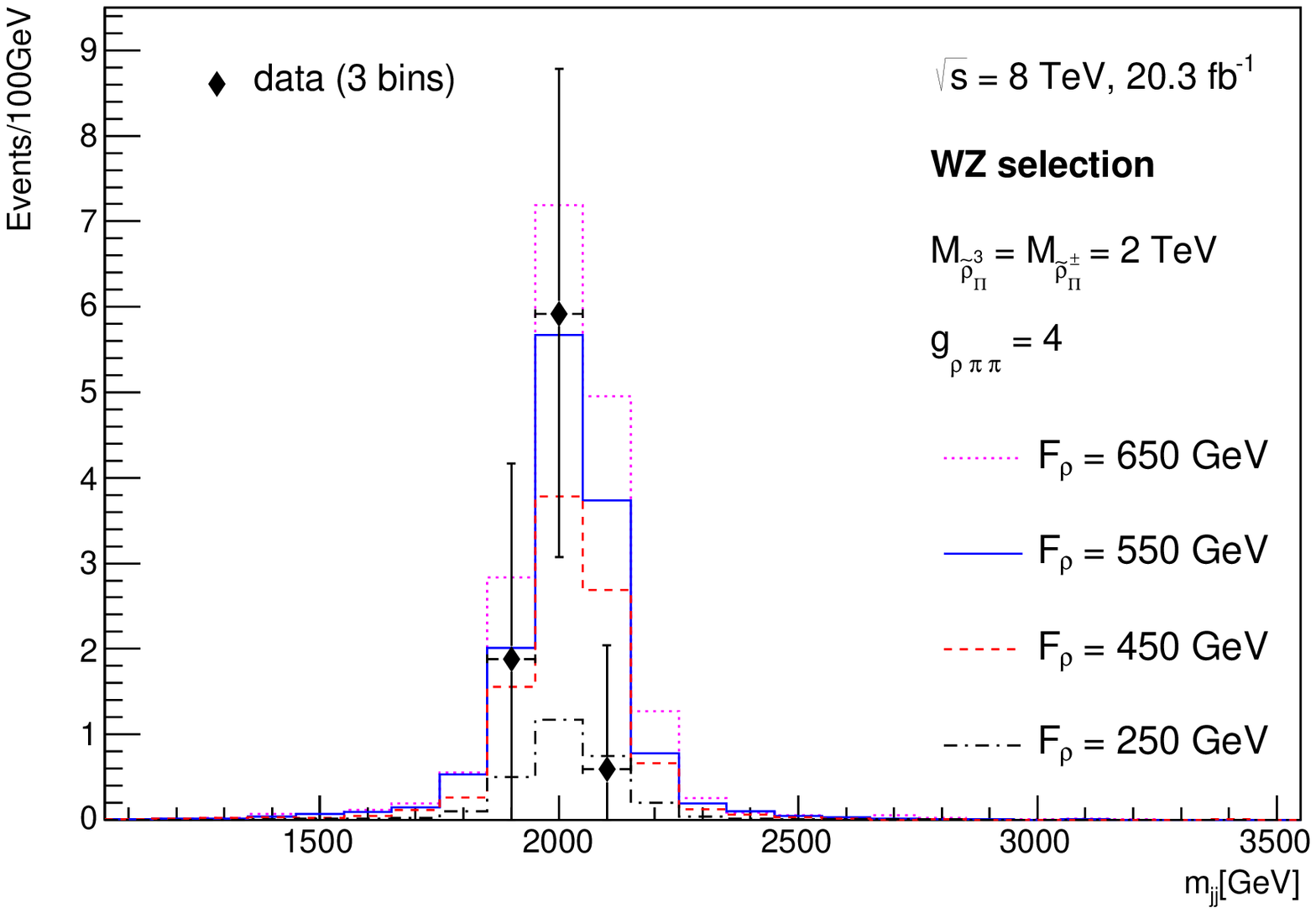} 
&
\includegraphics[scale=0.25]{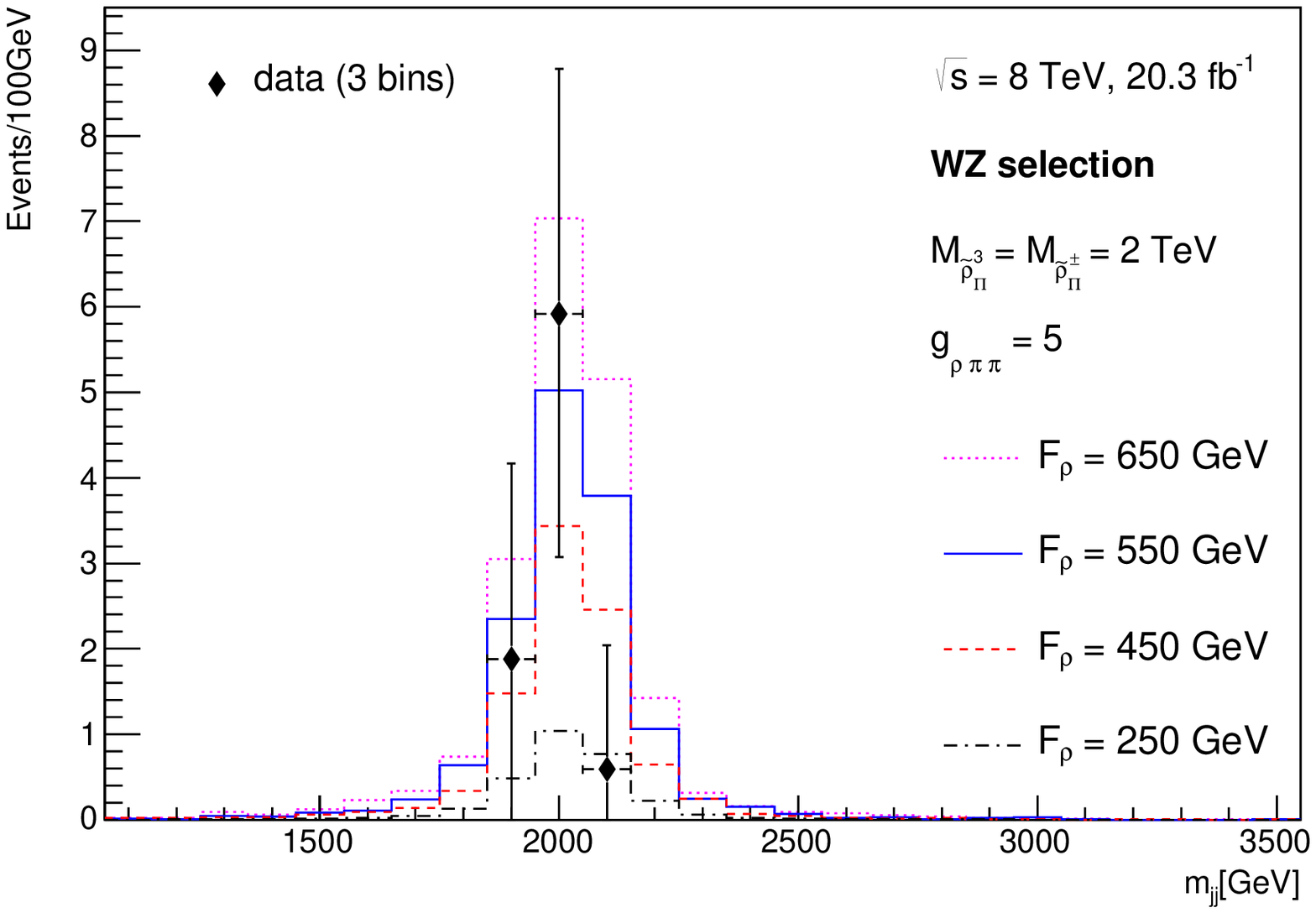} 
\\
\includegraphics[scale=0.25]{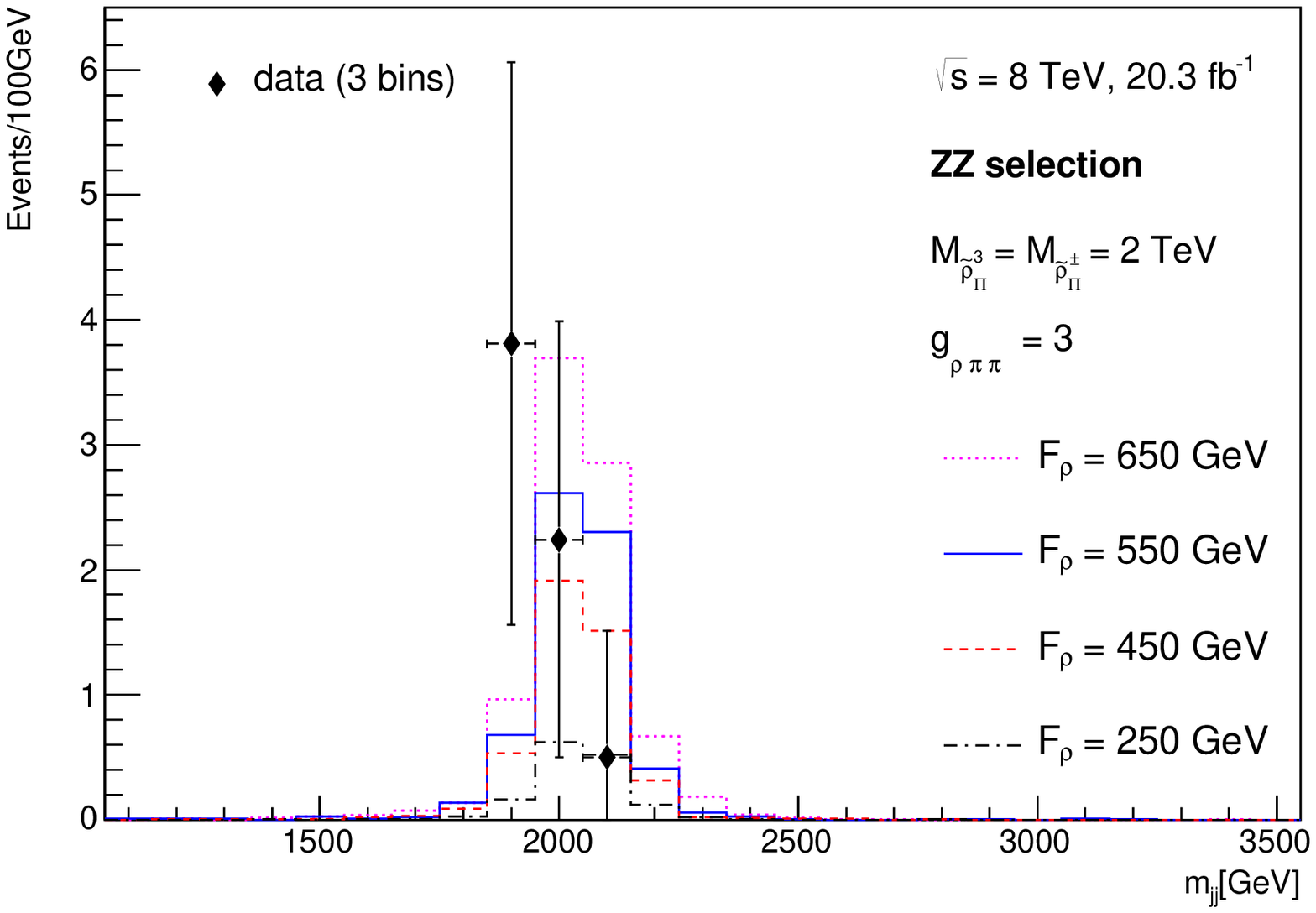} 
&
\includegraphics[scale=0.25]{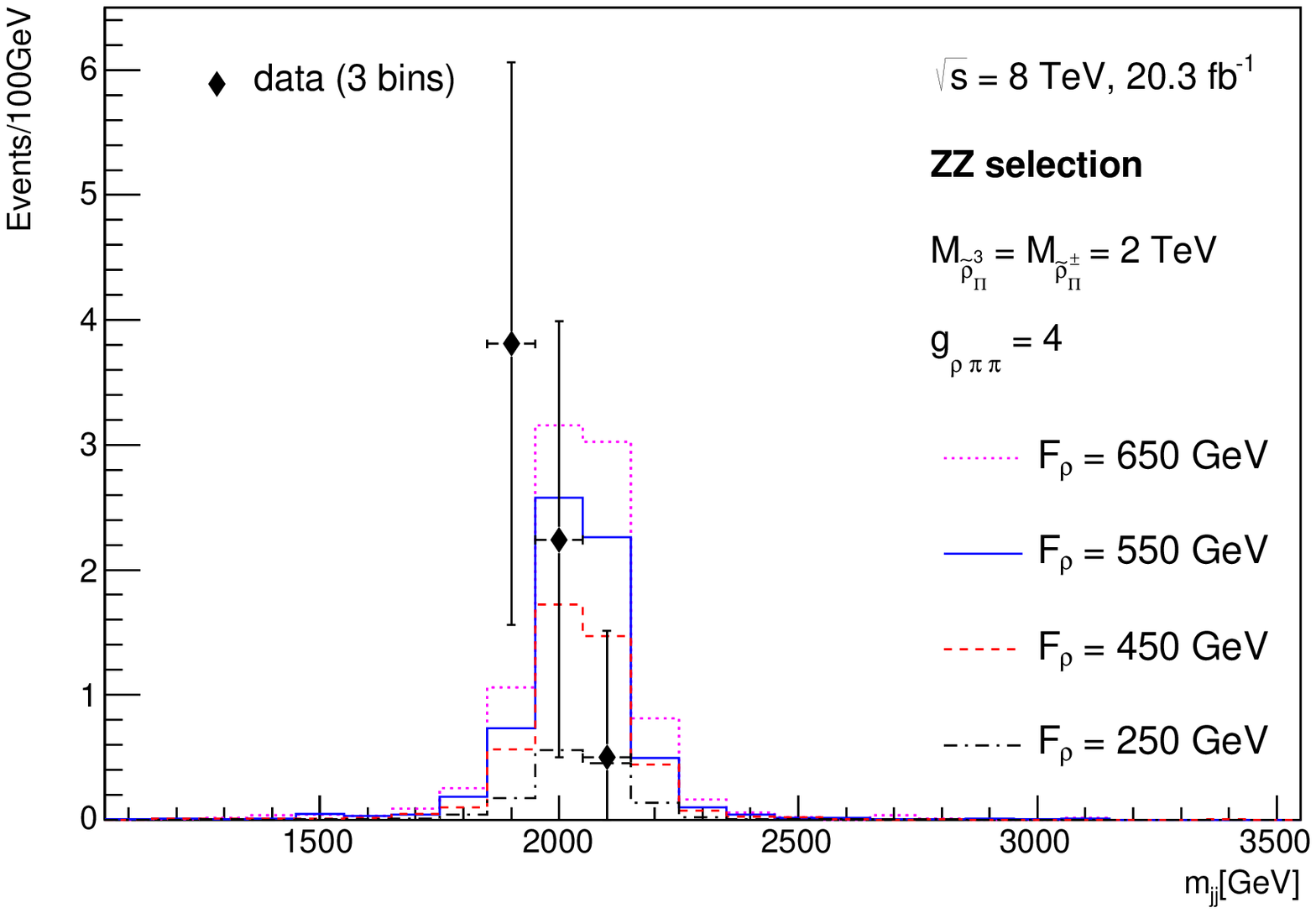} 
&
\includegraphics[scale=0.25]{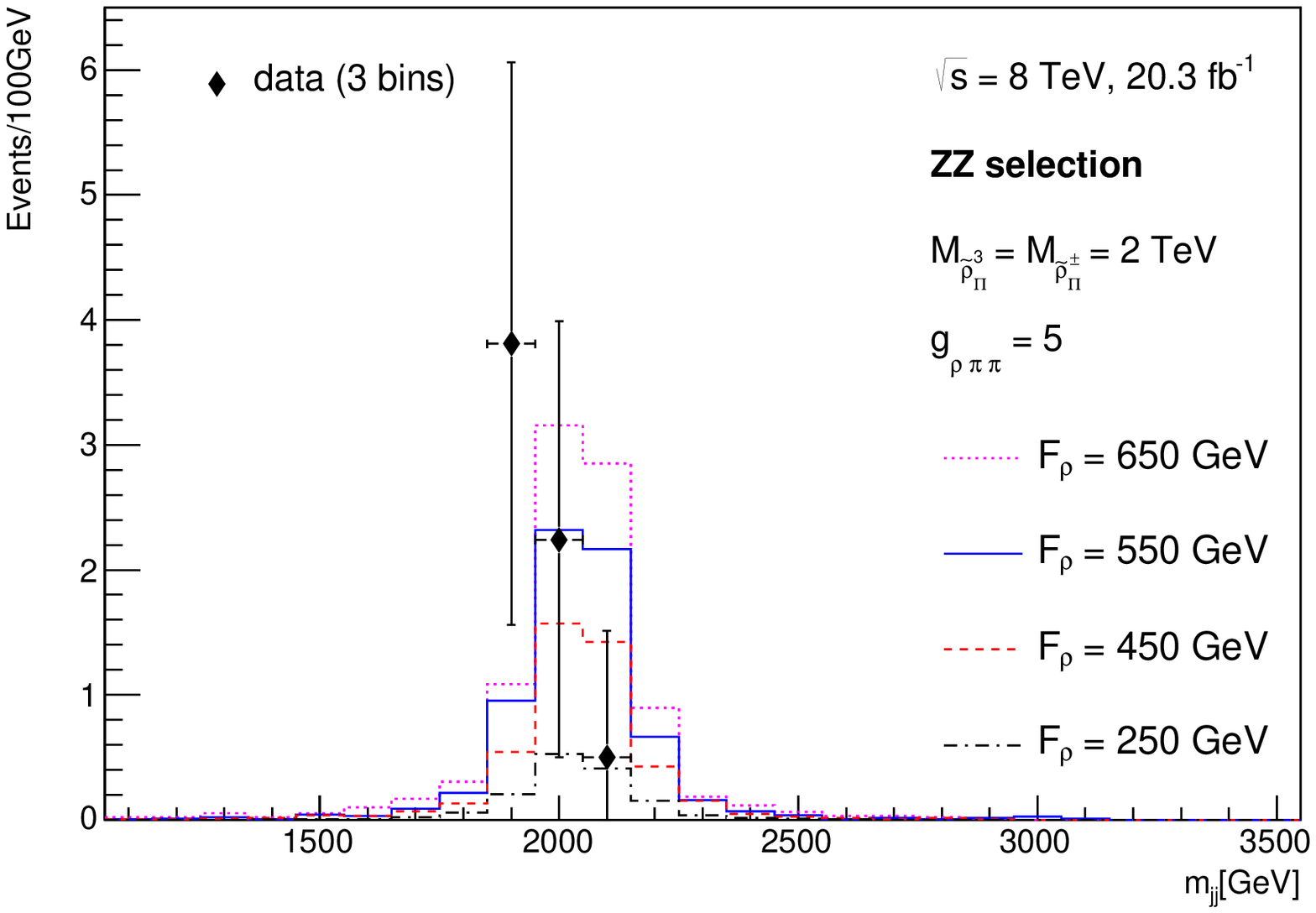} 
\end{tabular}
\caption{
Results for the signal events after all cuts selected in 
$WW$ (top row), $WZ$ (middle row) and $ZZ$ (bottom row) selections  
for technirhos with 
$g_{\rho \pi \pi} =3$ (left column), $4$ (middle column) and $5$ (right column),
$F_\rho = 650$ (dotted line), $550$ (solid line), $450$ (dashed line) and $250$ (dashed-dotted line)
at $M_{\tilde{\rho}^{\pm,3}_{_\Pi}} = 2\,\TeV$.
For comparison, 
the data given in Table.\ref{ATLAS_results} are also shown. 
All the histograms include the $\tilde{\rho}_P^0$ contributions with the small mass difference and width 
explained in the text. 
\label{results_technirhos}}
\end{center}
\end{figure}%

\section{Summary and discussion} 
\label{summary}

In this paper, 
we discussed in details the consequences of the conformal barrier, 
including the higher order effects through mixing and transverse $W, Z$ effects,  
which were not considered in the previous analysis.  
We also demonstrated another characteristic feature of our model, 
the gauge invariance of the HLS, 
the exact symmetry (though spontaneously broken): 
the HLS forbids a possible mixing between the isospin-triplet one-family walking technirho's, $\rho_\Pi^i, \rho_P^i$, 
which, were it not for the HLS, would mix each other by the explicit  breaking of the global $SU(8)_L\times SU(8)_R$ symmetry 
by the SM gauge interaction, 
thereby affecting the previous analysis.  
The $\rho^i_{\Pi}$ are produced by the Drell-Yan process, 
while the $\rho_P^i$ orthogonal to $\rho_{\Pi}^i$ are not produced by the Drell-Yan process and 
is totally irrelevant to the diboson processes in the absence of the mixing. 
Instead, we newly investigated 
 the small mixing effect of $\rho_\Pi^3$ with $\rho^0_P$ (isospin-singlet) 
through the {\it transverse modes} of the $W/Z$ bosons (via $W/Z$ kinetic term mixing 
after mass diagonalization), being of higher order term of ${\cal O} (p^4)$ in the s-HLS chiral perturbation theory, 
which was not considered in the previous studies dealing with only the longitudinal $W/Z$ modes to treat the $W/Z$ as the ``external fields'' (not dynamical). 
We showed that
such higher order term are substantially negligible, and the essential features of our previous results, 
including the characteristic smallness of the decay width, remain unchanged thanks to the power of the conformal barrier and the HLS, 
after all the phenomenological analyses were newly performed under new setting and inputs.

At the ongoing Run II 
the excesses could grow in the diboson channels, but not in the $VH$ channel.  
If that happens, the new resonance is strongly suggested to be the 2 TeV walking technirho  
protected by the conformal barrier and the HLS. 
\\

Several comments are in order: 
\\

As noted in Eq.(\ref{S-cancel}), 
due to the $S$ parameter constraint, 
the large $F_\rho$ requires a large $F_{a_1}$, DY coupling of techni-a1
having the same mass $\simeq 2$ TeV as the technirho. 
This implies that the techni-$a_1$ meson at 2 TeV can be sufficiently produced through the DY process, 
to be discovered at the LHC Run-II. 
LHC discovery channels of the techni-$a_1$ mesons will be pursued in another publication. 
\\

Our model includes charged vector bosons $(\rho_\Pi^\pm, \rho_P^\pm)$ which are allowed to couple
 to electroweak gauge bosons and the Higgs (technidilaton $\phi$) such as $\rho_{\Pi,P}^+ - \rho_{\Pi,P}^- - Z/\gamma$ and 
$\rho_{\Pi,P}^+ - \rho_{\Pi,P}^- - \phi$, 
so one might suspect that
their radiative corrections 
could 
affect the Higgs coupling property, say, the Higgs coupling 
to diphoton. 
 However, it is not the case, since 
 the relevant process receives non-decoupling contributions, the effective theory estimate should be equivalent to
 the estimate based on the ultraviolet (UV) completion, namely the underlying walking technicolor,
  which has actually been already done in Refs.~\cite{Matsuzaki:2012gd,Matsuzaki:2012xx}, shown to be consistent with 
the current Higgs data.  In fact, explicit evaluation shows that the techni-$\rho$ loop contribution is
 negligibly small compared with the UV completed estimate~\cite{Matsuzaki:2012gd,Matsuzaki:2012xx}, due to the suppression of 
the $\rho_{\Pi, P}^+ - \rho_{\Pi, P}^- - \gamma$ coupling by a factor of $x \sim (m_W/M_\rho) \sim {\cal O}(10^{-2}-10^{-1})$ 
(See Eq.(\ref{x:val})), thanks to the HLS invariance. 
Of course, the effective theory estimate of such non-decoupling quantities
should include not just the techni-$\rho$ loop but also all the possible composite states contributions, which, if
done properly, would match the UV completed estimate of the Refs.~\cite{Matsuzaki:2012gd,Matsuzaki:2012xx}
anyway, as it 
naturally be expected. 
\\

As seen from Fig.~\ref{constraining-Frho-grhopipi-updated} 
a large region of the model-parameter space has already been 
strongly constrained by the present 8 TeV data. 
The upcoming LHC-Run II data will further squeeze 
the parameter space. 
Among constraints, the most severe bound has presently come from the $WZ(l\nu J)$ channel, 
$\sigma_{\rm 8TeV} \times {\rm Br} \lesssim 9.5 \, {\rm fb}$, 
which constrains the DY coupling $F_\rho$ to be $\lesssim 650$ GeV 
(See Eq.(140)). 
Recently ATLAS and CMS collaborations released preliminary 
results from 13 TeV data collected in LHC Run 2 on the diboson $WZ (l \nu J)$ channel~\cite{ATLAS13,CMS:2015nmz}. 
The resulting $\sigma \times {\rm BR}$ limits from 3.2 ${\rm fb}^{-1}$ 
(ATLAS)~\cite{ATLAS13} and 2.2 ${\rm fb}^{-1}$ (CMS)~\cite{CMS:2015nmz} data are 
similar between the two experiments (given the difference of the collected luminosities). 
The ATLAS result shows improved sensitivity with respect to the $\sigma_{\rm 8TeV} \times {\rm Br}$ 
limit scaled by the parton luminosity ratio between 8 and 13 TeV for the DY processes. 
More data at 13 (or future 14) TeV LHC could be enough to prove or disprove the present model.

\acknowledgments

This work was supported in part by 
the JSPS Grant-in-Aid for Young Scientists (B) \#15K17645 (S.M.).

\appendix

\section{Expanding the s-HLS Lagrangian in terms of the mass-eigenstates} 
\label{diagonalization}

In this Appendix we diagonalize the gauge sector in the s-HLS Lagrangian 
including the dynamical SM gauges and possible higher order corrections 
in the derivative expansion.

The starting Lagrangian relevant to the present analysis is: 
\begin{eqnarray}
{\cal L}_{\rm s-HLS}
&=& \chi^2 F_\pi^2 \left(
 {\rm tr}[\hat{\alpha}_{\perp \mu}^2] 
 + 
 a 
 {\rm tr}[\hat{\alpha}_{|| \mu}^2]
 \right) 
 - \frac{1}{2g^2} {\rm tr}[V_{\mu\nu}^2] 
 - \frac{1}{2g_W^2} {\rm tr}[W_{\mu\nu}^2] 
 - \frac{1}{2g_Y^2} {\rm tr}[B_{\mu\nu}^2] 
+ {\cal L}_{4} 
\,, \label{Lag:start} \\ 
{\cal L}_4 
&=& z_3 {\rm tr}[\hat{\cal V}_{\mu\nu} V^{\mu\nu}] 
- i z_4 {\rm tr}[V_{\mu\nu} \hat{\alpha}_\perp^\mu \hat{\alpha}_\perp^\nu] 
+ i z_5 {\rm tr} [V_{\mu\nu} \hat{\alpha}_{||}^\mu \hat{\alpha}_{||}^\nu] 
\nonumber \\ 
&& 
+ i z_6 {\rm tr}[\hat{\cal V}_{\mu\nu} \hat{\alpha}_\perp^\mu \hat{\alpha}_\perp^\nu] 
+ i z_7 {\rm tr}[\hat{\cal V}_{\mu\nu} \hat{\alpha}_{||}^\mu \hat{\alpha}_{||}^\nu] 
- i z_8 {\rm tr}[\hat{\cal A}_{\mu\nu} (\hat{\alpha}_\perp^\mu \hat{\alpha}_{||}^\nu 
+ \hat{\alpha}_{||}^\mu \hat{\alpha}_{\perp}^\nu 
)]  
\,.  \label{Lag:start:4}
\end{eqnarray}
In the unitary gauges for the HLS and chiral symmetry, we have 
($\tau^i=\sigma^i/2$)
\begin{eqnarray} 
&&  W_{\mu} = W_\mu^i \tau^i\,, \qquad B_\mu = B_\mu \tau^3 
\,,   \nonumber \\ 
&& 
V_\mu = 
\frac{g}{2}\left( 
\begin{array}{c|c} 
 (\rho_{\Pi \mu}^i \tau^i + \frac{1}{2 \sqrt{3}} \rho_{P \mu}^0 \cdot {\bf 1}_{2\times 2}) \times {\bf 1}_{3\times 3} & 0 \\ 
\hline 
 0 & \rho_{\Pi \mu}^i \tau^i - \frac{3}{2\sqrt{3}} \rho_{P \mu}^0 \cdot {\bf 1}_{2\times 2} 
\end{array}
\right)
\,, 
\end{eqnarray}    
\begin{eqnarray}    
\hat{\alpha}_{\perp \mu} 
&=& {\cal A}_\mu 
=  
\frac{1}{2}\left( 
\begin{array}{c|c} 
 (g_Y B_\mu \tau^3 - g_W W_\mu^i \tau^i) \times {\bf 1}_{3\times 3} & 0 \\ 
\hline 
 0 & (g_Y B_\mu \tau^3 - g_W W_\mu^i \tau^i)
\end{array}
\right) 
\,, \\ 
\hat{\alpha}_{|| \mu} 
&=& {\cal V}_\mu - V_\mu 
=  
\frac{1}{2}\left( 
\begin{array}{c|c} 
 (g_W W_\mu^i \tau^i + g_Y B_\mu (\tau^3 + \frac{1}{3} \, {\bf 1}_{2 \times 2}) ) \times {\bf 1}_{3\times 3} & 0 \\ 
\hline 
 0 & g_W W_\mu^i \tau^i - g_Y B_\mu (1 - \tau^3) 
\end{array}
\right) 
- V_\mu 
\,, 
\end{eqnarray}          
where we have focused only on the color-singlet vector and axialvector fields relevant to the LHC diboson study. 
 and discarded 
terms involving the color-singlet isospin-triplet $\rho_P^i$ since they do not couple to 
the electroweak sector up to the cubic order in fields, due to the orthogonality reflected by the HLS gauge invariance 
(See text).

\subsection{Diagonalization at ${\cal O}(p^2)$}

At the leading order of the derivative expansion, ${\cal O}(p^2)$, 
substituting the above expressions into the Lagrangian Eq.(\ref{Lag:start}),  
we find the mass matrices for the charged and neutral vector boson sectors (${\cal M}_{CC}^2$ and ${\cal M}_{NC}^2$), 
\begin{eqnarray} 
 {\cal L}_{\rm mass}^{{\cal O}(p^2)} 
&=&  
 \left( 
 \begin{array}{c} 
 W_\mu^+ \\
 \rho_{\Pi \mu}^+ 
\end{array}
\right)^T 
\cdot {\cal M}_{CC}^2  
\cdot 
   \left( 
 \begin{array}{c} 
 W_\mu^+ \\
 \rho_{\Pi \mu}^+ 
\end{array}
\right) 
+ \frac{1}{2} 
 \left( 
 \begin{array}{c} 
 W_\mu^3 \\
 \rho_{\Pi \mu}^3 \\ 
B_\mu \\ 
\rho_{P \mu}^0  
\end{array}
\right)^T 
\cdot {\cal M}_{NC}^2  
\cdot 
   \left( 
 \begin{array}{c} 
 W_\mu^3 \\
 \rho_{\Pi \mu}^3 \\ 
B_\mu \\ 
\rho_{P \mu}^0 
\end{array}
\right) 
\,, 
\end{eqnarray}
\begin{eqnarray} 
 {\cal M}_{CC}^2 
 &=& 
 \frac{g^2 v_{\rm EW}^2}{4} 
 \left[
 \begin{array}{c|c} 
 (1 + a) x^2 & - a x \\ 
\hline 
- a x & a   
 \end{array}
 \right]
 \,, \\ 
  {\cal M}_{NC}^2 
 &=& 
 \frac{g^2 v_{\rm EW}^2}{4} 
 \left[
 \begin{array}{c|c|c|c} 
 (1 + a) x^2 & - a x & - (1 -a) t x^2 & 0 \\ 
\hline 
- a x & a & - at x & 0 \\    
\hline 
- t (1 -a) x^2 & - a t x & (1 + \frac{7}{3}a) t^2 x^2 & - \frac{2}{\sqrt{3}} a t x \\ 
\hline 
 0 & 0 & - \frac{2}{\sqrt{3}} a t x & a 
 \end{array}
 \right] 
\end{eqnarray} 
Since $x = g_W/g \sim m_W/M_\rho \ll 1$, one may diagonalize the mass matrices 
by expanding terms in powers of $x$.

Thus the charged sector can be diagonalized by the orthogonal rotation $R_{CC}^{(2)}$, 
\begin{equation}
  R_{CC}^{(2)} 
  = 
  \left[ 
\begin{array}{c|c}    
 1 
& x \\ 
\hline 
 - x & 1 
\end{array}
\right] 
+ {\cal O}(x^2)   
\, , 
\qquad 
 \left( 
 \begin{array}{c} 
 W_\mu^+ \\
 \rho_{\Pi \mu}^+ 
\end{array}
\right) 
=  
  R_{CC}^{(2)} \cdot 
\left( 
 \begin{array}{c} 
 \bar{W}_\mu^+ \\
 \bar{\rho}_{\Pi \mu}^+ 
\end{array}
\right) 
\,, \label{RCC:2}
\end{equation}
with the corresponding eigenvalues 
\begin{eqnarray} 
 m_{\bar{W}}^2 
 &=&  
\frac{g_W^2 v_{\rm EW}^2}{4} 
\left (1 + {\cal O}(x^2) \right) 
= m_W^2 \left (1 + {\cal O}(x^2) \right) 
\,, \label{mass:W} \\ 
M_{\bar{\rho}_\Pi^\pm}^2 
&=& 
 \frac{a g^2 v_{\rm EW}^2}{4} 
\left (1 + {\cal O}(x^2) \right) 
= M_\rho^2 (1 + {\cal O}(x^2) ) 
\,,  \label{mass:rhoc}
\end{eqnarray} 
where we have defined 
\begin{equation} 
m_W^2 = \frac{g_W^2 v_{\rm EW}^2}{4} 
\,, \qquad 
M_\rho^2 = \frac{a g^2 v_{\rm EW}^2}{4}
\,.   
\end{equation}

Similarly for the neutral sector, 
one can easily find 
\begin{equation}
  R_{NC}^{(2)} 
  = 
  \left[ 
\begin{array}{c|c|c|c}    
 c & s & - c_\rho x & s_\rho x \\ 
\hline 
 c(1-t^2) x & 2 sx & c_\rho & - s_\rho \\ 
 \hline 
 - s_\rho & c_\rho & - \left(c_\rho +  \frac{2}{\sqrt{3}} s_\rho \right) t x & - \left( \frac{2}{\sqrt{3}} c_\rho - s_\rho \right) t x \\ 
\hline 
- \frac{2}{\sqrt{3}} ct^2 x & \frac{2}{\sqrt{3}} s x & s_\rho & c_\rho  
\end{array}
\right] 
+ {\cal O}(x^2)   
\, , 
\qquad 
\left( 
 \begin{array}{c} 
 W_\mu^3 \\
 \rho_{\Pi \mu}^3 \\ 
B_\mu \\ 
\rho_{P \mu}^0  
\end{array}
\right)
=  
  R_{NC}^{(2)} \cdot 
\left( 
 \begin{array}{c} 
 \bar{Z}_\mu^3 \\
 \bar{A}_\mu \\ 
 \bar{\rho}_{\Pi \mu}^3 \\ 
\bar{\rho}_{P \mu}^0  
\end{array}
\right) 
\,, \label{RNC:2}
\end{equation}
with the corresponding eigenvalues: 
\begin{eqnarray} 
 m_{\bar{Z}}^2 
 &=&  
\frac{g_W^2 v_{\rm EW}^2}{4c^2} 
\left (1 + {\cal O}(x^2) \right) 
= m_Z^2 \left (1 + {\cal O}(x^2) \right) 
\,, \label{mass:Z:app} 
\\ 
m_{\bar{A}}^2 
&=& 0 
\,, \\ 
M_{\bar{\rho}_\Pi^3}^2 
&=& 
 \frac{a g^2 v_{\rm EW}^2}{4} 
\left (1 + {\cal O}(x^2) \right) 
= M_\rho^2 (1 + {\cal O}(x^2) ) 
\,, 
\\
M_{\bar{\rho}_P^0}^2 
&=& 
 \frac{a g^2 v_{\rm EW}^2}{4} 
\left (1 + {\cal O}(x^2) \right) 
= M_\rho^2 (1 + {\cal O}(x^2) ) 
\,, 
\label{EV_rhoP0}
\end{eqnarray}
where we defined 
\begin{equation} 
 m_Z^2 = \frac{m_W^2}{c^2} = \frac{g_W^2 v_{\rm EW}^2}{4 c^2}
\,. 
\end{equation}
In Eq.(\ref{RNC:2}) 
we have defined the mixing angles as 
\begin{eqnarray}
t &=& \frac{g_W}{g_Y} 
\,, \qquad 
c = \frac{1}{\sqrt{1+t^2}}
\,, \qquad 
s = \sqrt{1 - c^2} 
\,, \\  
c_\rho &=& \frac{1}{\sqrt{2}} \sqrt{ \frac{(3-t^2) + \sqrt{ (3-t^2)^2 + 48 t^4 }}{\sqrt{(3 -t^2)^2 + 48t^4}} }
\,, \qquad 
s_\rho = \sqrt{1 - c_\rho^2}
\,. 
\end{eqnarray}

\subsection{Diagonalization at ${\cal O}(p^4)$}

Inclusion of the ${\cal O}(p^4)$ terms gives rise to 
the kinetic term mixing through the $z_3$ term in Eq.(\ref{Lag:start:4}) as 
\begin{eqnarray}
 {\cal L}_{\rm kin}^{(2)+(4)}
 &=& 
 - 
  \left( 
 \begin{array}{c} 
 \bar{W}_\mu^+ \\
 \bar{\rho}_{\Pi \mu}^+ 
\end{array}
\right)^T \cdot D^{\mu\nu} \cdot 
{\cal K}_{CC} 
\cdot 
 \left( 
 \begin{array}{c} 
 \bar{W}_\nu^- \\
 \bar{\rho}_{\Pi \nu}^- 
\end{array}
\right) 
- 
\frac{1}{2}
\left( 
 \begin{array}{c} 
 \bar{A}_\mu \\ 
 \bar{Z}_\mu \\
 \bar{\rho}_{\Pi \mu}^3 \\ 
\bar{\rho}_{P \mu}^0  
\end{array}
\right)^T \cdot D^{\mu\nu} \cdot 
{\cal K}_{NC} 
\cdot 
\left( 
 \begin{array}{c} 
 \bar{A}_\nu \\ 
 \bar{Z}_\nu  \\
 \bar{\rho}_{\Pi \nu}^3 \\ 
\bar{\rho}_{P \nu}^0  
\end{array}
\right) 
\,, 
\end{eqnarray} 
where $D_{\mu\nu} = - \partial^2 g_{\mu\nu} + \partial_\mu\partial_\nu $ and 
\begin{eqnarray} 
{\cal K}_{CC} 
&=& 
\left[ 
\begin{array}{c|c} 
1   & - g^2 z_3 x \\ 
\hline 
-g^2 z_3 x & 1  
\end{array}
\right] 
+ {\cal O}(x^2) 
\,, 
\\ 
{\cal K}_{NC} 
&=& 
\left[ 
\begin{array}{c|c|c|c} 
1  &  0 & K_{A \rho_\Pi} & K_{A \rho_P} \\ 
\hline 
 0  & 1 & K_{Z \rho_\Pi} & K_{Z \rho_P} \\ 
\hline 
 K_{A \rho_\Pi}  & K_{Z \rho_\Pi}  & 1 & 0 \\
\hline 
 K_{A \rho_P}  & K_{Z\rho_P}  & 0  & 1    
\end{array}
\right] 
+ {\cal O}(x^2) 
\,, 
\end{eqnarray}
 with 
\begin{eqnarray} 
 K_{A\rho_\Pi} &=& 
- 2s \left( c_\rho + \frac{1}{\sqrt{3}} s_\rho  \right) x  
\,, \\ 
 K_{A\rho_P} &=& 
 2s \left( s_\rho - \frac{1}{\sqrt{3}} c_\rho  \right) x  
\,, \\ 
 K_{Z \rho_\Pi} &=& 
- c \left( c_\rho - \left( c_\rho + \frac{2}{\sqrt{3}} s_\rho \right) t^2  \right) x  
\,, \\ 
 K_{Z \rho_P} &=& 
 c \left( s_\rho - \left( s_\rho - \frac{2}{\sqrt{3}} c_\rho \right) t^2  \right) x  
\,. 
\end{eqnarray}

The kinetic term mixing for the charged sector can be diagonalized by 
the orthogonal rotation ${\cal O}_{CC}$: 
\begin{equation} 
{\cal O}_{CC} = \frac{1}{\sqrt{2}} 
\left[
\begin{array}{c|c} 
 1 + \frac{x}{2} & 1 - \frac{x}{2} \\
\hline 
- 1 + \frac{x}{2} & 1 + \frac{x}{2} 
\end{array} 
\right] 
+ {\cal O}(x^2) 
\,, \qquad 
 \left( 
 \begin{array}{c} 
 \bar{W}_\nu^- \\
 \bar{\rho}_{\Pi \nu}^- 
\end{array}
\right) 
= 
{\cal O}_{CC} 
   \left( 
 \begin{array}{c} 
 \hat{W}_\nu^- \\
 \hat{\rho}_{\Pi \nu}^- 
\end{array}
\right) 
\,,\label{OCC}  
\end{equation}
with the  eigenvalues 
\begin{equation} 
 k_W = 1 - g^2 z_3 x
 \,, \qquad 
 k_{\rho_\Pi^\pm} = 1 + g^2 z_3 x 
\,.  
\end{equation}  
 After canonically normalizing the fields as $\hat{W}(\hat{\rho}_\Pi) \to \sqrt{k_{W(\rho_\Pi)}} \hat{W}(\hat{\rho}_\Pi) $,  
 one gets the mass matrix for the charged sector at the ${\cal O}(p^4)$, 
\begin{eqnarray} 
 {\cal M}_{CC}^2 \Bigg|_{{\cal O}(p^4)} 
 &=& \frac{g^2 v_{\rm EW}^2}{4} 
\left( 
{\cal O}_{CC}^T \cdot 
\left( 
\begin{array}{c|c} 
 1+ \frac{g^2 z_3}{2}x & 0 \\ 
\hline 
 0 & 1 - \frac{g^2 z_3}{2}x 
\end{array}
\right) 
\right)^T 
\cdot \left[
\begin{array}{c|c} 
 x^2 & 0 \\  
\hline 
0 & (1+x^2) a 
\end{array}
\right]  
\nonumber \\ 
&& 
\times 
\left( 
{\cal O}_{CC}^T \cdot 
\left( 
\begin{array}{c|c} 
 1+ \frac{g^2 z_3}{2}x & 0 \\ 
\hline 
 0 & 1 - \frac{g^2 z_3}{2} x 
\end{array}
\right) 
\right) 
\,.  
\end{eqnarray} 
 This can be diagonalized by the orthogonal rotation, 
\begin{equation} 
U_{CC} = 
\frac{1}{\sqrt{2}}
\left( 
\begin{array}{c|c} 
  -1 + \frac{g^2 z_3 - 1}{2} x & 1 + \frac{g^2z_3 -1}{2} x \\ 
\hline 
1 + \frac{g^2z_3 -1}{2} x & 1- \frac{g^2 z_3 -1}{2} x 
\end{array}
\right) 
+ {\cal O}(x^2) 
\,, \qquad 
\left( 
\begin{array}{c}
\hat{W}^\pm_\mu  \\ 
\hat{\rho}^\pm_{\Pi \mu}
\end{array} 
\right) 
= 
U_{CC} \cdot 
\left( 
\begin{array}{c}
\tilde{W}^\pm_\mu  \\ 
\tilde{\rho}^\pm_{\Pi \mu}
\end{array} 
\right)
\,,  \label{UCC}
\end{equation} 
 with the mass eigenvalues which are the same as in Eqs.(\ref{mass:W}) and (\ref{mass:rhoc}) up to corrections of ${\cal O}(x^2)$. 
Using Eqs.(\ref{RCC:2}), (\ref{OCC}) and (\ref{UCC}), 
we see that in the charged sector the gauge eigenstates $(W_\mu^\pm, \rho_{\Pi \mu}^\pm)$ 
are related to the mass eigenstates  $(\tilde{W}_\mu^\pm, \tilde{\rho}_{\Pi \mu}^\pm)$  at ${\cal O}(p^4)$ as 
\begin{equation} 
\left( 
\begin{array}{c} 
W_\mu^\pm \\ 
\rho_{\Pi \mu}^\pm
\end{array}
\right) 
= 
\left[ 
\begin{array}{c|c}  
-1 & - (1 - g^2 z_3) x \\ 
\hline 
- x & 1 
\end{array} 
\right] 
\left( 
\begin{array}{c} 
\tilde{W}_\mu^\pm \\ 
\tilde{\rho}_{\Pi \mu}^\pm
\end{array}
\right) 
\,, \label{wavefunc:CC}
\end{equation} 
up to components of ${\cal O}(x^2)$. 
Note that this transformation is not orthogonal because both 
the kinetic and mass mixing cannot simultaneously be diagonalized by the same 
orthogonal rotation matrix.


Similarly for the neutral sector, 
we see that the gauge eigenstates $(W_\mu^3, \rho_{\Pi \mu}^3, B_\mu, \rho_P^0)$ 
are related to the mass eigenstates  $(\tilde{Z}_\mu, \tilde{A}_{\mu}, \tilde{\rho}_{\Pi \mu}^3, \tilde{\rho}_{P \mu}^0)$  at ${\cal O}(p^4)$ as 
\begin{equation} 
\left( 
\begin{array}{c} 
W_\mu^3 \\ 
\rho_{\Pi \mu}^3 \\ 
B_\mu \\ 
\rho_{P \mu}^0 
\end{array}
\right) 
= 
\left[ 
\begin{array}{c|c|c|c}  
c & s & -(1 - g^2 z_3)c_\rho x & (1 - g^2 z_3) s_\rho x \\  
\hline 
c (1 - t^2) x & 2 s x & c_\rho & - s_\rho \\ 
\hline 
- s & c & - (1 - g^2 z_3) (c_\rho + \frac{2}{\sqrt{3}} s_\rho) t x & (1 - g^2 z_3) (s_\rho - \frac{2}{\sqrt{3}} c_\rho) t x \\ 
\hline 
- \frac{2}{\sqrt{3}} c t^2 x & \frac{2}{\sqrt{3}} s x & s_\rho & c_\rho  
\end{array} 
\right] 
\left( 
\begin{array}{c} 
\tilde{Z}_\mu \\ 
\tilde{A}_\mu \\ 
\tilde{\rho}_{\Pi \mu}^3 \\ 
\tilde{\rho}_{P \mu}^0 
\end{array}
\right) 
\,, \label{wavefunc:NC}
\end{equation} 
up to components of ${\cal O}(x^2)$. 
The mass eigenvalues are the same as those given in Eqs.(\ref{mass:Z:app}) - (\ref{EV_rhoP0}) up to corrections of ${\cal O}(x^2)$. 
Note, again, that the transformation in Eq.(\ref{wavefunc:NC})  is not orthogonal because both 
the kinetic and mass mixing cannot simultaneously be diagonalized by the same 
orthogonal rotation matrix.


\section{The partial decay widths} 
\label{decay}

In this Appendix we give the analytic formulae for the $\tilde{\rho}_\Pi^{\pm, 3}$ 
and $\tilde{\rho}_P^0$ decay widths.

\begin{itemize} 

\item 
The decays to the SM fermions:  
\begin{eqnarray} 
 \Gamma(\tilde{\rho}_\Pi \to \psi \bar{\psi}) 
 &=& \frac{N_{c}^{(\psi)}}{12 \pi} \left[
(A_V^\psi)^2 \left( 1 + \frac{2 m_{\psi}^2}{M_\rho^2} \right) 
+ 
(A_A^\psi)^2 \left( 1 - \frac{4 m_{\psi}^2}{M_\rho^2} \right) 
\right] 
\sqrt{ M_\rho^2 - 4 m_\psi^2} 
\,, \\ 
 \Gamma(\tilde{\rho}_P \to \psi \bar{\psi}) 
 &=& \frac{N_{c}^{(\psi)}}{12 \pi} \left[
(B_V^\psi)^2 \left( 1 + \frac{2 m_{\psi}^2}{M_\rho^2} \right) 
+ 
(B_A^\psi)^2 \left( 1 - \frac{4 m_{\psi}^2}{M_\rho^2} \right) 
\right] 
\sqrt{ M_\rho^2 - 4 m_\psi^2} 
\,, \\ 
\Gamma(\tilde{\rho}_\Pi^+ \to \psi_{u} \bar{\psi}_d) 
&=& \frac{N_c^{(\psi)}}{48 \pi} 
|c_L|^2 
\frac{2 M_\rho^4 - (m_{\psi u}^2 + m_{\psi_d}^2) M_\rho^2 - (m_{\psi_u}^2 - m_{\psi_d}^2)^2}{M_\rho^4} 
\nonumber \\ 
&& 
\times 
\frac{\sqrt{ (M_\rho^2 - (m_{\psi_u} + m_{\psi_d})^2)(M_\rho^2 - (m_{\psi_u} - m_{\psi_d})^2) }}{M_\rho} 
\,, 
\end{eqnarray}

\item 
The decays to weak bosons: 
\begin{eqnarray} 
\Gamma(\tilde{\rho}_{\Pi,P} \to \tilde{W}\tilde{W}) 
&=& 
\frac{1}{192 \pi} \left( \frac{M_\rho}{m_W} \right)^4 \frac{( M_\rho^2  - 4 m_W^2)^{3/2}}{M_\rho^2} 
\nonumber \\ 
&&
\times 
\Bigg[ 
 12 g_{\tilde{\rho} \tilde{W}\tilde{W}} g_{\tilde{W}\tilde{W}\tilde{\rho}} \left( \frac{m_W}{M_\rho}  \right)^2 
 + 
 g_{\tilde{\rho} \tilde{W}\tilde{W}}^2 
 \left( 1 + 4 \left( \frac{m_W}{M_\rho} \right)^2  \right) 
\nonumber\\ 
&& + 
 4 g_{\tilde{W}\tilde{W} \tilde{\rho}}^2 \left\{ 
 \left( \frac{m_W}{M_\rho} \right)^2 + 3 \left( \frac{m_W}{M_\rho} \right)^4 
 \right\}
\Bigg] \,, 
\\
\Gamma(\tilde{\rho}_\Pi^\pm \to \tilde{W}^\pm \tilde{Z}) 
&=& 
\frac{1}{192 \pi} \left( \frac{M_\rho^2}{m_W m_Z} \right)^2 \frac{ \left[ (M_\rho^2 - (m_W+m_Z)^2)(M_\rho^2 - (m_W - m_Z)^2) \right]^{3/2}}{M_\rho^5}
\nonumber \\ 
&& 
\times 
\Bigg[ 
 6 \Bigg\{  g_{\tilde{\rho}_\Pi \tilde{W}\tilde{Z}} g_{\tilde{Z}\tilde{W}\tilde{\rho}_\Pi} \left( \frac{m_Z}{M_\rho} \right)^2  
 - g_{\tilde{\rho}_\Pi \tilde{W}\tilde{Z}} g_{\tilde{W}\tilde{Z}\tilde{\rho}_\Pi} \left( \frac{m_W}{M_\rho} \right)^2 
\nonumber \\  
&&
- g_{\tilde{W} \tilde{Z}\tilde{\rho}_\Pi} g_{\tilde{Z} \tilde{W}\tilde{\rho}_\Pi} \left( \frac{m_W}{M_\rho} \right)^2 \left( \frac{m_Z}{M_\rho} \right)^2
\Bigg\} 
\nonumber \\ 
&& 
+ 
g_{\tilde{\rho}_\Pi \tilde{W}\tilde{Z}}^2 
\left\{ 
 1 + 2 \left( \frac{m_W}{M_\rho} \right)^2 + 2 \left( \frac{m_Z}{M_\rho} \right)^2
\right\}  
\nonumber \\ 
&& + 
g_{\tilde{W} \tilde{Z}\tilde{\rho}_\Pi}^2 
\left( \frac{m_W}{M_\rho} \right)^2
\left\{ 
 2 +  \left( \frac{m_W}{M_\rho} \right)^2 + 2 \left( \frac{m_Z}{M_\rho} \right)^2
\right\} 
\nonumber \\ 
&& 
+ 
g_{\tilde{Z} \tilde{W}\tilde{\rho}_\Pi}^2 
\left( \frac{m_Z}{M_\rho} \right)^2
\left\{ 
 2 +  2 \left( \frac{m_W}{M_\rho} \right)^2 +  \left( \frac{m_Z}{M_\rho} \right)^2
\right\} 
\Bigg] 
\,, 
\end{eqnarray}

\end{itemize}

\section{The breakdown of signal events}
\label{breakdown_Signal}

In this Appendix we give a list of tables (\ref{cut_eff_grhopipi3}, \ref{cut_eff_grhopipi4} and \ref{cut_eff_grhopipi5})  
to present the number of events for the 2 TeV walking technirho signals 
after each cut selection applied described in the text.

\begin{table}[ht]
\centering
\begin{tabular}[t]{|c||c|c||c|c||c|c||c|c|}
\hline
\parbox[c][4ex][c]{0ex}{}
$g_{\rho \pi \pi} =3$
& \multicolumn{2}{ c| |}{$F_{\rho} = 650 \,\GeV$}
& \multicolumn{2}{ c| |}{$F_{\rho} = 550 \,\GeV$}
& \multicolumn{2}{ c| |}{$F_{\rho} = 450 \,\GeV$}
& \multicolumn{2}{ c| }{$F_{\rho} = 250 \,\GeV$} 
\\ \hline \hline
\parbox[c][4ex][c]{0ex}{}
&$\tilde{\rho}_{_\Pi}^\pm$ &  $\tilde{\rho}_{_\Pi}^3+ \tilde{\rho}^0_{_P}$ 
&$\tilde{\rho}_{_\Pi}^\pm$ &  $\tilde{\rho}_{_\Pi}^3+ \tilde{\rho}^0_{_P}$ 
&$\tilde{\rho}_{_\Pi}^\pm$ &  $\tilde{\rho}_{_\Pi}^3+ \tilde{\rho}^0_{_P}$
&$\tilde{\rho}_{_\Pi}^\pm$ &  $\tilde{\rho}_{_\Pi}^3+ \tilde{\rho}^0_{_P}$ \\ \hline 
\parbox[c][4ex][c]{0ex}{}
Total events 
& $68.77$ & $38.30$ & $50.38$ & $28.28$ & $34.41$ & $19.46$ & $10.95$ & $6.28$
\\ \hline \hline
\parbox[c][4ex][c]{0ex}{}
After Cut 1
& $48.50$ & $27.13$ & $35.57$ & $20.00$ & $24.29$ & $13.66$ & $7.79$ & $4.45$
\\ \hline 
\parbox[c][4ex][c]{0ex}{}
After Cut 2
& $47.87$ & $26.83$ & $35.12$ & $19.75$ & $24.03$ & $13.48$ & $7.69$ & $4.40$
\\ \hline
\parbox[c][4ex][c]{0ex}{}
After Cut 3 (i)
& $15.93$ & $9.74$ & $11.71$ & $7.31$ & $8.00$ & $4.87$ & $2.72$ & $1.58$
\\ \hline \hline
\parbox[c][4ex][c]{0ex}{}
$1050 \,\GeV \leq m_{jj} \leq 3550\,\GeV$
& \multicolumn{8}{ c |}{}
\\ \hline 
\parbox[c][4ex][c]{0ex}{}
After Cut 3 (ii) ($WW = $ A+B+C )
& $8.60$ & $6.76$ & $6.28$ & $5.10$ & $4.19$ & $3.47$ & $1.46$ & $1.12$
\\ \hline 
\parbox[c][4ex][c]{0ex}{}
After Cut 3 (ii) ($WZ = $ B+C+D+E ) 
& $11.66$ & $5.89$ & $8.67$ & $4.56$ & $5.85$ & $3.05$ & $2.00$ & $0.98$
\\ \hline 
\parbox[c][4ex][c]{0ex}{}
After Cut 3 (ii) ($ZZ = $ C+E+F) 
& $6.37$ & $2.37$ & $4.60$ & $1.74$ & $3.22$ & $1.27$ & $1.12$ & $0.38$
\\ \hline \hline
\parbox[c][4ex][c]{0ex}{}
$1850 \,\GeV \leq m_{jj} \leq 2150\,\GeV$
& \multicolumn{8}{ c |}{}
\\ \hline
\parbox[c][4ex][c]{0ex}{}
After Cut 3 (ii) ($WW = $ A+B+C )
& $7.49$ & $6.19$ & $5.63$ & $4.71$ & $3.73$ & $3.19$ & $1.32$ & $1.03$
\\ \hline 
\parbox[c][4ex][c]{0ex}{}
After Cut 3 (ii) ($WZ = $ B+C+D+E ) 
& $10.25$ & $5.41$ & $7.77$ & $4.18$ & $5.25$ & $2.80$ & $1.80$ & $0.90$
\\ \hline 
\parbox[c][4ex][c]{0ex}{}
After Cut 3 (ii) ($ZZ = $ C+E+F) 
& $5.64$ & $2.13$ & $4.12$ & $1.59$ & $2.89$ & $1.18$ & $1.01$ & $0.33$
\\ \hline
\end{tabular}
\caption{The number of events for $g_{\rho \pi \pi} =3$.}\label{cut_eff_grhopipi3}
\end{table}

\begin{table}[ht]
\centering
\begin{tabular}[t]{|c||c|c||c|c||c|c||c|c|}
\hline
\parbox[c][4ex][c]{0ex}{}
$g_{\rho \pi \pi} =4$
& \multicolumn{2}{ c| |}{$F_{\rho} = 650 \,\GeV$}
& \multicolumn{2}{ c| |}{$F_{\rho} = 550 \,\GeV$}
& \multicolumn{2}{ c| |}{$F_{\rho} = 450 \,\GeV$}
& \multicolumn{2}{ c| }{$F_{\rho} = 250 \,\GeV$} 
\\ \hline \hline
\parbox[c][4ex][c]{0ex}{}
&$\tilde{\rho}_{_\Pi}^\pm$ &  $\tilde{\rho}_{_\Pi}^3+ \tilde{\rho}^0_{_P}$ 
&$\tilde{\rho}_{_\Pi}^\pm$ &  $\tilde{\rho}_{_\Pi}^3+ \tilde{\rho}^0_{_P}$ 
&$\tilde{\rho}_{_\Pi}^\pm$ &  $\tilde{\rho}_{_\Pi}^3+ \tilde{\rho}^0_{_P}$
&$\tilde{\rho}_{_\Pi}^\pm$ &  $\tilde{\rho}_{_\Pi}^3+ \tilde{\rho}^0_{_P}$ \\ \hline 
\parbox[c][4ex][c]{0ex}{}
Total events 
& $70.47$ & $39.80$ & $51.18$ & $29.05$ & $34.63$ & $19.77$ & $10.91$ & $6.28$
\\ \hline \hline
\parbox[c][4ex][c]{0ex}{}
After Cut 1
& $49.49$ & $28.45$ & $36.70$ & $20.79$ & $24.60$ & $14.04$ & $7.76$ & $4.44$
\\ \hline 
\parbox[c][4ex][c]{0ex}{}
After Cut 2
& $48.97$ & $28.06$ & $36.31$ & $20.51$ & $24.30$ & $13.87$ & $7.66$ & $4.39$
\\ \hline
\parbox[c][4ex][c]{0ex}{}
After Cut 3 (i)
& $16.21$ & $10.09$ & $12.27$ & $7.52$ & $8.49$ & $5.14$ & $2.60$ & $1.58$
\\ \hline \hline
\parbox[c][4ex][c]{0ex}{}
$1050 \,\GeV \leq m_{jj} \leq 3550\,\GeV$
& \multicolumn{8}{ c |}{}
\\ \hline
\parbox[c][4ex][c]{0ex}{}
After Cut 3 (ii) ($WW = $ A+B+C )
& $8.56$ & $7.18$ & $6.36$ & $5.26$ & $4.55$ & $3.60$ & $1.36$ & $1.13$
\\ \hline 
\parbox[c][4ex][c]{0ex}{}
After Cut 3 (ii) ($WZ = $ B+C+D+E ) 
& $11.66$ & $6.13$ & $8.82$ & $4.69$ & $6.26$ & $3.18$ & $1.87$ & $0.98$
\\ \hline 
\parbox[c][4ex][c]{0ex}{}
After Cut 3 (ii) ($ZZ = $ C+E+F) 
& $6.39$ & $2.43$ & $4.80$ & $1.81$ & $3.25$ & $1.29$ & $1.04$ & $0.40$
\\ \hline \hline
\parbox[c][4ex][c]{0ex}{}
$1850 \,\GeV \leq m_{jj} \leq 2150\,\GeV$
& \multicolumn{8}{ c |}{}
\\ \hline
\parbox[c][4ex][c]{0ex}{}
After Cut 3 (ii) ($WW = $ A+B+C )
& $7.13$ & $6.30$ & $5.36$ & $4.56$ & $3.92$ & $3.12$ & $1.18$ & $0.97$
\\ \hline 
\parbox[c][4ex][c]{0ex}{}
After Cut 3 (ii) ($WZ = $ B+C+D+E ) 
& $9.84$ & $5.39$ & $7.48$ & $4.08$ & $5.44$ & $2.76$ & $1.61$ & $0.85$
\\ \hline 
\parbox[c][4ex][c]{0ex}{}
After Cut 3 (ii) ($ZZ = $ C+E+F) 
& $5.41$ & $2.11$ & $4.12$ & $1.58$ & $2.82$ & $1.08$ & $0.89$ & $0.34$
\\ \hline
\end{tabular}
\caption{The number of events for $g_{\rho \pi \pi} =4$.}\label{cut_eff_grhopipi4}
\end{table}

\begin{table}[ht]
\centering
\begin{tabular}[t]{|c||c|c||c|c||c|c||c|c|}
\hline
\parbox[c][4ex][c]{0ex}{}
$g_{\rho \pi \pi} =5$
& \multicolumn{2}{ c| |}{$F_{\rho} = 650 \,\GeV$}
& \multicolumn{2}{ c| |}{$F_{\rho} = 550 \,\GeV$}
& \multicolumn{2}{ c| |}{$F_{\rho} = 450 \,\GeV$}
& \multicolumn{2}{ c| }{$F_{\rho} = 250 \,\GeV$} 
\\ \hline \hline
\parbox[c][4ex][c]{0ex}{}
&$\tilde{\rho}_{_\Pi}^\pm$ &  $\tilde{\rho}_{_\Pi}^3+ \tilde{\rho}^0_{_P}$ 
&$\tilde{\rho}_{_\Pi}^\pm$ &  $\tilde{\rho}_{_\Pi}^3+ \tilde{\rho}^0_{_P}$ 
&$\tilde{\rho}_{_\Pi}^\pm$ &  $\tilde{\rho}_{_\Pi}^3+ \tilde{\rho}^0_{_P}$
&$\tilde{\rho}_{_\Pi}^\pm$ &  $\tilde{\rho}_{_\Pi}^3+ \tilde{\rho}^0_{_P}$ \\ \hline 
\parbox[c][4ex][c]{0ex}{}
Total events 
& $70.85$ & $40.34$ & $51.15$ & $29.24$ & $34.41$ & $19.79$ & $10.78$ & $6.23$
\\ \hline \hline
\parbox[c][4ex][c]{0ex}{}
After Cut 1
& $50.73$ & $28.72$ & $36.60$ & $21.09$ & $24.44$ & $14.06$ & $7.64$ & $4.45$
\\ \hline 
\parbox[c][4ex][c]{0ex}{}
After Cut 2
& $50.05$ & $28.26$ & $36.14$ & $20.81$ & $24.14$ & $13.89$ & $7.54$ & $4.39$
\\ \hline
\parbox[c][4ex][c]{0ex}{}
After Cut 3 (i)
& $17.17$ & $10.55$ & $12.57$ & $7.60$ & $8.29$ & $4.96$ & $2.59$ & $1.56$
\\ \hline \hline
\parbox[c][4ex][c]{0ex}{}
$1050 \,\GeV \leq m_{jj} \leq 3550\,\GeV$
& \multicolumn{8}{ c |}{}
\\ \hline
\parbox[c][4ex][c]{0ex}{}
After Cut 3 (ii) ($WW = $ A+B+C )
& $8.95$ & $7.42$ & $6.27$ & $5.28$ & $4.29$ & $3.48$ & $1.36$ & $1.07$
\\ \hline 
\parbox[c][4ex][c]{0ex}{}
After Cut 3 (ii) ($WZ = $ B+C+D+E ) 
& $12.48$ & $6.59$ & $9.16$ & $4.84$ & $6.16$ & $3.03$ & $1.92$ & $0.96$
\\ \hline 
\parbox[c][4ex][c]{0ex}{}
After Cut 3 (ii) ($ZZ = $ C+E+F) 
& $6.73$ & $2.46$ & $4.87$ & $1.99$ & $3.34$ & $1.19$ & $1.08$ & $0.38$
\\ \hline \hline
\parbox[c][4ex][c]{0ex}{}
$1850 \,\GeV \leq m_{jj} \leq 2150\,\GeV$
& \multicolumn{8}{ c |}{}
\\ \hline
\parbox[c][4ex][c]{0ex}{}
After Cut 3 (ii) ($WW = $ A+B+C )
& $7.17$ & $6.24$ & $5.11$ & $4.36$ & $3.49$ & $2.88$ & $1.09$ & $0.89$
\\ \hline 
\parbox[c][4ex][c]{0ex}{}
After Cut 3 (ii) ($WZ = $ B+C+D+E ) 
& $10.04$ & $5.57$ & $7.29$ & $3.98$ & $4.97$ & $2.50$ & $1.53$ & $0.79$
\\ \hline 
\parbox[c][4ex][c]{0ex}{}
After Cut 3 (ii) ($ZZ = $ C+E+F) 
& $5.30$ & $2.03$ & $3.90$ & $1.61$ & $2.64$ & $1.00$ & $0.86$ & $0.31$
\\ \hline
\end{tabular}
\caption{The number of events for $g_{\rho \pi \pi} =5$.}\label{cut_eff_grhopipi5}
\end{table}

\end{document}